\documentclass[equations,11pt]{article}

\renewcommand{\theequation}{\arabic{equation}}
\usepackage{amssymb,amsmath}
\usepackage{color}
\usepackage{accents}
\usepackage{eucal}
\usepackage{slashed}

\begin{document}
\bibliographystyle{plain}
\def\m@th{\mathsurround=0pt}
\mathchardef\bracell="0365
\def\upbrall{$\m@th\bracell$}
\def\undertilde#1{\mathop{\vtop{\ialign{##\crcr
    $\hfil\displaystyle{#1}\hfil$\crcr
     \noalign
     {\kern1.5pt\nointerlineskip}
     \upbrall\crcr\noalign{\kern1pt
   }}}}\limits}
\def\theequation{\arabic{section}.\arabic{equation}}
\newcommand{\ar}{\alpha}
\newcommand{\aar}{\bar{a}}
\newcommand{\bb}{\beta}
\newcommand{\gm}{\gamma}
\newcommand{\Gm}{\Gamma}
\newcommand{\en}{\ven}
\newcommand{\ven}{\varepsilon}
\newcommand{\dd}{\delta}
\newcommand{\sg}{\sigma}
\newcommand{\kp}{\kappa}
\newcommand{\ld}{\lambda}
\newcommand{\oa}{\omega}
\newcommand{\be}{\begin{equation}}
\newcommand{\ee}{\end{equation}}
\newcommand{\bea}{\begin{eqnarray}}
\newcommand{\eea}{\end{eqnarray}}
\newcommand{\bse}{\begin{subequations}}
\newcommand{\ese}{\end{subequations}}
\newcommand{\nn}{\nonumber}
\newcommand{\bR}{\bar{R}}
\newcommand{\bP}{\bar{\Phi}}
\newcommand{\bS}{\bar{S}}
\newcommand{\bW}{\bar{W}}
\newcommand{\vf}{\varphi}
\newcommand{\sn}{{\rm sn}}
\newcommand{\wh}{\widehat}
\newcommand{\ol}{\overline}
\newcommand{\wt}{\widetilde}
\newcommand{\ut}{\undertilde}
\newcommand{\ip}{{i^\prime}}
\newcommand{\jp}{{j^\prime}}
\newcommand{\cn}{{\rm cn}}
\newcommand{\dn}{{\rm dn}}
\newcommand{\bun}{\boldsymbol{1}}
\newcommand{\Ld}{{\boldsymbol \Lambda}}
\newcommand{\tLd}{\,^{t\!}{\boldsymbol \Lambda}}
\newcommand{\bI}{{\boldsymbol I}}
\newcommand{\bO}{{\boldsymbol O}}
\newcommand{\bU}{{\boldsymbol U}}
\newcommand{\bC}{{\boldsymbol C}}
\newcommand{\bOm}{{\boldsymbol \Omega}}
\newcommand{\buk}{{\boldsymbol u}_\kappa}
\newcommand{\bul}{{\boldsymbol u}_\ell}
\newcommand{\tII}{\,^{t\!}{\boldsymbol I}}
\newcommand{\tuk}{\,^{t\!}{\boldsymbol u}_{\kappa^\prime}}
\newcommand{\thh}{\,^{t\!}h}
\newcommand{\tul}{\,^{t\!}{\boldsymbol u}_{\ell^\prime}}
\newcommand{\tull}{\,^{t\!}{\boldsymbol u}_{-\ell+\ld}}
\newcommand{\tuq}{\,^{t\!}{\boldsymbol u}_{-q_j+\ld}}
\newcommand{\tcl}{\,^{t\!}{\boldsymbol c}_{\ell}}
\newcommand{\tclp}{\,^{t\!}{\boldsymbol c}_{\ell^\prime}}
\newcommand{\tck}{\,^{t\!}{\boldsymbol c}_{\kp^\prime}}
\newcommand{\ssk}{\sigma_{\kappa^\prime}}
\newcommand{\ssl}{\sigma_{\ell^\prime}}
\newcommand{\pte}{(\partial_t-\partial_\eta)}
\newcommand{\pxe}{(\partial_x-\partial_\eta)}
\newcommand{\dint}{\int_\Gamma d\mu(\ell) }
\def\hypotilde#1#2{\vrule depth #1 pt width 0pt{\smash{{\mathop{#2}
\limits_{\displaystyle\widetilde{}}}}}}
\def\hypohat#1#2{\vrule depth #1 pt width 0pt{\smash{{\mathop{#2}
\limits_{\displaystyle\widehat{}}}}}}
\def\hypo#1#2{\vrule depth #1 pt width 0pt{\smash{{\mathop{#2}
\limits_{\displaystyle{}}}}}}

\newcommand{\bblu}{\begin{color}{blue}}
\newcommand{\bred}{\begin{color}{red}}
\newcommand{\bgreen}{\begin{color}{green}}
\newcommand{\ecl}{\end{color}}

\newcommand{\Ups}{\Upsilon}
\newcommand{\bA}{\boldsymbol{A}}
\newcommand{\bB}{\boldsymbol{B}}
\newcommand{\bD}{\boldsymbol{D}}
\newcommand{\bE}{\boldsymbol{E}}
\newcommand{\bF}{\boldsymbol{F}}
\newcommand{\bG}{\boldsymbol{G}}
\newcommand{\bH}{\boldsymbol{H}}
\newcommand{\bJ}{\boldsymbol{J}}
\newcommand{\bK}{\boldsymbol{K}}
\newcommand{\bL}{\boldsymbol{L}}
\newcommand{\bM}{\boldsymbol{M}}
\newcommand{\bN}{\boldsymbol{N}}
\newcommand{\bQ}{\boldsymbol{Q}}
\newcommand{\bT}{\boldsymbol{T}}
\newcommand{\bV}{\boldsymbol{V}}
\newcommand{\bsX}{\boldsymbol{X}}
\newcommand{\bsY}{\boldsymbol{Y}}
\newcommand{\bZ}{\boldsymbol{Z}}

\newcommand{\mbe}{{\boldsymbol e}}
\newcommand{\pl}{\partial}
\newcommand{\bnab}{{\boldsymbol \nabla}}
\newcommand{\bu}{\boldsymbol{u}}
\newcommand{\bv}{{\boldsymbol v}}
\newcommand{\ba}{{\boldsymbol a}}
\newcommand{\bbb}{\boldsymbol{b}}
\newcommand{\bc}{\boldsymbol{c}}
\newcommand{\bd}{\boldsymbol{d}}
\newcommand{\bme}{\boldsymbol{e}}
\newcommand{\bff}{\boldsymbol{f}}
\newcommand{\bk}{\boldsymbol{k}}
\newcommand{\bl}{\boldsymbol{l}}
\newcommand{\brr}{\boldsymbol{r}}
\newcommand{\bw}{{\boldsymbol w}}
\newcommand{\mbx}{{\boldsymbol x}}
\newcommand{\mby}{{\boldsymbol y}}
\newcommand{\bz}{{\boldsymbol z}}
\newcommand{\bp}{{\boldsymbol p}}
\newcommand{\bs}{{\boldsymbol s}}
\newcommand{\btt}{{\boldsymbol t}}
\newcommand{\bmm}{{\boldsymbol m}}
\newcommand{\bdd}{{\boldsymbol \delta}}
\newcommand{\bze}{{\boldsymbol 0}}
\newcommand{\boma}{{\boldsymbol \omega}}
\newcommand{\bet}{{\boldsymbol \eta}}
\newcommand{\bphi}{{\boldsymbol \phi}}
\newcommand{\bpsi}{{\boldsymbol \psi}}
\newcommand{\bkp}{{\boldsymbol \kappa}}
\newcommand{\bxi}{{\boldsymbol \xi}}
 \newcommand{\mbv}{\boldmath{v}}
 \newcommand{\mbxi}{\boldmath{\xi}}
 \newcommand{\mbeta}{\boldmath{\eta}}
 \newcommand{\mbw}{\boldmath{w}}
 \newcommand{\mbu}{\boldmath{u}}
\newcommand{\tbphi}{\,^{t\!}\boldsymbol{\bphi}}

\begin{center}
{\Large{\bf Elliptic solutions of Boussinesq type lattice equations
and the elliptic $N^{\rm th}$ root of unity} } \\
\vspace{.2cm}
\emph{ Frank W Nijhoff, Ying-Ying Sun and  Da-Jun Zhang}

\it{
Frank W Nijhoff, School of Mathematics, University of Leeds, Leeds LS2 9JT, United Kingdom\\
Ying-Ying Sun, Department of Mathematics, University of Shanghai for Science and Technology, Shanghai 200093, China\\
Da-Jun Zhang, Department of Mathematics,  Shanghai University, Shanghai 200444, China}
\end{center}

{\bf Abstract:} We establish an infinite family of solutions in terms of elliptic functions of the lattice Boussinesq  
systems by setting up a direct linearisation scheme, which provides the solution structure for those equations in the 
elliptic case. The latter, which contains as main structural element a Cauchy kernel on the torus, is obtained from 
a dimensional reduction of the elliptic direct linearisation scheme of the lattice Kadomtsev-Petviashvili equation, which requires 
the introduction 
of a novel technical concept, namely the ``elliptic cube root of unity''. Thus, in order to implement the reduction we define, 
more generally, the notion of {\em elliptic $N^{\rm th}$ root of unity}, and discuss some of its properties in connection with 
a special class of elliptic addition formulae. As a particular concrete application we present the class of elliptic 
$N$-soliton solutions of the lattice Boussinesq systems.

\section{Introduction}\label{sec-1}

Many integrable equations, be it partial differential or difference equations, or autonomous ordinary differential equations 
or the difference equations describing integrable dynamical mappings, admit special solutions in terms of elliptic functions 
(or in terms of the associated quasi-periodic functions). A prototypical example is the Hirota equation
\be\label{eq:Hirota}
A\tau_{n+1,m,l}\tau_{n,m+1,l+1}+B\tau_{n,m+1,l}\tau_{n+1,m,l+1}+C\tau_{n,m,l+1}\tau_{n+1,m+1,l}=0 \ ,
\ee 
which is an integrable 3-dimensional lattice equation, i.e., a partial difference equation (P$\Delta$E), with arbitrary fixed coefficients $A,B,C$ for the function
$\tau_{n,m,l}$ of three discrete variables $n,m,l\in\mathbb{Z}$. This equations allows for elliptic (type) solutions in the form 
\[ \tau_{n,m,l}=\sigma(\xi_0+n\delta+m\ven+l\nu)\  , \] 
provided the coefficients take the form:
\[  A=\frac{\sigma(\ven-\nu)}{\sigma(\ven)\,\sigma(\nu)}\  , \quad B=\frac{\sigma(\nu-\delta)}{\sigma(\delta)\,\sigma(\nu)}\  , 
\quad C=\frac{\sigma(\delta-\ven)}{\sigma(\delta)\,\sigma(\ven)}\  ,  \] 
where $\sg(x)$ is the Weierstrass $\sigma$-function, and where $\delta$, $\ven$ and $\nu$ are fixed parameters, called 
lattice parameters, while $\xi_0$ is an initial value. The fact that the Weierstrass $\sigma$ function solves the Hirota 
bilinear equation is a direct consequence of the three-term addition formula for the latter function, (cf. 
eq. \eqref{eq:8} of Appendix A). Typically, this is not the only elliptic type solution of \eqref{eq:Hirota}, but a special 
solution from which infinite families of such solutions 
can be constructed involving determinants with elliptic entries, cf. e.g. \cite{Y-KN}. In the continuous case of integrable PDEs, e.g. Korteweg-de Vries (KdV) or Boussinesq (BSQ) type 
equations, such elliptic solutions exist as well, e.g. the well-known cnoidal wave solution (in terms of the Jacobi cn-function) of the KdV equation, or 
in the case of ODEs the Lam\'e-Baker function solution of the Lam\'e equation. 

In recent years the theory of integrable difference equations has grown into a substantive body of new insights, cf. e.g. \cite{HJN-book}. In this context, certain classes of 
integrable partial difference equations (otherwise known as {\it lattice equations}) have gained prominence, e.g. the class of quadrilateral lattice 
equations integrable through the so-called multidimensional consistency (MDC) property.  The scalar affine-linear family of such equations was classified 
by Adler, Bobenko and Suris (ABS) in \cite{ABS} and subsequently their elliptic solutions were established in \cite{AN,NA}. Whereas the ABS equations 
are for single-component functions, elliptic solutions for higher rank equations, such as the lattice equations in the Gel'fand-Dikii hierarchy of \cite{GD} 
remained outstanding. A particular case of such equations is the (rank 3) lattice Boussinesq (BSQ) system, \cite{GD}, which reads:
\begin{eqnarray} \label{eq:BSQ}
&&\frac{P-Q}{u_{n+1,m+1}-u_{n+2,m}}- 
\frac{P-Q}{u_{n,m+2}-u_{n+1,m+1}}= \nonumber  \\ 
&& \quad = (u_{n+1,m+2}-u_{n+2,m+1})
(u_{n,m+1}-u_{n+2,m+2}) -(u_{n,m+1}-u_{n+1,m})(u_{n,m}-u_{n+2,m+1}) \ ,\nonumber \\ 
\end{eqnarray} 
with $P$ and $Q$ being lattice parameters, and which can be visualised as a relation between the values of the solution function on the vertices on a 9-point stencil, or can be reformulated as  
a 2-component quadrilateral lattice system. The latter type of systems have yet to be classified, and our knowledge about both the structure of the equations 
(e.g. in terms of the polynomial structure of the function defining the equation, which is no longer affine-linear, as in the case of \cite{ABS}) as well 
as the solution structure is as yet incomplete. In \cite{GD}, cf. also \cite{N1,DIGP,W}, eq.  \eqref{eq:BSQ} and its companion equations (lattice modified BSQ 
equation and lattice Schwarzian BSQ equation), and their extensions (in the sense if the unfolding of the dispersion curves, cf. \cite{ZZN}) were constructed, and 
the integrability aspects (Lax pair, reductions, Poisson structure) were established, on the basis of a framework called {\it direct linearisation} (DL). The latter 
framework has the advantage that it treats all these equations simultaneously within one framework based on singular integral equations involving well-chosen Cauchy 
kernels. Furthermore, although the DL approach is formal, it nonetheless covers large classes of solutions, including solutions that are otherwise obtained through inverse 
scattering techniques. However, to obtain algebra-geometric solutions the DL has to be extended in a nontrivial way. 

In the present paper we perform a first step towards the latter goal, namely to set up the framework for elliptic type solutions of the lattice BSQ systems. This 
includes both `elementary' elliptic solutions, as well as $N$-soliton towers of elliptic solutions and in principle also the inverse scattering type solutions 
based on elliptic asymptotic behaviour. Thus, the paper extends to the higher rank case the results obtained in \cite{NA} for the ABS list (excluding the case of the 
so-called Q4 equation, which was covered in \cite{AN}, but an analogue of which has not yet been established in the higher-rank case\footnote{We note that in 
\cite{DNY-K} a higher rank elliptic system was proposed generalising the 3-leg form of Q4, but it is not clear yet what is the rational form of that system.}). 
The strategy in the present paper, to set up the DL scheme is to consider realise the structure as a reduction of the one for the lattice Kadomtsev-Petviashvili (KP) 
system, whose elliptic type solutions were constructed in \cite{Y-KN}. However, a technical problem is that that in the rational case, the reduction from KP to BSQ 
(or to other systems in the Gel'fand-Dikii family) involves cube- or higher roots if unity. In order to perform the reduction from KP to such a system in the 
elliptic case, it turns out that an elliptic analogue of the roots of unity are needed. To our knowledge such objects were never before considered, and, thus, they 
had to be newly introduced in the present context, requiring higher-order elliptic identities which are given in Appendix B, cf. also \cite{DN}). Once these 
elliptic roots of unity are defined the reduction from the KP system to the BSQ and the higher-rank systems are obtained straightforwardly, and the solution structure 
is revealed through the corresponding DL formalism. 

The organisation of the paper is as follows. In section 2 we review the DL scheme for elliptic type solutions of the lattice KP class with the aim to make dimensional 
reductions to the KdV and BSQ cases. For the sake of the latter reduction we need an elliptic analogue of the cube root of unity. Thus, we introduce in section 3 the 
notion of elliptic $N^{\rm th}$ root of unity, and examine some of its properties. In section 4 we employ this novel concept to set up the DL system for the elliptic type  
solutions for the BSQ lattice equations,  while in section 5 we derive the actual nonlinear equations in closed form. In section 6 we demonstrate how the DL scheme leads to explicit solutions, namely elliptic seed and elliptic 
$N$-soliton solutions for the equations in standard form. Some Conclusions follow in section 7. For completeness 
we have given the elliptic parametrisation for the Lax pairs in an Appendix , although we don't need them for the purpose 
of presenting the elliptic solutions in this paper.

\paragraph{Warning:} Some standard facts on elliptic functions are introduced in Appendix A and we 
make heavy use of the various addition formulae throughout the paper, but often in a rather 
idiosyncratic way. Thus, we find that, as the story unfolds, several layers of ad-hoc notation are needed in 
order to write the equations and the underpinning relations in the most transparent and convenient way.

\section{Infinite Matrix Structure of elliptic solutions of KP lattice equations}\label{sec-2}
\setcounter{equation}{0}

In this section we review the DL structure for elliptic type solutions of KP lattice equations. The structure is based on a unifying 
framework of formal linear integral equations, from which the various lattice KP equations can be derived. The set-up is that we start from a large 
solution class in terms of which a multitude of functions on the lattice can be defined, each of which solves an equation (or rather a parameter-family of equations) in the 
KP class. Thus, the DL structure not only provides solutions of the individual KP equations, but also supplies us with the (Miura/B\"acklund) relations between the 
various equations.

\subsection{DL frame work}\label{sec-2-1}

\newcommand{\ddint}{\iint_{D} d\mu(\ell,\ell^\prime)}

\noindent In an early paper \cite{NCWQ-1984} we derived a system of
integrable lattice equations (i.e. partial difference equations)
associated with the KP equation from a system of linear integral
equations. In the present paper we generalise the integral equation
to one for a vector-valued function $\buk$, depending on a spectral
parameter $\kp$, of the following form
 \bse
 \label{eq:inteq}
 \be
\buk + \rho_\kp \ddint \ssl \bul
\Phi_\xi(\kp+\ell^\prime) = \rho_\kp \Phi_\xi(\Ld) {\boldsymbol c}_\kp\ ,
\label{eq:inteqa}
 \ee
 and one for its adjoint vector $\tuk$,
depending on another spectral variable $\kp^\prime$, of the form
\be
\tuk + \ssk \ddint \rho_\ell
\tul \Phi_\xi(\kp^\prime+\ell) = \ssk\tck \Phi_\xi(\tLd) \ .
\label{eq:inteqb}
\ee
\ese
 In eqs. (\ref{eq:inteq}) it is not necessary at this point to specify
the integrations over a general measure $d\mu(\ell,\ell^\prime)$ and
hypersurface in the space of both spectral variables $\ell$ and
$\ell^\prime$. we merely need some general assumptions (such as
uniqueness of the solution). The kernel is given in terms of the
Lam\'e function $\Phi_\xi$ defined by
\be\label{eq:Phi}
\Phi_\xi(x):= \frac{\sg(x+\xi)}{\sg(x)\sg(\xi)}\   ,
\ee
 in which $\sg(x)$ is the Weierstrass sigma-function, \cite{Akh}, (for
the pertinent definitions see Appendix A). The choice of the kernel
implies that the spectral parameters $\ell$, $\ell^\prime$ live on
the torus. This is not unlike the situation in the Landau-Lifschitz
equations, where the Riemann-Hilbert problem is defined on the torus
and where integral kernels in terms of elliptic functions naturally
arise, cf. e.g. \cite{DJKM-1983,Rodin-1983,Mik-PLA-1982}.
However, in this paper we do not investigate ``elliptic integrable equations'' but seek elliptic solutions
of otherwise non-elliptic equations.

The integral in eq. (\ref{eq:inteq}) is performed over an arbitrary
region in the complex hyperplane ${\mathbb C}^2$ of $\ell$ and
$\ell^\prime$, with integrations over suitable measure
$d\mu(\ell,\ell^\prime)$. Measure and integration region $D$ must be
chosen such that some general conditions (for instance being such
that the integral operator defined by (\ref{eq:inteq}) has trivial
kernel) hold, but are otherwise arbitrary. The integral equations
will be treated merely as a formal tool,  and we will postpone the
issue of specifying measure and contour for the time being. Thus,
under general conditions, we will derive from (\ref{eq:inteq}) an
interesting algebraic structure in terms of infinite matrices, from
which all relevant equations can be derived.

The inhomogeneous term, containing the `plane-wave factors'
$\rho_\kp$ respectively $\ssk$, determines the dependence of the
solutions $\buk$ and $\tuk$ on additional variables such as lattice
parameters and the discrete dynamical variables (measure and contour
are taken to be independent on those variables). Thus, in the
discrete case, we will impose on the plane-wave factors the
following translation properties under shifts along the lattice:
\bse\label{eq:rho}\be \label{eq:rhoa}
 \rho_\kp\ \ \to\ \
\wt{\rho}_\kp=T_\delta\rho_\kp=\Phi_\delta(\kp)\rho_\kp\    \ ,\    \ \ssk\ \
\to\ \
\wt{\sg}_{\kp^\prime}=T_\delta\sg_{\kp^\prime}=(\Phi_\delta(-\kp^\prime))^{-1}\ssk\   ,
\ee
associated with the lattice parameter $\delta$, and similarly for the
other lattice directions, namely 
\bea
 && \rho_\kp\ \ \to\ \
\wh{\rho}_\kp=T_\varepsilon\rho_\kp=\Phi_\varepsilon(\kp)\rho_\kp\    \ ,\    \ \ssk\ \
\to\ \
\wh{\sg}_{\kp^\prime}=T_\varepsilon\sg_{\kp^\prime}=(\Phi_\varepsilon(-\kp^\prime))^{-1}\ssk\   ,  \\ 
 && \rho_\kp\ \ \to\ \
\ol{\rho}_\kp=T_\nu\rho_\kp=\Phi_\nu(\kp)\rho_\kp\    \ ,\    \ \ssk\ \
\to\ \
\ol{\sg}_{\kp^\prime}=T_\nu\sg_{\kp^\prime}=(\Phi_\nu(-\kp^\prime))^{-1}\ssk\   ,
\eea\ese 
with lattice parameters $\varepsilon$ and $\nu$ respectively. Here the $T_\delta$, $T_\ven$ and $T_\nu$ denote 
the lattice shift operators, i.e. $\rho_\kp$ and $\sigma_{\kp'}$ should be considered as functions of associated 
lattice variables $n,m,l\in\mathbb Z$, associated with the three lattice parameters respectively, namely 
\[ \rho_\kp=(\Phi_\delta(\kp))^n (\Phi_\ven(\kp))^m(\Phi_\nu(\kp))^l\rho_\kp^0\ , \quad 
\sigma_{\kp'}=(\Phi_\delta(-\kp'))^{-n} (\Phi_\ven(-\kp'))^{-m}(\Phi_\nu(-\kp'))^{-l}\sigma_{\kp'}^0\ , \] 
with $\rho_\kp^0$, $\sigma_{\kp'}^0$ independent of these variables, and where the operators  $T_\delta$, $T_\ven$ and $T_\nu$ 
act by shifting these discrete variables by one unit respectively. By definition these operators commute when acting on proper
functions of $n$, $m$ and $l$ and distribute over products of functions.

The algebraic structure that (\ref{eq:inteq}) gives rise to appears as
a consequence of the object ${\boldsymbol c}_\kappa$ on the right-hand side
in (\ref{eq:inteq}), which is an infinite-component vector whose
components are just powers of $\kappa$, i.e.
${\boldsymbol c}_\kappa=(c_\kappa^{(i)})_{i\in {\mathbb{Z}}}$, where $c_\kappa^{(i)} =\kappa^i$. The
solutions of (\ref{eq:inteq}) are necessarily also
infinite-component vectors ${\boldsymbol u}_\kappa=(u_\kappa^{(i)})_{i\in {\mathbb{Z}}}$,
where each $u_\kp^{(i)}$ is supposed to be solved from (\ref{eq:inteq})
with the corresponding factor $\kappa^i$ in the inhomogeneous term.
This gradation by powers of the spectral parameter now gives rise to
an infinite matrix structure in terms of these infinite-component
vectors in terms of which the basic equations become very
transparent. 

Therefore, we will introduce index-raising operators $\Ld$ and $\tLd$ acting from the left and right on the 
power vectors as follows: 
\be \label{c:operator}
 (\Ld\bc_\kp)_i=(\bc_\kp)_{i+1}=\kp^{i+1}\  , \quad (\tck\tLd)_{i'}=(\tck)_{i'+1}=(\kp')^{i'+1}\  . 
\ee
In the present context, however, since we will use these 
vectors $\bc_\kp$ and $\tck$ merely as a basis in terms of powers to expand the elliptic functions, 
the structure will give rise to a quasi-grading in terms of \textit{elliptic matrices}, where the 
gradation is defined by index raising/lowering operators associated with an elliptic curve, cf. \cite{JN}. Thus, we will use formal operators, such as 
\[ \Phi_\xi(\Ld), \quad \Phi_\xi(\tLd), \quad \wp(\Ld), \quad \wp(\tLd), \quad \zeta(\Ld), \quad \zeta(\tLd)\ , \] 
as the elliptic analogues of raising/lowering operators, and they are defined through the formal power 
series expansions of the Weierstrass $\sigma$, $\zeta$ and $\wp$ functions. Consequently, they are assumed to obey the same addition formulae as the usual 
elliptic functions (see Appendix A), but in terms of arguments containing the operators $\Ld$ and 
$\tLd$. This is justified by thinking of these operators in terms of their symbols, i.e. in terms of their 
Fourier representation.  

In \cite{NCWQ-1984}, integral equations of the form (\ref{eq:inteq}) were
investigated in the case of the rational limit of the elliptic
functions, and it has been shown that its solutions give rise to
solutions of the lattice KP equation. In the elliptic case the
derivation is somewhat more complicated, and we will analyse the
relevant structures that emerge in this paper. For this purpose,
let us give a slightly more abstract presentation of the main
ingredients needed for the calculations.

We will now set up this formal structure.  We need the following three ingredients to
formulate the entire scheme:
\begin{itemize}
\item An infinite ( ${\mathbb Z}\times {\mathbb Z}$) matrix ${\boldsymbol C}$  which we
can take of the form
\be\label{eq:C}
{\boldsymbol C}=\ddint\,\rho_\ell\ssl {\boldsymbol
c}_{\ell} \tclp\   ,
\ee
in which ${\boldsymbol c}_\ell$ and $\tcl$ are infinite vectors with
components $({\boldsymbol c}_\ell)_j=(\tcl)_j=\ell^j$, and
$\rho_\ell$ depends on additional variables that are to be
determined later. The integrations over region $D$ and measure $d
\mu$ need not be specified at this point but we will loosely
assume that they can be chosen such that the objects to be
introduced below are well-defined.
\item
The matrices $\Ld$ and $\tLd$ as explained above, that define the operations of index-raising
when multiplied at the left respectively the right, acting on
${\boldsymbol c}_\kappa$ respectively $\tck$ by ~$\Ld {\boldsymbol c}_\kappa=\kappa {\boldsymbol c}_\kappa $~,
resp. ~$\tck\tLd=\kappa^\prime\tck$~.
\item
A formal \textit{Cauchy matrix}, i.e., an infinite matrix $\bOm_\xi$ obeying the equations:
\be\label{eq:Omega}
\bOm_\xi\Phi_\gm(\Ld) - \Phi_\gm(-\tLd)\bOm_{\xi+\gm} =
\Phi_\xi(\tLd)\bO\Phi_{\xi+\gm}(\Ld)\   ,
\ee
in which $\bO$ is the projection matrix on the central element, i.e.
$({\boldsymbol O} {\boldsymbol C}_k)_{i,j}=\dd_{i,0}{\boldsymbol C}_{0,j}$, etc.
\end{itemize}

As a consequence of the behaviour of the plane-wave factors $\rho_k$
and $\ssk$ under translations along the lattice, we have the following
linear relations for ${\boldsymbol C}$
\bse\label{eq:CC}\bea
\wt{\boldsymbol C}\Phi_\dd(-\tLd) &=& \Phi_\dd(\Ld) {\boldsymbol C}\   ,  \label{eq:CCa}\\
\wh{\boldsymbol C}\Phi_\ven(-\tLd) &=& \Phi_\ven(\Ld) {\boldsymbol C}\   , \label{eq:CCb} \\
\ol{\boldsymbol C}\Phi_\nu(-\tLd) &=& \Phi_\nu(\Ld) {\boldsymbol C}\   ,\label{eq:CCc}
\eea\ese
for the various lattice directions associated with the lattice
parameters $p=\zeta(\dd)$, $q=\zeta(\ven)$, resp. $r=\zeta(\nu)$.
In principle, since these relations are linear, we can select an
arbitrary set of lattice parameters $p_\nu$, each associated with
its own lattice shift, and in view of the linearity
of the corresponding equations for ${\boldsymbol C}$  we can impose all
the discrete evolutions for all chosen values $p_\nu$ simultaneously.

The true objects of interest, however, obey nonlinear equations,
and these objects are the following:
\begin{itemize}
\item an infinite matrix
\be \label{eq:U-al}
\bU_\xi := \Phi_\xi(\Ld){\boldsymbol C} \left( {\boldsymbol 1} +
\bOm_\xi\bC\right)^{-1}\Phi_\xi(\tLd)\   ,
\ee
\item
the $\tau$-function given by
the infinite determinant\footnote{
To make sense of the infinite determinant in (\ref{eq:tau})
we can use the well-known expansion
\[
\det({\boldsymbol 1}+A)=1+\sum_i A_{ii} + \sum_{i<j} \left| \begin{array}{cc}A_{ii}&A_{ij}\\
A_{ji}&A_{jj}\end{array}\right| + \dots \   ,
\]
which is valid if $A=\bOm_\xi\cdot\bC$ is of finite rank, which
imposes some conditions on the integrations in (\ref{eq:C}), in view of
the fact that all terms in the expansion
are of the form
\[ {\rm tr}_{\mathbb Z}\left( (\bOm_\xi\cdot\bC)^n\right)\  .   \] }
\be\label{eq:tau}
\tau_\xi := \det\,_{{\mathbb Z}\times {\mathbb Z}}\left(
{\boldsymbol 1}+\bOm_\xi\cdot\bC\right)\  .
\ee
\item  the ``wave vectors'' $\buk$ and $\tuk$ defined by
\bse\label{eq:uk} \bea
\buk(\xi) &=& \left( \Phi_\xi(\Ld) - \bU_\xi \Phi_\xi^{-1}(\tLd) \bOm_\xi\right)
{\boldsymbol c}_\kappa \rho_\kp\ ,  \label{eq:buk} \\
\tuk(\xi) &=& \sigma_{\kp'}\tck\left( \Phi_\xi(\tLd) -
\bOm_\xi \Phi_\xi^{-1}(\Ld) \bU_\xi \right)\   , \label{eq:tuk}
\eea\ese
which can be identified with the solutions of the integral equations
(\ref{eq:inteqa}) resp. (\ref{eq:inteqb}). ${}^T$ in \eqref{eq:tuk} denotes the transpose of matrix $\bU_\xi.$
\end{itemize}

We note that the infinite-component  matrix $\bU_\xi$ \eqref{eq:U-al}, 
with entries $u_{i,j}(\xi)$, ($i,j\in {\mathbb Z}$), which are the main quantities of interest in what follows, 
also has the integral representation 
\be\label{eq:U-in}
{\boldsymbol U}_\xi\ :=\ \ddint\,{\boldsymbol
u}_{\ell}(\xi)\,^{t\!}{\boldsymbol c}_{\ell^\prime}\ssl \Phi_\xi(\tLd)\  .
\ee
(For notational convenience, here and elsewhere we indicate the dependence on the variable $\xi$ by a suffix, rather than 
writing it as an argument, e.g. rather than writing $\bU(\xi)$.) 
The entire structure can now be reformulated in the following way.
For the purpose of brevity, let us formulate the structure in a
rather abstract way.

Let us now give the basic equations resulting from this scheme.
From the above ingredients it is elementary to derive
the basic relations describing the behaviour of $\tau_\xi$,
$\bU_\xi$ and $\buk$ and $\tuk$ under lattice translations.
Thus, for the $\tau$-function we have (see Appendix B for the derivation): 
\be\label{eq:taurel}
\frac{\wt{\tau}_\xi}{\tau_{\xi+\delta}}
= 1 - \left(\bU_{\xi+\delta}\left[
\zeta(\xi+\delta)-\zeta(\delta)+\zeta(\tLd)-\zeta(\xi+\tLd)\right]^{-1}
\right)_{0,0}\   ,
\ee
and similarly for the other lattice directions. For $\bU_\xi$
we can derive the discrete matrix Riccati type of relations
\bea \label{eq:Urel}
&&-\wt{\bU}_\xi\left[ \zeta(\xi+\delta)-\zeta(\delta)+\zeta(\tLd)-
\zeta(\xi+\tLd)\right] =  \nn \\
&& ~~~= \left[ \zeta(\xi)+\zeta(\delta)+\zeta(\Ld)-
\zeta(\xi+\delta+\Ld) - \wt{\bU}_\xi\bO\right]\bU_{\xi+\delta}
\   .  \eea
Eq. (\ref{eq:Urel}) together with its counterparts for the other
lattice directions forms the starting point for the construction of a
number of integrable three-dimensional lattice equations. By combining
the different lattice translations associated with the different
lattice parameters $\dd$, $\varepsilon$, $\nu$ one can actually derive all relevant
discrete equations within the KP family, as we shall show in the
next section. Let us finish here by giving the linear equations
for $\buk$ and $\tuk$ that form the basis for the derivation of the
Lax pairs for the above-mentioned equations derived from
(\ref{eq:Urel}). These relations read
\bse\label{eq:ukrels}
\bea
\wt{\bu}_\kappa(\xi) &=& \left[ \zeta(\xi)+\zeta(\delta)+\zeta(\Ld)-
\zeta(\xi+\delta+\Ld) - \wt{\bU}_\xi\bO\right]\buk(\xi+\delta)\   , \label{eq:bukrel} \\
\tuk(\xi+\delta) &=& -\,^{t\!}\wt{\boldsymbol u}_{\kappa^\prime}(\xi)
\left[ \zeta(\xi+\delta)-\zeta(\delta)+\zeta(\tLd)-\zeta(\xi+\tLd)
-\bO\bU_{\xi+\delta}\right]\   .  \label{eq:tukrel}
\eea\ese
Eqs. (\ref{eq:taurel}), (\ref{eq:Urel}) and eqs. (\ref{eq:ukrels}), the derivation of 
which is given in Appendix B, 
form the starting point for the subsequent derivation of nonlinear
equations in closed form. For this purpose we have to specify
special elements (or combinations of elements) of the infinite
matrix $\bU_\xi$ and to combine the different lattice shifts in
such a way that closed-form equations for such these preferred
elements are obtained. This we will do in the next subsection.

\subsection{Lattice KP systems from elliptic DL}\label{sec-2-2}

We will now derive closed-form difference equations for special
quantities defined in terms of the entries of the matrix $\bU_\xi$.
From the relation (\ref{eq:taurel}) it is suggestive to introduce
the quantities
\bse\label{eq:vw}\bea
v_\ar(\xi) &:=& 1 - \left( \left[ \zeta(\xi)+\zeta(\ar)+
\zeta(\Ld)- \zeta(\xi+\ar+\Ld)\right]^{-1}\bU_\xi\right)_{0,0}
\   ,  \label{eq:v} \\
w_\ar(\xi) &:=& 1 - \left( \bU_\xi\left[ \zeta(\xi)+
\zeta(\ar)+ \zeta(\tLd)-\zeta(\xi+\ar+\tLd)\right]^{-1}\right)_{0,0}
\   ,  \label{eq:w}
\eea \ese
for which from (\ref{eq:Urel}) we immediately have
\be\label{eq:vwrel}
\wt{v}_\delta(\xi-\delta)w_{-\delta}(\xi)=1\   .
\ee
It should be remarked that we can introduce quantities $v_\ar$ for
any value of the parameter $\ar$, not necessarily one of the values
associated with the lattice parameters $\delta$, $\ven$ or $\nu$ which
represent the lattice shifts. For a generic value, $\ar$ say,
we can now derive the relation:
\bea
&& \zeta(\delta)-\zeta(\ven)+\zeta(\xi-\delta)-\zeta(\xi-\ven)
+ \wh{u}(\xi-\ven) -\wt{u}(\xi-\delta) = \nn \\
&& ~~ = \left[\zeta(\delta)-\zeta(\xi-\ven)-\zeta(\alpha)
+\zeta(\xi+\alpha-\delta-\ven)\right]
\frac{\wh{v}_\ar(\xi-\ven)}{\wh{\wt{v}}_\ar(\xi-\delta-\ven)} \nn \\
&& ~~~ - \left[\zeta(\ven)-\zeta(\xi-\delta)-\zeta(\alpha)
+\zeta(\xi+\alpha-\delta-\ven)\right]
\frac{\wt{v}_\ar(\xi-\delta)}{\wh{\wt{v}}_\ar(\xi-\delta-\ven)}\   ,
\label{eq:uvwrel}
\eea
in which ~$u(\xi):= (\bU_\xi)_{0,0}$~. To derive eq.
(\ref{eq:uvwrel}) extensive use has been made of a special
elliptic relation, namely eq. (\ref{eq:special}) given in
appendix A. Eq. (\ref{eq:uvwrel})
leads to equations for both $u(\xi)$ and $v_\alpha(\xi)$, namely
\bea\label{eq:veq}
&&\left[\zeta(\dd)-\zeta(\xi-\ven)-\zeta(\ar)
+\zeta(\xi+\ar-\dd-\ven)\right]
\frac{\wh{v}_\ar(\xi-\ven)}{\wh{\wt{v}}_\ar(\xi-\dd-\ven)} \nn \\
&& - \left[\zeta(\ven)-\zeta(\xi-\dd)-\zeta(\ar)
+\zeta(\xi+\ar-\dd-\ven)\right]
\frac{\wt{v}_\ar(\xi-\dd)}{\wh{\wt{v}}_\ar(\xi-\dd-\ven)}\
+\ {\rm  cycl.}\ = 0\   ,
\eea
in which ``+ cycl.'' means the addition of similar terms obtained by
cyclic permutation of the action of the shifts $v\mapsto\wt{v}$,
$v\mapsto\wh{v}$, $v\mapsto\ol{v}$ together with the three parameters
$\dd$,$\ven$ and $\nu$. A four-term equation is obtained by
taking $\alpha$ to be equal to either one of the three parameters
$\delta$,$\ven$ or $\nu$. Using the identifications 
\begin{equation}\label{eq:vwtau} 
\wt{v}_\dd(\xi-\dd)= \frac{\tau_\xi}{\wt{\tau}_{\xi-\dd}}\  \quad \Leftrightarrow \quad 
w_{-\delta}(\xi) =\frac{\wt{\tau}_{\xi-\delta}}{\tau_\xi}\  , 
\end{equation} 
cf. (\ref{eq:taurel}), and similarly for the other lattice directions, 
we then obtain the Hirota bilinear KP equation
 in the following form
\bea\label{eq:taueq}
&&\left[\zeta(\dd)-\zeta(\ven)+\zeta(\xi-\dd)
-\zeta(\xi-\ven)\right]\wh{\wt{\tau}}_{\xi-\dd-\ven}
\ol{\tau}_{\xi-\nu}\nn \\
&& ~~+\left[\zeta(\ven)-\zeta(\nu)+\zeta(\xi-\ven)
-\zeta(\xi-\nu)\right]\wh{\ol{\tau}}_{\xi-\ven-\nu}
\wt{\tau}_{\xi-\dd}\nn \\
&& ~~+\left[\zeta(\nu)-\zeta(\dd)+\zeta(\xi-\nu)
-\zeta(\xi-\dd)\right]\wt{\ol{\tau}}_{\xi-\dd-\nu}
\wh{\tau}_{\xi-\ven}\ = 0   \ .
\eea
Finally, an equation for $u(\xi)$ is obtained as
\bea\label{eq:ueq}
&&\left[\zeta(\dd)-\zeta(\ven)+\zeta(\xi-\dd)
-\zeta(\xi-\ven)\right]\wh{\wt{u}}(\xi-\dd-\ven) \nn \\
&& ~~+\left[\zeta(\nu)-\zeta(\dd)+\zeta(\xi-\ven-\nu)
-\zeta(\xi-\dd-\ven)\right]\wh{u}(\xi-\ven) \nn \\
&& ~~+\left( \wh{\wt{u}}(\xi-\dd-\ven) -
\wh{\ol{u}}(\xi-\ven-\nu)\right)\wh{u}(\xi-\ven)\ +\ {\rm cycl.}\ = 0\  ,
\eea
which we prefer to call the lattice KP equation, cf. \cite{NCWQ-1984}.

Another object that is of interest within the scheme is the following
\bea
s_{\ar,\bb}(\xi) &:=& \left( \left[ \zeta(\xi)+\zeta(\ar)+
\zeta(\Ld)- \zeta(\xi+\ar+\Ld)\right]^{-1}\cdot\bU_\xi\right.    \nn \\
&&\left.  ~~~~ \cdot \left[ \zeta(\xi)+ \zeta(\bb)+ \zeta(\tLd)-
\zeta(\xi+\bb+\tLd)\right]^{-1}\right)_{0,0} \   ,  \label{eq:s}
\eea
for which we can derive the following important identity
\bea \label{eq:srel}
&& \wt{v}_\ar(\xi)\;w_\bb(\xi+\dd) =  \nn \\
&& ~~ = 1 -
\left[ \zeta(\xi+\dd)+\zeta(\ar)-\zeta(\dd)- \zeta(\xi+\ar)\right]
s_{\ar,\bb}(\xi+\dd) \   \nn \\
&& ~~~~~ -
\left[ \zeta(\xi)+\zeta(\bb)+ \zeta(\dd)- \zeta(\xi+\dd+\bb)\right]
\wt{s}_{\ar,\bb}(\xi) \   ,
\eea
and similar relations for the other lattice shifts $\wh{\phantom{a}}$ and $\ol{\phantom{a}}$ involving the
parameters $\ven$ and $\nu$ instead of $\delta$. In the derivation of
(\ref{eq:srel}) use has been made again of the special relation
(\ref{eq:special}) of Appendix A.  Eliminating the variables
$v_\alpha$ and $w_\beta$ by combining three lattice shifts, we arrive at
closed-form equation for $s_{\ar,\bb}$ which reads
\bea\label{eq:seq}
&&\frac{ 1-\chi^{(1)}_{\ar,-\dd}(\xi-\nu)\ol{s}_{\ar,\bb}(\xi-\nu)-
\chi^{(1)}_{\bb,\dd}(\xi-\dd-\nu)\wt{\ol{s}}_{\ar,\bb}(\xi-\dd-\nu)}
{ 1-\chi^{(1)}_{\ar,-\ven}(\xi-\nu)\ol{s}_{\ar,\bb}(\xi-\nu)-
\chi^{(1)}_{\bb,\ven}(\xi-\ven-\nu)\wh{\ol{s}}_{\ar,\bb}(\xi-\ven-\nu)} =\nn \\
&&~~=\frac{ 1-\chi^{(1)}_{\ar,-\dd}(\xi-\ven)\wh{s}_{\ar,\bb}(\xi-\ven)-
\chi^{(1)}_{\bb,\dd}(\xi-\dd-\ven)\wh{\wt{s}}_{\ar,\bb}(\xi-\dd-\ven)}
{ 1-\chi^{(1)}_{\ar,-\nu}(\xi-\ven)\wh{s}_{\ar,\bb}(\xi-\ven)-
\chi^{(1)}_{\bb,\nu}(\xi-\ven-\nu)\wh{\ol{s}}_{\ar,\bb}(\xi-\ven-\nu)}\times
\nn \\
&&~~\times\frac{ 1-\chi^{(1)}_{\ar,-\nu}(\xi-\dd)\wt{s}_{\ar,\bb}(\xi-\dd)-
\chi^{(1)}_{\bb,\nu}(\xi-\dd-\nu)\wt{\ol{s}}_{\ar,\bb}(\xi-\dd-\nu)}
{ 1-\chi^{(1)}_{\ar,-\ven}(\xi-\dd)\wt{s}_{\ar,\bb}(\xi-\dd)-
\chi^{(1)}_{\bb,\ven}(\xi-\dd-\ven)\wh{\wt{s}}_{\ar,\bb}(\xi-\dd-\ven)}\    .
\eea
in which we have used the abbreviation
\be \chi^{(1)}_{\alpha,\beta}(x):= \zeta(\alpha)+\zeta(\beta)+\zeta(x)-\zeta(\alpha+\beta+x)
=\frac{\Phi_{\alpha}(x)\,\Phi_\beta(x)}{\Phi_{\alpha+\beta}(x)}\  . \label{chi-1}
\ee
For arbitrary $\xi$ the equations \eqref{eq:ueq}, \eqref{eq:veq} and \eqref{eq:seq} form a complicated system of non-autonomous difference equations for effectively 
a set of four-variable functions, depending on the discrete variables $n,m,l$ and a continuous variable $\xi$, as partial delay-difference equations. 
However, if we allow the variable $\xi$ to depend on the discrete variables as follows:
\be\label{eq:xi}
\xi=\xi_0-n\delta-m\varepsilon-l\nu \ , 
\ee
where $\xi_0$ is fixed, i.e., the $\xi$ becomes dynamical, then those equations are nothing else than the lattice KP, lattice modified KP 
and lattice Schwarzian KP equations, cf. \cite{NCWQ-1984}, respectively, and can be transformed into their standard forms by 
an appropriate gauge transformation, thereby removing the 
elliptic coefficients. We will not do so here, as they are not the main objects of study in the present paper, but explicit soliton type solutions were given along that line in \cite{Y-KN}. 
We now consider the problem of their dimensional reduction to equations in the GD hierarchy, which in the elliptic case, requires the 
notion of the elliptic $N^{\rm th}$ root of unity.

\section{Elliptic $N^{\rm th}$ root of unity}\label{sec-3}
\setcounter{equation}{0}

In the rational case, to perform the dimensional reduction from three-dimensional KP systems to two-dimensional lattice equations of KdV or 
BSQ type we need to choose integration measures of the integral equations \eqref{eq:inteq} where the spectral variable $\ell'$ is identified 
with $-\oa\ell$, where $\oa$ is a $N^{\rm th}$ root of unity. However, this prescription no longer works, for $N>2$ in the elliptic case, and 
in fact we need to revisit our notion of what is meant by the $N^{\rm th}$ root of unity to get a sensible reduction. The main problem is how 
the multiplication by roots of unity {\it within the arguments of the elliptic functions} is working through in the formulae for the reduction. 

For $N=2$, however, the usual square root of unity $-1$ still can be implemented within the arguments of the elliptic functions without a problem, 
but we need to reexamine what we mean by this elliptic root. In this context, we redefine it as follows:  
The elliptic square roots of unity are given by the condition
\be\label{eq:ellsqroot}
\Phi_\kp(\dd)\Phi_\kp(\oa(\dd))=\wp(\kp)-\wp(\dd)\  , \quad \forall\ \kappa\  . 
\ee
The condition \eqref{eq:ellsqroot} obviously 
has the simple solution $\oa(\dd)=-\dd$, which coincides with what
one has in the usual case of a square root of unity. Note that in this case the elliptic square root $-\delta$ can be ``normalised'' by 
dividing by $\delta$, and hence we can extract a notion of square root (namely -1) which is independent of the parameter 
$\delta$. However, this is no longer the case when $N>2$, and in those cases the elliptic roots will essentially depend on a parameter.  

Thus already for $N=3$ we see an essential change of this notion, when we define the elliptic cube root of unity as follows: 

\paragraph{Definition 3.1:} \textit{ We call the parameter-dependent quantities $\oa_j(\delta)$ ($j=0,1,2$),  
where $\oa_0(\delta)=\delta$, the elliptic cube roots of unity, if they obey the following relation } 
\be\label{eq:ell3root}
\Phi_\kp(\delta)\Phi_\kp(\oa_1(\delta))\Phi_\kp(\oa_2(\delta))
=-\tfrac{1}{2}\left(\wp'(\kp)+\wp'(\delta)\right)\  , \quad \forall \kp\  ,  
\ee
\vspace{.2cm} 
\textit{subject to the relation} 
$\delta+\oa_1(\delta)+\oa_2(\delta)\equiv 0 (\textrm{mod\ \ period\ \ lattice})$. 

\noindent
The main thing to assert about this definition is that the quantities $\oa_i(\delta)$, albeit they depend on $\delta$, are 
independent of $\kp$. This is confirmed by the following Lemma. 

\paragraph{Lemma 3.1:} \textit{The condition \eqref{eq:ell3root} is equivalent to the following 
set of relations:}
\bse \label{eq:ell3rootrels}
\bea
&& \zeta(\delta)+\zeta(\oa_1(\delta))+\zeta(\oa_2(\delta))=0\  ,
\label{eq:ell3rootrelsa} \\
&& \wp'(\delta)=\wp'(\oa_1(\delta))=\wp'(\oa_2(\delta))\  .
\label{eq:ell3rootrelsb} 
\eea 
\textit{Furthermore, we also have} 
\be
\wp(\delta)+\wp(\oa_1(\delta))+\wp(\oa_2(\delta))=0\  . \ee \ese
{\bf Proof:} Using \eqref{eq:8} of Appendix A, and the relation 
$\oa_1(\delta)+\oa_2(\delta)\equiv-\delta (\textrm{mod\ \ period\ \ lattice})$, and assuming 
\eqref{eq:ell3rootrelsa} to hold, we have 
\begin{eqnarray*}
&& \Phi_\kp(\delta)\Phi_\kp(\oa_1(\delta))\Phi_\kp(\oa_2(\delta))= 
\Phi_\kp(\delta)\Phi_\kp(\oa_1(\delta)+\oa_2(\delta))\left[ \zeta (\kp)+\zeta(\oa_1(\delta))
+\zeta(\oa_2(\delta))-\zeta(\kp-\delta)\right] \\ 
&& = \Phi_\kp(\delta)\Phi_\kp(-\delta)\left[ \zeta(\kp)-\zeta(\delta)-\zeta(\kp-\delta)\right] \\ 
&& = - \left(\wp(\kp)-\wp(\delta)\right)\,\tfrac{1}{2}\frac{\wp'(\kp)+\wp'(\delta)}{\wp(\kp)-\wp(\delta)}
= -\tfrac{1}{2}\left(\wp'(\kp)+\wp'(\delta) \right) \  , 
\end{eqnarray*} 
which proves \eqref{eq:ell3root} for all values of $\kp$. Conversely, if we assume \eqref{eq:ell3root} to hold, 
with $\delta+\oa_1(\delta)+\oa_2(\delta)\equiv0 (\textrm{mod\ \ period\ \ lattice})$, then using the same 
identities we have   
\[ 0=\Phi_\kp(\delta)\Phi_\kp(\oa_1(\delta))\Phi_\kp(\oa_2(\delta))+\tfrac{1}{2}\left(\wp'(\kp)+\wp'(\delta) \right)
=-\left(\wp(\kp)-\wp(\delta)\right)\left[ \zeta(\oa_1(\delta))+\zeta(\oa_2(\delta))+\zeta(\delta)\right] \ , \] 
and since this holds for all $\kp$ we must have that the square bracket on the right hand side must vanish, which 
leads to \eqref{eq:ell3rootrelsa}. Furthermore, \eqref{eq:ell3rootrelsb} follows either from \eqref{eq:22} inserting 
the $\delta$, $\oa_1(\delta)$ and/or $\oa_2(\delta)$ in the arguments, or from setting $\kp=-\oa_i(\delta)$, ($i=1,2$) in 
\eqref{eq:ell3root}. 
$\square$

\paragraph{Remark:} The rational analogue of the relation \eqref{eq:ell3root} is the requirement that the 
third order polynomial of an indeterminate $k$ 
\[  (k-p_1)(k-p_2)(k-p_3)=k^3-p^3\  , \] 
is a pure cubic in $k$, which leads to the solution that the roots $p_1$, $p_2$, $p_3$ are related through 
cube roots of unity: $p_i=\oa^i p$ with $\oa=\exp\left(2\pi i/3\right)$ (up to permutations of the roots). 
Thus, the solutions of \eqref{eq:ell3root} as the elliptic analogue of a pure cubic and hence we can think of $\oa_1(\delta)$ and
$\oa_2(\delta)$ as generalisations of $e^{2\pi i/3}\delta$ and
$e^{4\pi i/3}\delta$, but here they are defined by the implicit
equation \eqref{eq:ell3root}. Note also that when $\oa_1$ and
$\oa_2$ correspond to the half roots of the root lattice, i.e. when
$\oa_1(\delta)=\oa$ and $\oa_2(\delta)=\oa'$, then the above
relations are all automatically satisfied by means of elliptic
identities, in the special case that $\wp'(\oa)=\wp'(\oa')=0$.

\paragraph{}
In the case of $N=4$ the elliptic quartic root of unity can be
defined using the identity \eqref{eq:tripleprod} of Appendix C, noting that here we need higher-order 
elliptic identites.

\paragraph{Definition 3.2:} \textit{ We call the parameter-dependent quantities $\oa_j(\delta)$ ($j=0,1,2,3$),  
where $\oa_0(\delta)=\delta$, the elliptic quartic roots of unity, if they obey the following relation } 
\be\label{eq:ell4root}
\Phi_\kp(\delta)\Phi_\kp(\oa_1(\delta))\Phi_\kp(\oa_2(\delta))\Phi_\kp(\oa_3(\delta))
=\tfrac{1}{6}\left(\wp''(\kp)-\wp''(\delta)\right)\  , \quad \forall \kp\  ,  
\ee
\vspace{.2cm} 
\textit{subject to the relation} 
$\delta+\oa_1(\delta)+\oa_2(\delta)+\oa_3(\delta)\equiv 0 (\textrm{mod\ \ period\ \ lattice})$. 

\noindent
Once again the independence of the quantities $\oa_i(\delta)$ of $\kp$ is asserted by the following Lemma: 

\paragraph{Lemma 3.2:} \textit{The condition \eqref{eq:ell4root} is equivalent to the following 
set of relations:}
\bse\label{eq:N4conds}
\begin{eqnarray}
&&  \zeta(\dd)+\zeta(\oa_1(\dd))+\zeta(\oa_2(\dd))+\zeta(\oa_3(\dd))=0\  , \\
&&  \wp(\dd)+\wp(\oa_1(\dd))+\wp(\oa_2(\dd))+\wp(\oa_3(\dd))=0 \ ,
\end{eqnarray}\ese 
\textit{Furthermore, we also have} 
\be\label{eq:N4condsb} 
\wp''(\delta)=\wp''(\oa_1(\delta))=\wp''(\oa_2(\delta)))=\wp''(\oa_3(\delta))\ . \ee 
{\bf Proof:} 
We have from \eqref{eq:tripleprod}  using $\dd+\oa_1(\delta)+\oa_2(\delta)+\oa_3(\delta)\equiv 0(\textrm{mod\
\ period\ \ lattice})$ and and assuming \eqref{eq:N4conds} to hold the following:
\begin{eqnarray*}
\Phi_\kp(\dd)\Phi_\kp(\oa_1(\dd))\Phi_\kp(\oa_2(\dd))\Phi_\kp(\oa_3(\dd)) &=& \Phi_\kp(\dd)\tfrac{1}{2}\Phi_\kp(-\dd)
\Big(  \left(\zeta(\kp)-\zeta(\dd)-\zeta(\kp-\dd)\right)^2 \\
&&~~~+\wp(\kp)+\wp(\dd)-\wp(\kp-\dd)\Big)  \\
&=& \tfrac{1}{2}\left( \wp(\kp)-\wp(\dd)\right)\left[ 2\wp(\kp)+2\wp(\dd)\right]  \\
&=& \left( \wp^2(\kp)-\wp^2(\dd)\right)=\tfrac{1}{6}\left(\wp''(\kp)-\wp''(\dd)  \right)\  ,
\end{eqnarray*}
identifying $\wp''=6\wp^2-g_2/2$. Thus, \eqref{eq:N4conds} lead to \eqref{eq:ell4root}. Conversely by 
back-engineering the above argument one can show that \eqref{eq:N4conds} follow from \eqref{eq:ell4root}.  
Furthermore, setting $\kp=-\oa_i(\delta$, ($i=1,2,3$) in \eqref{eq:ell4root} we find \eqref{eq:N4condsb}. $\square$

Similarly, proceeding along the same line, we can, from the higher order identity \eqref{eq:4ord3derrel} of Appendix C, derive the 
equation for the elliptic quintic root of unity. Thus, setting
\bse\label{eq:N5conds}
\begin{eqnarray}
&&  \zeta(\dd)+\zeta(\oa_1(\dd))+\zeta(\oa_2(\dd))+\zeta(\oa_3(\dd))+\zeta(\oa_4(\dd))=0\  , \\
&&  \wp(\dd)+\wp(\oa_1(\dd))+\wp(\oa_2(\dd))+\wp(\oa_3(\dd))+\wp(\oa_4(\dd))=0\  , \\
&&  \wp'(\dd)+\wp'(\oa_1(\dd))+\wp'(\oa_2(\dd))+\wp'(\oa_3(\dd))+\wp'(\oa_4(\dd))=0\  ,
\end{eqnarray}
\ese
together with $\dd+\oa_1(\delta)+\oa_2(\delta)+\oa_3(\delta)+\oa_4(\delta)\equiv 0(\textrm{mod\ \ period\ \ lattice})$,
we derive from \eqref{eq:4ord3derrel}
\begin{eqnarray*}
&& \Phi_\kp(\dd)\Phi_\kp(\oa_1(\delta))\Phi_\kp(\oa_2(\delta))\Phi_\kp(\oa_3(\delta)) \Phi_\kp(\oa_4(\delta))=  \\
&&\quad =  \Phi_\kp(\dd)\tfrac{1}{6}\Phi_\kp(-\dd)
\Big[  \Big(\zeta(\kp)-\zeta(\dd)-\zeta(\kp-\dd)\Big)^3  +\wp'(\kp-\dd)-\wp'(\kp)+\wp'(\dd) \\
&& \quad \qquad \qquad \qquad \qquad +3\Big(\zeta(\kp)-\zeta(\dd)
-\zeta(\kp-\dd)\Big)\Big(\wp(\kp)+\wp(\dd)-\wp(\kp-\dd)\Big)  \Big]   \\
&& \quad =\tfrac{1}{6}\Big(\wp(\kp)-\wp(\dd)\Big)\Big[ \Big(\zeta(\kp)-\zeta(\dd)
-\zeta(\kp-\dd)\Big)\Big( 4\wp(\kp)+4\wp(\dd)-2\wp(\kp-\dd)\Big) \\
&& \quad \quad \quad +\wp'(\kp-\dd)-\wp'(\kp)+\wp'(\dd)  \Big]   \\
&& \quad =\tfrac{1}{6}\Big[ -\tfrac{1}{2}\left(\wp'(\kp)+\wp'(\dd)\right)\Big( 4\wp(\kp)
+4\wp(\dd)-2\wp(\kp-\dd)\Big)\\
&& \quad \quad\quad +\left(\wp(\kp)-\wp(\dd)\right)
\left( \wp'(\kp-\dd)-\wp'(\kp)+\wp'(\dd)\right) \Big]\\
&& \quad =-\tfrac{1}{12}\left( 6\wp(\kp)\wp'(\kp)+6\wp(\dd)\wp'(\dd)\right)
=-\tfrac{1}{24}\left(\wp'''(\kp)+\wp'''(\dd)  \right)\  ,
\end{eqnarray*}
where use has been made of the addition formula \eqref{eq:addform}, the relation \eqref{eq:22}, which implies also
\[
\left((\wp'(\kp)+\wp'(\dd)\right)\left(\wp(\kp-\dd)-\wp(\dd)\right)=
-\left(\wp(\kp)-\wp(\dd)\right)\left( \wp'(\kp-\dd)-\wp'(\dd)\right)\  ,
\]
and the well-known relation $\wp'''=12\wp\wp'$.

Extrapolating from the latter cases of $N=4$ and $N=5$ cases, the general relation for the elliptic $N^{\rm th}$ root of unity can be conjectured, but 
the proof requires increasingly complicated elliptic identities along the lines given in Appendix C. Thus, we can define: 
\paragraph{Definition 2.3:} 
\textit{The elliptic $N^{\rm th}$ root of unity are, up to the periodicity of the period lattice, defined as the roots of the following equation} 
\be\label{eq:ellNroot}
\prod_{j=0}^{N-1} \Phi_\kp(\oa_j(\delta))=\tfrac{1}{(N-1)!}\left(\wp^{(N-2)}
(-\kp)-\wp^{(N-2)}(\dd)\right)\  ,
 \ee 
\textit{with $\oa_0(\dd):=\dd$, and subject to the relation} 
\be\label{eq:sumroots}
\sum_{j=0}^{N-1}\,\oa_j(\dd) \equiv 0(\textrm{mod\ \ period\ \ lattice})\  .
\ee

\paragraph{} It remains to be proven at this stage that these elliptic roots $\oa_i(\delta)$, ($i=0,\dots,N-1$) are in general independent of $\kp$, and that the relation 
\eqref{eq:ellNroot} is equivalent to the set of relations:  
\be\label{eq:Nrootrels}
\sum_{j=0}^{N-1}\,\zeta^{(l)}(\oa_j(\dd)) =0 \  ,  \quad l=0,\dots,N-3  \  ,
\ee
where $\zeta^{(l)}(x)=d^l\zeta(x)/dx^l$. 

\paragraph{} 
We don't present a formal proof of the above assertions at this stage, but merely indicate how this can be proven. In fact, the general form of
identities of the type \eqref{eq:tripleprod} and
\eqref{eq:4ord3derrel} which we need for the $N^{\rm th}$ root of unity can be described as follows:
\[
\prod_{j=1}^n \Phi_{\kp_j}(x) =\tfrac{1}{(n-1)!} \mathcal{F}(\kp_1,\dots,\kp_n;x)\  ,
\]
where the function $\mathcal{F}$ takes exactly the form of the
expansion of the $(N-1)^{\rm th}$ derivative of the Weierstrass
$\sg$-function:
\[
\frac{\sg'(x)}{\sg(x)}=\zeta(x)\  , \quad
\frac{\sg''(x)}{\sg(x)}=\zeta^2(x)-\wp(x)\  ,\quad
\frac{\sg'''(x)}{\sg(x)}=\zeta^3(x)-3\zeta(x)\wp(x)-\wp'(x)\  ,
\dots
\]
when expanded in $\zeta$, $\wp$ and $\wp'$. The expression 
$\mathcal{F}$ corresponds to this expansion, where whenever we have
an odd function in this expansion (like $\zeta$, $\wp'$, etc.) we substitute the combination
\[ \zeta(x)+\sum_{j=1}^N \zeta(\kp_j)-\zeta\left(\sum_{j=1}^N \kp_j+x\right)\  , \]
(and similar for $\wp'$),  and whenever we encounter $\wp$ (we have
to replace everywhere the higher derivatives of $\wp$ by their
expressions in terms of $\wp$ and $\wp'$ exclusively) we substitute the combination:
\[
\sum_{j=1}^N \wp(\kp_j)+\wp\left(\sum_{j=1}^N \kp_j+x\right)-\wp(x)\  .
\]
Subsequently, by substituting $\kp_j=\oa_j(\delta)$ and $x=\kp$ we obtain the identity \eqref{eq:ellNroot} subject to \eqref{eq:sumroots}. Conversely, we can revert the argument as in the above specific cases, and 
show that in return this is a sufficient condition for the set of relations \eqref{eq:Nrootrels} to hold.

\section{Reductions from the KP system to the KdV and BSQ systems}\label{sec-4}
\setcounter{equation}{0}

The fundamental KP relations can be summarized as follows. As a consequence of \eqref{eq:rho} the plane-wave factors 
$\rho_\kp(n,m,l)$ and $\sigma_{\kp'}(n,m,l)$ are of the form 
\bse\label{eq:rhosg}\bea
\rho_\kp &=& \Phi_\delta(\kp)^n\Phi_\varepsilon(\kp)^m\Phi_\nu(\kp)^l\rho_0  \ , \\
\sg_{\kp'} &=& \Phi_\delta(-\kp')^{-n}\Phi_\varepsilon(-\kp')^{-m}\Phi_\nu(-\kp')^{-l}\sg_0  \ ,
\eea\ese
and we set as in \eqref{eq:xi} the quantity $\xi=\xi(n,m,l)$ in the non-autonomous form $\xi=\xi_0-n\delta-m\varepsilon-l\nu$, 
then the fundamental KP system from the relations in subsection 2.1 takes the form:
\bse\label{eq:fundKPsyst}\bea
&& -\wt{\bU_\xi}\,\chi^{(1)}_{-\delta,\xi}(\tLd)=\chi^{(1)}_{\delta,\wt{\xi}}(\Ld)\,\bU_\xi-\wt{\bU_\xi}\,\bO\,\bU_\xi\  , \\
&& \wt{\buk(\xi)}=\left(\chi^{(1)}_{\delta,\wt{\xi}}(\Ld)-\wt{\bU_\xi}\,\bO\right)\buk(\xi)\  , \\
&& \tuk(\xi)=-\wt{\tuk(\xi)}\,\left( \chi^{(1)}_{-\delta,\xi}(\tLd)-\bO\,\bU_\xi\right) \  ,
\eea\ese
with \eqref{chi-1}.
(Note that here the shift on the object $\wt{\bU_\xi}$ also implies  a shift the variable $\xi$, i.e., 
$\wt{\bU_\xi}:=\wt{\bU}_{\wt{\xi}}$, and $\wt{\buk(\xi)}=\wt{\buk}(\wt{\xi})$, and similarly for the other quantitues depending on $\xi$, and 
for the other lattice shifts.)
Similar relations to \eqref{eq:fundKPsyst} hold for the other shifts on the multidimensional lattice, replacing
$\wt{\phantom{a}}$ by  $\wh{\phantom{a}}$ or $\ol{\phantom{a}}$ and $\delta$ by $\ven$ or $\nu$ respectively.

Applying multiple shifts we get higher order relations, e.g. the second order relation
\bse\label{eq:hiordKPrels}\bea 
&&  \wh{\wt{\bU_\xi}}\,\chi^{(2)}_{-\delta,-\ven,\xi}(\tLd)= \chi^{(2)}_{\delta,\ven,\wh{\wt{\xi}}}(\Ld)\cdot\bU_\xi -
\chi^{(1)}_{\delta,\ven}(\wh{\wt{\xi}})\wh{\wt{\bU_\xi}}\,\bO\,\bU_\xi
 + \wh{\wt{\bU_\xi}}\left( \bO\,\eta_\xi(\Ld)-\eta_{\wh{\wt{\xi}}}(\tLd)\,\bO\right)\bU_\xi\  , \nn \\
 && \\ 
&&\wh{\wt{\buk(\xi)}} =\left[ 
\chi^{(2)}_{\delta,\ven,\wh{\wt{\xi}}}(\Ld) -
\chi^{(1)}_{\delta,\ven}(\wh{\wt{\xi}})\wh{\wt{\bU_\xi}}\cdot\bO\
 + \wh{\wt{\bU_\xi}}\left( \bO\,\eta_\xi(\Ld)-\eta_{\wh{\wt{\xi}}}(\tLd)\,\bO\right)\right]\buk(\xi)\  , \\ 
&& \tuk(\xi) = 
\wh{\wt{\tuk(\xi)}}\cdot\left[   
\chi^{(2)}_{-\delta,-\ven,\xi}(\tLd) -
\chi^{(1)}_{-\delta,-\ven}(\xi)\bO\cdot \bU_\xi-
\left( \bO\,\eta_\xi(\Ld)-\eta_{\wh{\wt{\xi}}}(\tLd)\,\bO\right)\cdot \bU_\xi\right]\  , \nn \\ 
&& 
\eea\ese 
where, according to \eqref{eq:tripleprod}, we have the coefficients 
\begin{eqnarray*}
 \chi^{(2)}_{\alpha,\beta,\gamma}(x)&:=&
 \frac{1}{2}\left[\left( \zeta(\alpha)+\zeta(\beta)
 +\zeta(\gamma)+\zeta(x)-\zeta(\alpha+\beta+\gamma+x)\right)^2\right. \\
            && \left. \quad + \wp(x)-\left( \wp(\alpha)+\wp(\beta)
            +\wp(\gamma)+\wp(\alpha+\beta+\gamma+x)\right)\right] \  ,
\end{eqnarray*}
and recalling the quantity \eqref{eq:22} 
\[ \eta_\alpha(x):=\zeta(\alpha+x)-\zeta(\alpha)-\zeta(x)\  . \] 
Note that the relations \eqref{eq:hiordKPrels} do not depend on the intermediary steps (i.e., on single-shifted objects).

\paragraph{Remark:} In general the following functions appear in the coefficients of the fundamental relations: 
\[
\chi^{(n)}_{\alpha_1,\cdots,\alpha_n}(x)=\frac{\prod_{j=1}^n\,\Phi_{\alpha_j}(x)}{\Phi_{\alpha_1+\cdots+\alpha_n}(x)}\  , \] 
and they can be expressed in terms of $\zeta$-functions using the relations of Appendix C.

\paragraph{}
 Similarly the third-order shift relations read:
\bse\label{eq:tripleshiftrels}\bea
&& -\ol{\wh{\wt{\bU_\xi}}}\,\chi^{(3)}_{-\delta,-\ven,-\nu,\xi}(\tLd)
 = \chi^{(3)}_{\delta,\ven,\nu,\wh{\wt{\ol{\xi}}}}(\Ld)\,\bU_\xi
-\chi^{[2]}_{\delta,\ven,\nu}(\wt{\wh{\ol{\xi}}})\ol{\wh{\wt{\bU_\xi}}}\bO \bU_\xi \nn \\
&& \quad + \chi^{[1]}_{\delta,\ven,\nu}(\wt{\wh{\ol{\xi}}}) \ol{\wh{\wt{\bU_\xi}}}\left(\bO\eta_\xi(\Ld)
-\eta_{\wh{\wt{\ol{\xi}}}}(\tLd)\bO\right) \bU_\xi
-\ol{\wh{\wt{\bU_\xi}}}\left(\bO\wp(\Ld)+\wp(\tLd)\bO
-\eta_{\wh{\wt{\ol{\xi}}}}(\tLd)\bO\eta_\xi(\Ld)\right) \bU_\xi\  , \nn \\
&& \\
&& \ol{\wh{\wt{\buk(\xi)}}} = \chi^{(3)}_{\delta,\ven,\nu,\wh{\wt{\ol{\xi}}}}(\Ld)\,\buk(\xi)
-\chi^{[2]}_{\delta,\ven,\nu}(\wt{\wh{\ol{\xi}}})\ol{\wh{\wt{\bU_\xi}}}\bO \buk(\xi) \nn \\
&& \quad + \chi^{[1]}_{\delta,\ven,\nu}(\wt{\wh{\ol{\xi}}}) \ol{\wh{\wt{\bU_\xi}}}\left(\bO\eta_\xi(\Ld)
-\eta_{\wh{\wt{\ol{\xi}}}}(\tLd)\bO\right) \buk(\xi)
-\ol{\wh{\wt{\bU_\xi}}}\left(\bO\wp(\Ld)+\wp(\tLd)\bO
-\eta_{\wh{\wt{\ol{\xi}}}}(\tLd)\bO\eta_\xi(\Ld)\right) \buk(\xi)\  , \nn \\
&& \\ 
&& \tuk=-\ol{\wh{\wt{\tuk}}}\left[ \chi^{(3)}_{-\delta,-\ven,-\nu,\xi}(\tLd)  
-\chi^{[2]}_{-\delta,-\ven,-\nu}(\xi)\bO \bU_\xi \right. \nn \\
&& \quad - \left. \chi^{[1]}_{-\delta,-\ven,-\nu}(\xi)\left(\bO\eta_\xi(\Ld)
-\eta_{\wh{\wt{\ol{\xi}}}}(\tLd)\bO\right) \bU_\xi
-\left(\bO\wp(\Ld)+\wp(\tLd)\bO
-\eta_{\wh{\wt{\ol{\xi}}}}(\tLd)\bO\eta_\xi(\Ld)\right) \bU_\xi\right]   , \nn \\
\eea \ese
where
\[ \chi^{[1]}_{\delta,\ven,\nu}(\wt{\wh{\ol{\xi}}}):= \zeta(\delta)+\zeta(\ven)+\zeta(\nu)+\zeta(\wh{\wt{\ol{\xi}}})-\zeta(\xi)\  , \]
and
\begin{eqnarray*}
\chi^{[2]}_{\delta,\ven,\nu}(\wt{\wh{\ol{\xi}}})&
=&\frac{1}{2}\left[ \left(\zeta(\delta)+\zeta(\ven)+\zeta(\wt{\wh{\ol{\xi}}})-\zeta(\ol{\xi})\right)^2
-\wp(\delta)-\wp(\ven)-\wp(\ol{\xi}) \right. \\
&& +\left( \zeta(\ven)+\zeta(\nu)+\zeta(\wh{\ol{\xi}})-\zeta(\xi)\right)^2
-\wp(\ven)-\wp(\nu)-\wp(\wh{\ol{\xi}}) \\
&& \left. -2\left(\zeta(\wh{\ol{\xi}})-\zeta(\wh{\wt{\ol{\xi}}})
-\zeta(\delta)\right)\left(\zeta(\nu)+\zeta(\ol{\xi})-\zeta(\xi)\right)
+\wp(\wh{\wt{\ol{\xi}}})+\wp(\xi)\right]  \\
&=& \left(\zeta(\wh{\wt{\ol{\xi}}})-\zeta(\xi) +\zeta(\dd)
+\zeta(\ven+\nu)\right)\Big(\zeta(\dd)+\zeta(\ven)+\zeta(\nu)-\zeta(\dd+\ven+\nu)\Big)  \\
&&  + ~\wp(\wh{\wt{\ol{\xi}}})+\wp(\xi)-\wp(\dd)=\chi^{[2]}_{-\delta,-\ven,-\nu}(\xi)\ ,
\end{eqnarray*}
(which remarkably does not depend on the intermediate values $\wh{\ol{\xi}}$ or $\ol{\xi}$). 
Note that the latter expression does not seem manifestly symmetric w.r.t. permutation of $\dd$,~$\ven$,~$\nu$, 
but nevertheless it is, as can be shown using \eqref{eq:21} from the Appendix C. Furthermore,  we recall 
\[ \chi^{(3)}_{\alpha,\beta,\gamma,\delta}(x):=
\frac{\Phi_\alpha(x)\,\Phi_\beta(x)\Phi_\gamma(x)\,\Phi_\delta(x)}{\Phi_{\alpha+\beta+\gamma+\delta}(x)}\  ,
\]
which can be expressed in terms of $\zeta$ functions using the identities in Appendix C.

Finally, the relevant basic relations for the $\tau$-function read: 
\be\label{eq:taurels}
\frac{\wt{\tau_\xi}}{\tau_\xi}=1-\left(U_\xi\,\frac{1}{\chi^{(1)}_{-\delta,\xi}(\tLd)}\right)_{0,0} \quad \Leftrightarrow\quad 
\frac{\tau_\xi}{\wt{\tau_\xi}}=1-\left(\frac{1}{\chi^{(1)}_{\delta,\wt{\xi}}(\Ld)}\,\wt{\bU_\xi}\right)_{0,0}  \ ,
\ee 
as derived in Appendix B. 

We can now use these higher order shift relations to obtain the higher order dimensional reductions of the lattice 
KP system, by imposing periodicity w.r.t. the combined shifts where the shift parameters are related through the 
elliptic roots of unity. In this way we get for $N=2$ (square root of unity) the reduction to the lattice 
KdV system, while for $N=3$ (elliptic cube root of unity) we attain the reduction to the lattice BSQ system, 
and hence to the elliptic solutions of the latter.

\subsection{Reduction to the lattice KdV}\label{sec-4-1} 

The reduction
from KP to KdV type systems is imposed by requiring that the
following condition holds on the plane-wave factors:
\be\label{eq:rscond} 
T_{-\delta}\circ T_\delta (\rho_\kp\,\sg_{\kp'})=\rho_\kp\,\sg_{\kp'}\  .
\ee 
Here effectively we have employed the elliptic square root of unity $\oa(\delta)=-\delta$ as explained in section \ref{sec-3}, and  
\eqref{eq:rscond} amounts to the assertion that the reverse operation to any shift in the multidimensional lattice generated by shifts $T_\delta$ (for all 
$\delta$) 
corresponding to a parameter $\dd$ acts in the same way on the dependent variables as the shift corresponding to the
parameter ~$-\dd$~. In fact, the above condition then leads via \eqref{eq:rhoa} to the
identification $\kp'=\oa(\kp)$, and hence in the KP integral
equations \eqref{eq:inteq} we need to restrict ourselves to
integration over measures and contours where this identification is
implemented. Thus, the integral equation for the KdV class, reads:
\be\label{eq:inteqKdV}
\buk + \rho_\kp \sum_{j=0}^1\int_{\Gamma_j} d\mu_j(\ell)\,
\sg_{\oa_j(\ell)} \bul \Phi_\xi(\kp+\oa_j(\ell)) =
\rho_\kp\Phi_\xi(\Ld){\boldsymbol c}_\kp\ ,
\ee
 with the corresponding
expression for the matrix $\bU_\xi$: \be\label{eq:UKdV} \bU_\xi=
\sum_{j=0}^1\int_{\Gamma_j} d\mu_j(\ell)\, \sg_{\oa_j(\ell)}
\bul\,^{t\!}{\boldsymbol c}_{\oa_j(\ell)}\Phi_\xi(\tLd)\    . \ee
Thus, the reduction to the lattice KdV system is obtained by
implementing the relations \be\label{eq:KdVredcond} T_\dd\circ
T_{-\dd}\bU_\xi=\bU_\xi\  , \quad T_\dd\circ
T_{-\dd}\buk(\xi)=\left(\wp(\kp)-\wp(\dd)\right)\buk(\xi)\  , \ee
which follow as additional conditions on the fundamental system of
KP equations given in section 2. The relations for the KdV class can
be obtained in this way, and comprise:
\bse
\label{eq:fundKdVsyst}
\bea
&& -\wt{\bU_\xi}\,\chi^{(1)}_{-\delta,\xi}(\tLd)
=\chi^{(1)}_{\delta,\wt{\xi}}(\Ld)\,\bU_\xi-\wt{\bU_\xi}\,\bO\,\bU_\xi\  , \\
&& \wt{\buk(\xi)}=\left(\chi^{(1)}_{\delta,\wt{\xi}}(\Ld)
-\wt{\bU_\xi}\,\bO\right)\buk(\xi)\  , \\
&& \left(\wp(\kp)-\wp(\dd)\right)\buk(\xi)
=\left( \chi^{(1)}_{-\delta,\xi}(\Ld)-\bU_\xi\bO\right)\wt{\buk(\xi)} \  ,
\eea
\ese
where the first two relations are exactly the same as in the KP case, but where the last additional relation is obtained
from the condition \eqref{eq:KdVredcond}. We note that the reduction to the KdV case can also be interpreted as the condition
that the infinite matrix $\bU_\xi$ is symmetric under transposition, i.e. $\ ^{t\!}\bU_\xi=\bU_\xi$~.
Furthermore, in this reduction we have purely algebraic relations (i.e., without involving any lattice shifts), which are
obtained from the double-shift relations \eqref{eq:hiordKPrels} by taking $\ven=\oa(\dd)$ and taking into account the condition
\eqref{eq:KdVredcond}. Thus, we get
\bse
\label{eq:KdValgrels}
\bea
\bU_{\xi}\cdot\wp(\tLd) &=& \wp(\Ld)\cdot \bU_{\xi}
+\bU_{\xi}\left[\bO\eta_\xi(\Ld)-\eta_\xi(\tLd)\bO \right]\bU_{\xi}\   ,    \\
\wp(\kp)\buk(\xi) &=& \wp(\Ld)\cdot \buk(\xi)
+\bU_{\xi}\left[\bO\eta_\xi(\Ld)-\eta_\xi(\tLd)\bO \right]\buk(\xi)\  . \
\eea
\ese

The resulting equations in this case are the equations in the KdV class of lattice equations, namely two-dimensional lattice equations
involving only two lattice directions. Thus, as an example, we mention here only the lattice (potential)
KdV equation, cf. \cite{NQC}, in the form
\bea\label{eq:KdV}
&&\left[\zeta(\ar)-\zeta(\bb)+\zeta(\xi-\ar)
-\zeta(\xi-\bb)+\wh{u}(\xi-\bb) - \wt{u}(\xi-\ar)\right]\times \\
&& ~~ \times \left[\zeta(\ar)+\zeta(\bb)-\zeta(\xi) + \zeta(\xi-\ar-\bb)
+ u(\xi)-\wh{\wt{u}}(\xi-\ar-\bb)\right] = \wp(\ar) - \wp(\bb)\   .  \nn
\eea
In \cite{NA} it was shown how to obtain, from an elliptic Cauchy matrix scheme, elliptic $N$-soliton solutions for all
ABS equations \cite{ABS}, apart from Q4 (solitons of the latter were obtained by a different approach in \cite{AN}). In principle the
DL scheme covers an even wider class of  solutions. Since in this paper we concentrate on the higher-rank elliptic reductions,
we will refrain here from deriving all the lattice equations from the above scheme and refer to \cite{NA} for details.

\subsection{Reduction to the lattice BSQ system}\label{sec-4-2} 

The dimensional reduction to the BSQ case (i.e., the case $N=3$) from the lattice KP is obtained by imposing the condition
\be\label{eq:3dimred}
T_\delta\circ T_{\oa_1(\delta)}\circ T_{\oa_2(\delta)}\left(\rho_\kp\,\sg_{\kp'}\right)=\rho_\kp\,\sg_{\kp'}\  ,
\ee
which leads to the possible identifications $\kp'=-\oa_j(\kp)$ ($j=0,1,2$),
where we have set for convenience $\oa_0(\kp):=\kp$.
In fact, computing
\begin{eqnarray*}
T_\delta\circ T_{\oa_1(\delta)}\circ T_{\oa_2(\delta)}\,\rho_\kp &
=& \Phi_\delta(\kp)\,\Phi_{\oa_1(\delta)}(\kp)\,\Phi_{\oa_2(\delta)}(\kp)\,\rho_\kp \\
&=& \Phi_\delta(\kp)\Phi_{\oa_1(\delta)
+\oa_2(\delta)}(\kp)\,\Big[\zeta(\oa_1(\delta))+\zeta(\oa_2(\delta))
+\zeta(\kp) \\
&& - \zeta(\oa_1(\delta)+\oa_2(\delta)+\kp) \Big]\rho_\kp \\
&=& \Phi_\delta(\kp)\Phi_{-\delta}(\kp)\left[ \zeta(\kp)-\zeta(\delta)
-\zeta(\kp-\delta)\right]\rho_\kp \\
&=& -\left(\wp(\kp)-\wp(\delta)\right)\cdot\frac{1}{2}\frac{\wp'(\kp)
+\wp'(\delta)}{\wp(\kp)-\wp(\delta)}\rho_\kp=-\frac{1}{2}\left(\wp'(\kp)+\wp'(\delta)\right)\rho_\kp\  ,
\end{eqnarray*}
where in the second step we have used $\delta+\oa_1(\delta)+\oa_2(\delta)\equiv 0(\textrm{mod\ \ period\ \ lattice})$ and $\zeta(\delta)+\zeta(\oa_1(\delta))+\zeta(\oa_2(\delta))=0$,
which is the first of \eqref{eq:ell3rootrels}, while the second formula follows from the fact that the above computation can be done in the same way (with the same result)
permuting the roots $\delta,\oa_1(\delta),\oa_2(\delta)$. Similarly, we have that
\[ 
T_\delta\circ T_{\oa_1(\delta)}\circ T_{\oa_2(\delta)}\sg_{\kp'}
=\left(\frac{1}{2}\left(\wp'(\kp')-\wp'(\delta) \right)\right)^{-1}\sg_{\kp'}\  ,
\]
and, hence, we get the condition \eqref{eq:3dimred} to hold provided that $\kp'=-\oa_j(\kp)$ 
($j=0,1,2$). The latter is the condition on the spectral variable which reduces the integral equations \eqref{eq:inteq} 
to the ones for the BSQ case.

Thus, eq. \eqref{eq:3dimred} when implemented on the KP integral equations
\eqref{eq:inteq}, e.g. by restricting the measure
$d\mu(\ell,\ell')$ to contain  $\delta$-functions of the form
$\delta(\ell'+\oa_j(\ell))$), leads to the following integral
equation for the BSQ elliptic solutions:
\be
\label{eq:inteqBSQ}
\buk(\xi) + \rho_\kp \sum_{j=0}^2\int_{\Gamma_j} d\mu_j(\ell)\,
\sg_{-\oa_j(\ell)}\Phi_\xi(\kp-\oa_j(\ell))\bul(\xi) =
\rho_\kp\Phi_\xi(\Ld){\boldsymbol c}_\kp\    ,
\ee
where $\xi$ is
given as in \eqref{eq:xi} (which guarantees that $T_\delta\circ
T_{\oa_1(\delta)}\circ T_{\oa_2(\delta)}\xi \equiv \xi(\textrm{mod\ \ period\
\ lattice})$), and where $\Gamma_j$ and $d\mu_j(\ell)$ are contours
and measures that need to be suitably chosen. 
Furthermore, as a consequence of the reduction the eigenvector $\bu_\kp$ obeys 
\be\label{eq:BSQredcond} T_\delta\circ T_{\oa_1(\delta)}\circ
T_{\oa_2(\delta)}\buk=-\frac{1}{2}\left(
\wp'(\delta)+\wp'(\kp)\right) \buk\  , \ee 
implementing the above mentioned 
identifications $\kp'=-\oa_j(\kp)$ ($j=0,1,2$). Together with the
formal integral equation \eqref{eq:inteqBSQ} we have in this reduced case the following
formula for the infinite matrix $\bU_\xi$:
\be
\label{eq:UBSQ}
\bU_\xi= \sum_{j=0}^2\int_{\Gamma_j} d\mu_j(\ell)\,
\sg_{-\oa_j(\ell)} \bul(\xi)\,^{t\!}{\boldsymbol
c}_{-\oa_j(\ell)}\Phi_\xi(\tLd)\    .
\ee

As a consequence of the analysis
of the KP system, which fully goes through modulo the reduction, we can now readily inherit all the fundamental 
relations for the infinite-matrix scheme
for the BSQ system by implementing the constraint \eqref{eq:BSQredcond} on the KP system \eqref{eq:fundKPsyst} in the multidimensional space
of lattice variables. By assuming that this multidimensional grid includes the lattice directions associated 
with the  parameters $\oa_1(\delta)$ and $\oa_2(\delta)$, 
subject to the condition \eqref{eq:ell3root}, 
we get the following fundamental infinite matrix system for the elliptic BSQ solution. First, the discrete 
dynamics for any direction indicated by the parameter $\delta$, is given by the relations:
\bse\label{eq:fundBSQsyst}
\bea
&& -\wt{\bU_\xi}\,\chi^{(1)}_{-\delta,\xi}(\tLd)
=\chi^{(1)}_{\delta,\wt{\xi}}(\Ld)\,\bU_\xi-\wt{\bU_\xi}\,\bO\,\bU_\xi\  ,  \label{eq:fundBSQsyst-a}\\
&& \wt{\buk(\xi)}=\left(\chi^{(1)}_{\delta,\wt{\xi}}(\Ld)
-\wt{\bU_\xi}\,\bO\right)\buk(\xi)\  ,  \label{eq:fundBSQsyst-b}
\eea
for the forward shift (which is the same relation as for the KP system), while for the reverse shift we have additional relations:
\bea
&& \bU_\xi\left[ \wp(\xi)+\wp(\tLd)+\wp(\delta)-\eta_{-\delta}(\xi)\eta_\xi(\tLd)\right]   \label{eq:fundBSQsyst-c} \\
&& =  \left[ \wp(\wt{\xi})+\wp(\Ld)+\wp(\delta)
-\eta_\delta(\wt{\xi})\eta_{\wt{\xi}}(\Ld)\right]\wt{\bU_\xi}
+\bU_\xi\left[\eta_{-\delta}(\xi)\bO+\bO\eta_{\wt{\xi}}(\Ld)
-\eta_\xi(\tLd)\bO\right]\wt{\bU_\xi}\  , \nn \\
&& \phantom{a} \nn \\
&& -\tfrac{1}{2}(\wp'(\delta)+\wp'(\kp))\buk(\xi) =
\left[ \wp(\wt{\xi})+\wp(\Ld)+\wp(\delta)
-\eta_\delta(\wt{\xi})\eta_{\wt{\xi}}(\Ld)\right]\wt{\buk(\xi)} \nn  \\
&& \qquad\qquad\qquad\qquad\qquad\qquad
+~\bU_\xi\left[\eta_{-\delta}(\xi)\bO+\bO\eta_{\wt{\xi}}(\Ld)
-\eta_\xi(\tLd)\bO\right]\wt{\buk(\xi)}\ ,  \label{eq:fundBSQsyst-d}
\eea
\ese
and similar relations for the other lattice directions. These
follow directly from the double-shift KP relations
\eqref{eq:hiordKPrels} implementing the triple shift conditions on
the reduction \eqref{eq:BSQredcond}.  Second, the
conditions \eqref{eq:BSQredcond} give rise to the following
algebraic constraints (which involve no lattice shifts):
\bse
\label{eq:BSQalgconstrs}
\begin{align}
& -\tfrac{1}{2}\bU_\xi\,\wp'(\tLd)=\tfrac{1}{2}\wp'(\Ld)\bU_\xi+\bU_\xi\left[ \bO\wp(\Ld)
+\wp(\tLd)\bO-\eta_\xi(\tLd)\bO\eta_\xi(\Ld)\right]\bU_\xi+\wp(\xi)\bU_\xi\bO\bU_\xi\  ,\nn \\
& \phantom{a} \label{eq:BSQalgconstrsa} \\
& \tfrac{1}{2}\wp'(\kp)\buk(\xi)=\tfrac{1}{2}\wp'(\Ld)\buk(\xi)+\bU_\xi\left[ \bO\wp(\Ld)+\wp(\tLd)\bO
-\eta_\xi(\tLd)\bO\eta_\xi(\Ld)\right]\buk+\wp(\xi)\bU_\xi\bO\buk(\xi)\  , \nn \\
& \phantom{a} \label{eq:BSQalgconstrsb}
\end{align}
\ese
which follow from \eqref{eq:tripleshiftrels} implementing the
condition \eqref{eq:BSQredcond}, and noting that $\chi^{[2]}_{\delta,\oa_1(\delta),\oa_2(\delta)}(\xi)=\wp(\xi)$.  

The system of equations
\eqref{eq:fundBSQsyst} in all relevant lattice direction form the
fundamental system from which the BSQ lattice system, including the
corresponding Lax pairs, can be derived, thus implying that the
quantities we will extract from the infinite matrix $\bU_\xi$ of
\eqref{eq:UBSQ} form the elliptic class of solutions of those
lattice systems.

The next step, in order to derive from the set of relations \eqref{eq:fundBSQsyst} closed-form equations 
in terms of single elements of the matrix $\bU_\xi$ or combinations thereof, is to identify 
the relevant entries and to specify the relations involving those entries. Then closed-form equations are obtained 
by combining those relations for two different lattice shifts associated with two different lattice directions.  
For the sake of transparency of the structure, we will introduce yet another set 
of notations, which will make the relations as close as possible to the ones in the rational case. 
Thus, we introduce the notation: 
\[
\Ld_\xi:= \eta_\xi(\Ld)\  ,\quad \tLd_\xi:= \eta_\xi(\tLd)\  , \] 
i.e. $\xi$-dependent raising operators acting from the left and right, and  
\[
p_\xi:=\eta_{-\dd}(\xi)
=-\eta_\dd(\wt{\xi})\  ,\quad q_\xi:=\eta_{-\ven}(\xi)=-\eta_\ven(\wh{\xi})\ ,
\]
$\xi$-dependent lattice `parameters' (replacing in a sense the lattice parameters $p$ and $q$ 
of \cite{GD}). In terms of these the fundamental relations \eqref{eq:fundBSQsyst} can be rewritten as 
\bse\label{2ndfundBSQsyst}\bea 
&& \wt{\bU_\xi}(p_\xi+\tLd_{\wt{\xi}})= (p_\xi-\Ld_\xi)\bU_\xi-\wt{\bU_\xi}\bO\bU_\xi\  ,  \label{2ndfundBSQsyst-a}\\ 
&& \wt{\bu_\kp(\xi)}=\left[ p_\xi-\Ld_\xi-\wt{\bU_\xi}\bO\right]\bu_\kp(\xi)\  , \label{2ndfundBSQsyst-b}\\ 
&& \frac{1}{2}\bU_\xi \frac{\wp'(\delta)-\wp'(\tLd)}{p_\xi+\tLd_{\wt{\xi}}}  = 
\frac{1}{2} \frac{\wp'(\delta)+\wp'(\Ld)}{p_\xi-\Ld_\xi} \wt{\bU_\xi}
-\bU_\xi\left(p_\xi\bO+\bO\Ld_{\wt{\xi}}-\tLd_\xi\bO\right)\wt{\bU_\xi}\  , \label{2ndfundBSQsyst-c}\\ 
&& \tfrac{1}{2}(\wp'(\delta)+\wp'(\kp))\buk(\xi) =\left[ 
\frac{1}{2} \frac{\wp'(\delta)+\wp'(\Ld)}{p_\xi-\Ld_\xi}
-\bU_\xi\left(p_\xi\bO+\bO\Ld_{\wt{\xi}}-\tLd_\xi\bO\right)\right]\wt{\buk(\xi)}\  , \label{2ndfundBSQsyst-d}
\eea\ese 
(and a similar set of relations with $p_\xi\to q_\xi$ and $\wt{~}\to \wh{~}$), 
with 
\[ \chi^{(1)}_{-\delta,\xi}(\tLd)=-(p_\xi+\tLd_{\wt{\xi}})\ , \quad 
\chi^{(1)}_{\delta,\wt{\xi}}(\Ld)=p_\xi-\Ld_\xi\  ,   \]
and similarly for the other lattice directions.  
Some useful properties of these quantities are given by 
\[ \chi^{(1)}_{\delta,\wt{\xi}}(\Ld)\,\chi^{(1)}_{-\delta,\xi}(\Ld)= -(p_\xi-\Ld_\xi) (p_\xi+\Ld_{\wt{\xi}})=\wp(\Ld)-\wp(\delta)\  ,  \] 
and similarly for the quantities $q_\xi$ with $\Ld_{\wt{\xi}}$ replaced by $\Ld_{\wh{\xi}}$ and $\delta$ by $\ven$.

\section{Elliptic solutions of lattice BSQ systems}\label{sec-5}
\setcounter{equation}{0}

In this section we derive some concrete closed-form lattice equations within the BSQ class from the fundamental 
relations in terms of the infinite matrix $\bU_\xi$ presented in subsection \ref{sec-4-2}. The corresponding Lax 
pairs for those equations will be constructed in Appendix D.  

\subsection{Derivation of the regular lattice BSQ equation}\label{sec-5-1}

We proceed by constructing a system of closed-form equations for specific elements of the matrix $\bU_\xi$ 
by combining the relations from the system \eqref{eq:fundBSQsyst} for two different lattice 
shifts with parameters $\delta$ and $\ven$. Setting by definition: 
\bse
\label{def:u}
\begin{align}
& u_{0,0} :=   u_{0,0}(\xi)=\left(\bU_\xi \right)_{0,0}~, ~~~u_{1,0} := u_{1,0} (\xi)=\left(\Ld_\xi\bU_\xi \right)_{0,0}~, \\
&u_{0,1} :  = u_{0,1} (\xi)=\left(\bU_\xi \tLd_\xi \right)_{0,0}~,~~~
u_{1,1} :  = u_{1,1} (\xi)=\left(\Ld_\xi\bU_\xi \tLd_\xi \right)_{0,0}~,\\
&u_{2,0} := u_{2,0} (\xi)=  \left(\wp(\Ld)\bU_\xi \right)_{0,0}~, ~~~u_{0,2} := u_{0,2} (\xi)= \left(\bU_\xi \wp(\tLd) \right)_{0,0}~,
\end{align}
\ese
we can derive various relations from the system \eqref{2ndfundBSQsyst} by taking the $(~)_{0,0}$ element, either directly or 
after multiplication by factors of the form $p_\xi+\tLd_{\wt{\xi}}$ (from the right) or $p_\xi-\Ld_\xi$ (from the left). Thus, 
from \eqref{2ndfundBSQsyst-a} and \eqref{eq:fundBSQsyst-c} as well as \eqref{2ndfundBSQsyst-c}
we obtain the following set of relations (suppressing the arguments $\xi$ of the functions): 
\bse
\label{def:u-}
\begin{align}
& p_\xi \wt {u_{0,0}}+\wt {u_{0,1} }  = p_\xi u_{0,0} - u_{1,0} - \wt {u_{0,0}} u_{0,0}\ ,  \label{eq:u-t}\\
& p_\xi^2 \wt{u_{0,0}}+p_\xi(\wt{u_{1,0}}+\wt{u_{0,1}})+\wt{u_{1,1}} = \wp(\delta)u_{0,0}-u_{2,0}-(p_\xi \wt{u_{0,0}}+\wt{u_{1,0}})u_{0,0}\ , \label{eq:u-tt} \\ 
& \wp(\delta)\wt{u_{0,0}}-\wt{u_{0,2}}= p_\xi^2 u_{0,0} - p_\xi (u_{0,1} + u_{1,0}) + u_{1,1} -\wt{u_{0,0}}( p_\xi u_{0,0}-u_{0,1})\  ,  \label{eq:u-ttt}\\ 
& (\wp(\delta)+\wp(\xi))u_{0,0}+ u_{0,2}-p_\xi u_{0,1}  ~~~ \nn\\
& =  (\wp(\dd)+\wp(\wt \xi))\wt{u_{0,0}}+ \wt{u_{2,0}}+(p_\xi+u_{0,0})\wt{u_{1,0}} +(p_\xi u_{0,0}-u_{0,1}) \wt {u_{0,0}}\ .\label{eq:u-tttt}
\end{align}
\ese
Similarly from their counterparts for the other shift, i.e. replacing the shift $\wt{~}$~by the  shift $\wh{~}$~and $p_\xi$ by $q_\xi$, we obtain: 
\bse
\label{def:u--}
\begin{align}
& q_\xi \wh {u_{0,0}}+\wh {u_{0,1} }  = q_\xi u_{0,0} - u_{1,0} - \wh {u_{0,0}} u_{0,0}\ ,  \label{eq:u-h}\\
& q_\xi^2 \wh{u_{0,0}}+q_\xi(\wh{u_{1,0}}+\wh{u_{0,1}})+\wh{u_{1,1}} = \wp(\varepsilon)u_{0,0}-u_{2,0}-(q_\xi \wh{u_{0,0}}+\wh{u_{1,0}})u_{0,0}\ , \label{eq:u-hh} \\ 
& \wp(\varepsilon)\wh{u_{0,0}}-\wh{u_{0,2}}= q_\xi^2 u_{0,0} - q_\xi (u_{0,1} + u_{1,0}) + u_{1,1} -\wh{u_{0,0}}( q_\xi u_{0,0}-u_{0,1})\  ,  \label{eq:u-hhh}\\ 
& (\wp(\varepsilon)+\wp(\xi))u_{0,0}+ u_{0,2}-q_\xi u_{0,1}  ~~~ \nn\\
& =  (\wp(\varepsilon)+\wp(\wh \xi))\wh{u_{0,0}}+ \wh{u_{2,0}}+(q_\xi+u_{0,0})\wh{u_{1,0}} +(q_\xi u_{0,0}-u_{0,1}) \wh {u_{0,0}}\ .\label{eq:u-hhhh}
\end{align}
\ese
There exist several consistencies among these relations. For instance, subtracting \eqref{eq:u-hh} from \eqref{eq:u-tt} and using 
\eqref{eq:u-ttt} and  \eqref{eq:u-hhh} to eliminate $u_{1,1}$ we find a triviality provided the following relation 
\bse\begin{equation}\label{eq:pq1} 
p_{\wh{\xi}}+q_\xi= q_{\wt{\xi}}+p_\xi\ , 
\end{equation} 
which follows from the definition of these quantities, is being used. Another relation that is frequently needed in these 
computations is
\begin{equation}\label{eq:pq2} 
p_{\wh{\xi}}q_\xi+\wp(\wh{\xi})= q_{\wt{\xi}}p_\xi+\wp(\wt{\xi})\ , 
\end{equation} 
which is a consequence of 
\begin{equation}\label{eq:pq3}
\wp(\dd)+\wp(\varepsilon) +\wp(\wt \xi) +p_\xi q_{\wt \xi} = -\frac{1}{2}\frac{\wp'(\dd)-\wp'(\ven)}{ p_\xi-q_\xi} \ .
\end{equation}\ese 
which in turn follows from the general relation \eqref{eq:specialBSQ} of Appendix A.

Combining the above relations in a nontrivial way, while eliminating $u_{0,2}$ and $u_{2,0}$ and $u_{(1,1)}$, is obtained by subtracting 
\eqref{eq:u-hhh} from \eqref{eq:u-ttt} and using the $\wt{~}$-shift of \eqref{eq:u-hhhh} and the $\wh{~}$-shift of \eqref{eq:u-tttt}, 
which leads to the followin relation: 
\begin{align}\label{BSQ_uvw}
\frac{1}{2}\frac{\wp'(\dd)-\wp'(\ven)}{ p_\xi-q_\xi+\wh{u_{0,0}}-\wt{u_{0,0}}} 
=&~\frac{1}{2}\frac{\wp'(\dd)-\wp'(\ven)}{p_\xi-q_\xi} +\wh{ \wt{u_{1,0}}} +u_{0,1}+ u_{0,0}\wh{ \wt{u_{0,0}}}
+(p_{\wh{\xi}}+q_\xi) (\wh{ \wt{u_{0,0}}}- u_{0,0} )   \ ,
\end{align}
which together with \eqref{eq:u-t} and \eqref{eq:u-h} from a coupled 3-component system for the functions $u_{0,0},u_{1,0}$ and 
$u_{0,1}$  which is equivalent to the lattice BSQ system. In fact, 
by eliminate the quantities $u_{1,0}$ and $u_{0,1}$ in this three-component lattice system, by shifting \eqref{BSQ_uvw} in both 
lattice directions, we obtain the 9-point equation: 
\begin{eqnarray} \label{eq:lBSQ-u}
&&
\frac{1}{2} \frac{\wp'(\dd)-\wp'(\ven)}{p_{\wh{\xi}}-q_{\wh{\xi}}+\wh{\wh{u_{0,0}}}-\wh{\wt {u_{0,0}}}} - \frac{1}{2} \frac{\wp'(\dd)-\wp'(\ven)}{p_{\wt{\xi}}-q_{\wt{\xi}}+\wh{\wt {u_{0,0}}}-\wt{\wt {u_{0,0}}}} = \nn \\ 
&& =\quad \bigg(p_{\wh{\wt{\xi}}}-q_{\wh{\wt{\xi}}}+ \wh{\wh{\wt {u_{0,0}}}}-\wh{\wt{\wt {u_{0,0}}}}\bigg)
\bigg(p_{\wh{\wh{\wt{\xi}}}}+p_{\wh{\wh{\xi}}}++q_{\wh{\xi}}+\wh{u_{0,0}}-\wh{\wh{\wt{\wt {u_{0,0}}}}}\bigg) \quad \quad\quad \quad\nonumber  \\ 
&& \quad \quad -~\bigg(p_{\xi}-q_{\xi}+ \wh{ {u_{0,0}}}-\wt{{u_{0,0}}}\bigg) 
\bigg(p_{\wh{\wt \xi}}+p_{\wh{ \xi}}+q_{\xi}+ {u_{0,0}}-\wh{\wt{\wt {u_{0,0}}}} \bigg) \ .\quad \quad\quad \quad
\end{eqnarray} 
In section 7, we will relate \eqref{eq:lBSQ-u} to the 'normalised' form of the lattice BSQ equation \eqref{eq:BSQ} and show that the DL scheme leads to 
explicit solutions, namely of elliptic seed and soliton type.

\subsection{Parameter-dependent quantities and the lattice modified BSQ equation}\label{sec-5-2}
In order to obtain closed-form equations from the shift relations \eqref{eq:fundBSQsyst-a} and \eqref{eq:fundBSQsyst-c}, we 
introduce the following objects:
\bse
\label{BSQ_u}
\begin{align}
&s_{\alpha} (\xi)= \eta_{\alpha}(\xi) - \left(\left(\chi^{(1)}_{\alpha,\xi}(\Ld)\right) ^{-1} \bU_\xi \eta_\xi(\tLd)\right)_{0,0} \ ,  \\
&t_{\beta} (\xi)= \eta_{\beta}(\xi) - \left( \eta_\xi(\Ld) \bU_\xi  \left(\chi^{(1)}_{\beta,\xi}(\tLd)\right) ^{-1} \right)_{0,0} \ , \\
&r_{\alpha} (\xi)= \wp(\alpha) - \left(\left(\chi^{(1)}_{\alpha,\xi}(\Ld)\right) ^{-1} \bU_\xi \wp(\tLd)\right)_{0,0} \ , \\ 
&z_{\beta} (\xi)= \wp(\beta) - \left( \wp(\Ld) \bU_\xi  \left(\chi^{(1)}_{\beta,\xi}(\tLd)\right) ^{-1} \right)_{0,0}  \ ,
\end{align}
and recall $v_\alpha (\xi), w_\beta (\xi), s_{\alpha,\beta}(\xi)$ defined in subsection \ref{sec-2-2}~:
\begin{align}
&v_{\alpha} (\xi)= 1 - \left(\left(\chi^{(1)}_{\alpha,\xi}(\Ld)\right) ^{-1} \bU_\xi \right)_{0,0} \ ,  \label{BSQ_u-v}  \\
&w_{\beta} (\xi)=1 - \left( \bU_\xi  \left(\chi^{(1)}_{\beta,\xi}(\tLd)\right) ^{-1} \right)_{0,0} \ , \label{BSQ_u-w}  \\
&s_{\alpha,\beta}(\xi)=  \left( \left(\chi^{(1)}_{\alpha,\xi}(\Ld)\right) ^{-1}  \bU_\xi  \left(\chi^{(1)}_{\beta,\xi}(\tLd)\right) ^{-1} \right)_{0,0}  \ . \label{BSQ_u-s} 
\end{align}
\ese
Several identities are needed to derive the relevant single-shift relations for these quantities, namely 
\begin{equation}\label{eq:chichi} 
\frac{\chi^{(1)}_{\delta,\wt{\xi}}(\Ld)}{\chi^{(1)}_{\alpha,\wt{\xi}}(\Ld)}=
1-\frac{\eta_\delta(\wt{\xi})-\eta_\alpha(\wt{\xi})}{\chi^{(1)}_{\alpha,\xi}(\Ld)}\  , \quad 
\frac{\chi^{(1)}_{-\delta,\xi}(\tLd)}{\chi^{(1)}_{\beta,\xi}(\tLd)}=
1-\frac{\eta_{-\delta}(\xi)-\eta_\beta(\xi)}{\chi^{(1)}_{\beta,\wt{\xi}}(\tLd)}\  ,
\end{equation} 
which follow from \eqref{eq:special}, as well as  
\bse\label{eq:wpchi}\begin{eqnarray}
&&  \frac{\wp(\wt{\xi})+\wp(\Ld)+\wp(\delta)-\eta_{\delta}(\wt{\xi})\eta_{\wt{\xi}}(\Ld)}{\chi^{(1)}_{\alpha,\xi}(\Ld)}= \nn \\ 
&& \quad = \zeta(\alpha+\xi)-\zeta(\delta)-\zeta(\wt{\xi})-\zeta(\alpha+\Ld)+\zeta(\Ld)
-\eta_{\delta}(\wt{\xi})\frac{\eta_{-\delta}(\xi)-\eta_\alpha(\xi)}{\chi^{(1)}_{\alpha,\wt{\xi}}(\Ld)}\  , 
 \label{eq:wpchi-a} \\ 
&& \frac{\wp(\xi)+\wp(\tLd)+\wp(\delta)-\eta_{-\delta}(\xi)\eta_\xi(\tLd)}{\chi^{(1)}_{\beta,\wt{\xi}}(\tLd)}= \nn \\ 
&& \quad = \zeta(\beta+\wt{\xi})+\zeta(\delta)-\zeta(\xi)-\zeta(\beta+\tLd)+\zeta(\tLd)
-\eta_{-\delta}(\xi)\frac{\eta_\delta(\wt{\xi})-\eta_\beta(\wt{\xi})}{\chi^{(1)}_{\beta,\xi}(\tLd)}\  , 
\label{eq:wpchi-b}
\end{eqnarray}\ese 
which follow from \eqref{eq:addform} and \eqref{eq:21} in combination with \eqref{eq:chichi}. Using \eqref{eq:chichi} by multiplying \eqref{eq:fundBSQsyst-a} by 
factors $\left(\chi^{(1)}_{\alpha,\wt{\xi}}(\Ld)\right)^{-1}$ from the left and by $\left(\chi^{(1)}_{\beta,\xi}(\tLd)\right)^{-1}$ from 
the right and taking the $(~)_{0,0}$ element, and using the relations \eqref{eq:wpchi} by multiplying \eqref{eq:fundBSQsyst-c} by 
factors $\left(\chi^{(1)}_{\alpha,\xi}(\Ld)\right)^{-1}$ from the left and by $\left(\chi^{(1)}_{\beta,\wt{\xi}}(\tLd)\right)^{-1}$ from 
the right and taking the $(~)_{0,0}$ element, we obtain the following relations relating the 2-parameter quantity $s_{\alpha,\beta}(\xi)$ to the other 
quantities defined above:  
\bse
\label{relations_s}
\begin{align}
& 1+(p_\xi+a_{\wt{\xi}}) s_{\alpha,\beta}(\xi) - (p_\xi-b_\xi) \wt {s_{\alpha,\beta}(\xi)}  = \wt{v_\alpha (\xi) }  w_\beta (\xi)  \ , \label{eq:sa} \\
&  p_\xi-a_\xi+b_{\wt{\xi}} +B_{-\delta}(\xi)s_{\alpha,\beta}(\xi) - A_\delta(\wt{\xi}) \wt{s_{\alpha,\beta}(\xi)} = \nn \\ 
&~~~ = p_\xi v_\alpha(\xi) \wt{w_\beta(\xi)} + v_\alpha(\xi) \wt{t_\beta(\xi)} - s_\alpha(\xi) \wt{w_\beta(\xi)}\  , \label{eq:ss} 
\end{align}
\ese
respectively, in which, for convenience, we have introduced the notation: 
\bse\label{eq:abAB}
\begin{eqnarray}
&& a_\xi:= \eta_\alpha(\xi) \ ,\quad  \quad b_\xi:= \eta_\beta(\xi)\  , \\ 
&& A_\nu(\xi):= \frac{1}{2} \frac{\wp'(\nu)-\wp'(\alpha)}{\eta_\nu(\xi)-\eta_\alpha(\xi)}\  , \quad 
B_\nu(\xi):= \frac{1}{2} \frac{\wp'(\nu)-\wp'(\beta)}{\eta_\nu(\xi)-\eta_\beta(\xi)}\  . \label{eq:AB}
\end{eqnarray} \ese 
The latter quantities appear in \eqref{eq:ss} in the following form: 
\bse\begin{eqnarray}
A_\delta(\wt{\xi})&=&
p_\xi(p_\xi-a_\xi)-\wp(\wt{\xi})+\wp(\alpha)=\wp(\alpha)+\wp(\xi)+\wp(\delta)-\eta_{-\delta}(\xi)\eta_\alpha(\xi)     \\ 
B_{-\delta}(\xi) &=& p_\xi(p_\xi+b_{\wt{\xi}})-\wp(\xi)+\wp(\beta)=\wp(\beta)+\wp(\wt{\xi})+\wp(\delta)-\eta_{\delta}(\wt{\xi})\eta_\beta(\wt{\xi})  
\end{eqnarray}\ese  
(noting the curious interchange of the roles of $\xi$ and $\wt{\xi}$ when $\delta$ changes into $-\delta$). 
Note that we also have the relations: 
\begin{equation}\label{eq:alt-AB}
 A_\delta(\wt{\xi})=p_{-\alpha}(p_\xi-a_\xi)\ , \quad  B_{-\delta}(\xi)=p_\beta(p_\xi+b_{\wt{\xi}})\  .   
 \end{equation}  

For the single-parameter quantities $s_\alpha, t_\beta,r_\alpha,z_\beta$ we can derive a system of shift relations in a similar way, 
using single-parameter multiplying factors from the left or the right before taking the $(~)_{0,0}$ elelments. Thus, from 
\eqref{eq:fundBSQsyst-a} multiplying from the left by $\left(\chi^{(1)}_{\alpha,\wt{\xi}}(\Ld)\right)^{-1}$ or by $\left(\chi^{(1)}_{\beta,\xi}(\tLd)\right)^{-1}$ from the right and projecting on the $(~)_{0,0}$ element, we obtain: 
\bse\label{eq:st} 
\begin{align}
& \wt {s_\alpha (\xi)} = -(p_\xi+u_{0,0})\wt {v_\alpha (\xi)} + (p_\xi+a_{\wt{\xi}})v_\alpha (\xi) \ , \label{eq:st-a}\\
&{t_\beta (\xi)} =(p_\xi - \wt {u_{0,0}} ) {w_\beta (\xi)} - (p_\xi-b_\xi)\wt{ w_\beta (\xi)} \ .  \label{eq:st-b}
\end{align} 
\ese
Furthermore, by multiplying \eqref{eq:fundBSQsyst-a} from the left by $\left(\chi^{(1)}_{\alpha,\wt{\xi}}(\Ld)\right)^{-1}$ and from 
the right by $\chi^{(1)}_{\delta,\wt{\xi}}(\tLd)$ and taking the $(~)_{0,0}$ elelment, respectively my multiplying from the right 
by $\left(\chi^{(1)}_{\beta,\xi}(\tLd)\right)^{-1}$ and from the left by $\chi^{(1)}_{-\delta,\xi}(\Ld)$ and projecting, we obtain 
\bse
\begin{align}
& \wt {r_\alpha (\xi)} = (\wp(\delta)+p_\xi u_{0,0}-u_{0,1}) \wt {v_\alpha (\xi)} -(p_\xi+a_{\wt{\xi}})\left( 
p_\xi v_\alpha(\xi)-s_\alpha(\xi)\right)\   , \label{eq:rvs} \\ 
& z_\beta (\xi) = (\wp(\delta)-p_\xi \wt{u_{0,0}}-\wt{u_{1,0}})w_\beta(\xi) - (p_\xi-b_\xi) \left( p_\xi\wt{ w_\beta(\xi)} + 
\wt{t_\beta(\xi)} \right)\  . \label{eq:zwt}
\end{align} 
\ese
Finally, from \eqref{eq:fundBSQsyst-c} by multiplying from the left by $\left(\chi^{(1)}_{\alpha,\xi}(\Ld)\right)^{-1}$, or from 
the right by $\left(\chi^{(1)}_{\beta,\wt{\xi}}(\tLd)\right)^{-1}$ and taking the $(~)_{0,0}$ elements, while using 
the relations \eqref{eq:wpchi-a} and \eqref{eq:wpchi-b} respectively, we obtain: 
\bse\label{eq:rz} 
\begin{align}
&  {r_\alpha (\xi)} =  A_\delta(\wt{\xi}) \wt{v_\alpha (\xi)}  +(p_\xi-\wt{u_{0,0}}) {s_\alpha (\xi)} 
- \left(\wp( \xi)+ \wp(\dd)- p_\xi\wt{u_{0,0}}- \wt {u_{1,0}} \right) v_\alpha (\xi)   \ ,  \label{relations_strz-r}\\
& \wt{z_\beta (\xi)}=  B_{-\delta}(\xi) w_\beta (\xi)  -(p_\xi + {u_{0,0}}) \wt{t_\beta (\xi)} 
- (\wp(\wt \xi)+ \wp(\dd)+p_\xi {u_{0,0}}-  {u_{0,1}} )   \wt {w_\beta (\xi)}  
\ . \label{relations_strz-tz}
\end{align}
\ese
Henceforth we will, for brevity, suppress the arguments $\xi$ in these parameter functions, and simply 
write $v_\alpha$ for $v_\alpha(\xi)$, etc., where it is understood that shifts $\wt{v_\alpha}=\wt{v_\alpha(\xi)}$ act on the function as well as its argument, unless explicitly stated. 

Obviously the relations \eqref{eq:st}-\eqref{eq:rz} have their counterpart for the other lattice direction, replacing $\wt{~}$ by $\wh{~}$ and 
$\delta$ by $\varepsilon$. By combining relations of this type for the two lattice directions, and thereby eliminating some of the parameter 
quantities, various significant additional relations of Miura type can be derived. Here the relations \eqref{eq:u-t} and \eqref{eq:u-h} are 
needed as well to eliminate the quantities $u_{0,0}$, $u_{1,0}$ and $u_{0,1}$ when necessary. For instance, from \eqref{eq:rz} by eliminating the 
quantities $r_\alpha$ and $z_\beta$ and using \eqref{eq:u-t} and \eqref{eq:u-h} we find the 2-shift relations 
\bse\label{eq:vvus}\begin{eqnarray} 
&& A_\delta(\wt{\xi})\frac{\wt{v_\alpha}}{v_\alpha}-A_\ven(\wh{\xi})\frac{\wh{v_\alpha}}{v_\alpha}=(p_\xi-q_\xi+\wh{u_{0,0}}-\wt{u_{0,0}})\,\left( 
p_{\wh{\xi}}+q_\xi-\frac{s_\alpha}{v_\alpha}-\wh{\wt{u_{0,0}}}\right)\  , \label{eq:vvus-a}  \\  
&& B_{-\delta}(\wh{\xi})\frac{\wh{w_\beta}}{\wh{\wt{w_\beta}}}-B_{-\ven}(\wt{\xi})\frac{\wt{w_\beta}}{\wh{\wt{w_\beta}}}
=(p_\xi-q_\xi+\wh{u_{0,0}}-\wt{u_{0,0}})\,\left( p_{\wh{\xi}}+q_\xi+u_{0,0}+\frac{\wh{\wt{t_\beta}}}{\wh{\wt{w_\beta}}}\right)\  ,     \label{eq:vvus-b} 
\end{eqnarray} \ese 
where we have also used \eqref{eq:pq1} and the relation 
\begin{equation}\label{eq:pq4} 
(p_\xi-q_\xi)(p_{\wh{\xi}}+q_\xi)=\wp(\delta)-\wp(\ven)\  , 
\end{equation} 
which follows from the defintions of these quantities and the relation \eqref{eq:21}. 
At the same time, from \eqref{eq:st} and its counterparts with shifts in the other lattice direction, we can 
derive the relations: 
\bse\label{eq:wwuu}\begin{eqnarray}
p_\xi-q_\xi+\wh{u_{0,0}}-\wt{u_{0,0}} &=& (p_{\wh{\xi}}+a_{\wh{\wt{\xi}}}) 
\frac{\wh{v_\alpha}}{\wh{\wt{v_\alpha}}}- (q_{\wt{\xi}}+a_{\wh{\wt{\xi}}}) 
\frac{\wt{v_\alpha}}{\wh{\wt{v_\alpha}}}  \label{eq:wwuu-a}  \\ 
&=& (p_\xi-b_\xi) \frac{\wt{w_\beta}}{w_\beta}- (q_\xi-b_\xi) \frac{\wh{w_\beta}}{w_\beta}\  , \label{eq:wwuu-b} 
\end{eqnarray}\ese
as well as the relations
\bse\label{eq:wwss}\begin{eqnarray}
p_\xi-q_\xi-\frac{\wh{s_\alpha}}{\wh{v_\alpha}}+
\frac{\wt{s_\alpha}}{\wt{v_\alpha}} &=& (p_\xi+a_{\wt{\xi}}) 
\frac{v_\alpha}{\wt{v_\alpha}}- (q_\xi+a_{\wh{\xi}}) 
\frac{v_\alpha}{\wh{v_\alpha}}  \label{eq:wwss-a}  \\ 
p_{\wh{\xi}}-q_{\wt{\xi}}-\frac{\wh{t_\beta}}{\wh{w_\beta}}+\frac{\wt{t_\beta}}{\wt{w_\beta}} 
&=& (p_{\wh{\xi}}-b_{\wh{\xi}}) \frac{\wh{\wt{w_\beta}}}{\wh{w_\beta}}- 
(q_{\wt{\xi}}-b_{\wt{\xi}}) \frac{\wh{\wt{w_\beta}}}{\wt{w_\beta}}\  , \label{eq:wwss-b} 
\end{eqnarray}\ese 

We note at this points that we have now already two closed systems of equations, namely one for the 
quantities $u_{0,0}$, $v_\alpha$ and $s_\alpha$ given by  \eqref{eq:vvus-a}, and the two relations given by 
\eqref{eq:st-a} and its counterpart in the other lattice shift direction, and the other for $u_{0,0}$, 
$w_\beta$ and $t_\beta$ \eqref{eq:vvus-b}, and the two relations given by 
\eqref{eq:st-b} with its counterpart in the other lattice direction. 
However, one can also derive at this 
juncture a closed-form 9-point equation for $v_\alpha$, or alternatively for $w_\beta$, alone. In fact, by 
solving $s_\alpha/v_\alpha$ from \eqref{eq:vvus-a}, and inserting the result into \eqref{eq:st-a}, we get
\[ p_\xi+ p_{\wh{\wt{\xi}}}+q_{\wt{\xi}}+u_{0,0}-\wh{\wt{\wt{u_{0,0}}}}-\frac{\wh{\wt{\wt{v_\alpha}}}}{\wt{v_\alpha}}\, 
\frac{A_\delta(\wt{\wt{\xi}})\wt{\wt{v_\alpha}}-A_\ven(\wh{\wt{\xi}})\wh{\wt{v_\alpha}}}{(p_{\wh{\wt{\xi}}}+a_{\wh{\wt{\wt{\xi}}}}) 
\wh{\wt{v_\alpha}}- (q_{\wt{\wt{\xi}}}+a_{\wh{\wt{\wt{\xi}}}}) 
\wt{\wt{v_\alpha}} }=(p_\xi+a_{\wt{\xi}})\frac{v_\alpha}{\wt{v_\alpha}}\  ,  \] 
and by subtracting a second copy of this relation with $\delta$ and $\ven$ interchanged (interchanging also 
$p_\xi$ and $q_\xi$ and the $\wt{~}$- and $\wh{~}$ lattice shifts) we can use \eqref{eq:wwuu-a} 
to eliminate all the quantities $u_{0,0}$. The result is a 9-point equation for $v_\alpha$, namely 
\begin{eqnarray} \label{eq:lmodified BSQ}
&&\frac{-A_\delta ( \wh{\wt\xi}) \wh{\wt {v_\alpha (\xi)}}+A_\ven (  \wh{\wh \xi}) \wh{\wh {v_\alpha (\xi)}}}{(q_ 
{\wh{\wt{\xi}}}+a_{\wh{\wh{\wt{{\xi}}}}})\wh{\wt {v_\alpha (\xi)}}-(p_{ \wh{\wh{\xi}}}+a_{\wh{\wh{\wt{{\xi}}}}})\wh{\wh {v_
\alpha (\xi)}}}  \frac{\wh{\wh{\wt{{v_\alpha (\xi)}}}}}{\wh{{v_\alpha (\xi)}}} -
\frac{-A_\ven (  \wh{\wt \xi}) \wh{\wt {v_\alpha (\xi)}}+A_\delta ( \wt{\wt\xi})  \wt{\wt {v_\alpha (\xi)}}}{(p_{\wh{\wt 
\xi}}+a_{\wh{\wt{\wt{{\xi}}}}})\wh{\wt {v_\alpha (\xi)}}-(q_{\wt{\wt \xi}}+a_{\wh{\wt{\wt{{\xi}}}}})\wt{\wt {v_\alpha 
(\xi)}}} \frac{\wh{\wt{\wt{v_\alpha (\xi)}}}}{\wt{{v_\alpha (\xi)}}} =  \nonumber \\ 
&& \quad =~ \bigg(p_\xi+a_{\wt{\xi}}\bigg)\frac{{v_\alpha (\xi)}}{\wt {v_\alpha (\xi)}}- \bigg (q_{\xi}+a_{\wh{\xi}}\bigg)
\frac{{v_\alpha (\xi)}}{\wh {v_\alpha (\xi)}} ~ - ~ \bigg(p_ {\wh{\wh{\wt{\xi}}}}+a_{\wh{\wh{\wt{\wt{{\xi}}}}}}\bigg)
\frac{\wh{\wh{\wt{{v_\alpha (\xi)}}}}}{\wh{\wh{\wt{\wt {v_\alpha (\xi)}}}}}+\bigg(q_{\wh{\wt{\wt{{\xi}}}}}
+a_{\wh{\wh{\wt{\wt{{\xi}}}}}}\bigg)\frac{\wh{\wt{\wt{{v_\alpha (\xi)}}}}}{\wh{\wh{\wt{\wt {v_\alpha (\xi)}}}}}~, \nn \\ 
&& 
\end{eqnarray} 
which can be brought into the form of the lattice modified BSQ equation of \cite{GD}. 

Similarly, solving $\wh{\wt{t_\beta/w_\beta}}$ from \eqref{eq:vvus-b} and inserting this into \eqref{eq:st-b} we get 
\[ \frac{w_\beta}{\wh{\wt{w_\beta}}}\,
\frac{B_{-\delta}(\wh{\xi})\wh{w_\beta}-B_{-\ven}(\wt{\xi})\wt{w_\beta} }
{(p_\xi-b_\xi)\wt{w_\beta}-(q_\xi-b_\xi)\wh{w_\beta} }+
(p_{\wh{\wt{\xi}}}-b_{\wh{\wt{\xi}}})\frac{\wh{\wt{\wt{w_\beta}}}}{\wh{\wt{w_\beta}}}= 
q_\xi+p_{\wh{\xi}}+p_{\wh{\wt{\xi}}}+u_{0,0}-\wh{\wt{\wt{u_{0,0}}}}\  ,     \] 
and subtracting the shifted versions of this relation in both lattice directions, we can eliminate 
the quantities $u_{0,0}$ by using \eqref{eq:wwuu-b} to get a closed-form 9-point relation for $w_\beta$. 
The result is: 
\begin{align} \label{eq:modified BSQ-w}
 &\frac{\wh{w_\beta}}{\wh{\wh{\wt{w_\beta}}}}\,
\frac{B_{-\delta}(\wh{\wh{\xi}})\wh{\wh{w_\beta}}-B_{-\ven}(\wh{\wt{\xi}})\wh{\wt{w_\beta} }}
{(p_{\wh{\xi}}-b_{\wh{\xi}})\wh{\wt{w_\beta}}-(q_{\wh{\xi}}-b_{\wh{\xi}})\wh{\wh{w_\beta}} }+
(p_{\wh{\wh{\wt{\xi}}}}-b_{\wh{\wh{\wt{\xi}}}})\frac{\wh{\wh{\wt{\wt{w_\beta}}}}}{\wh{\wh{\wt{w_\beta}}}}  -\frac{\wt{w_\beta}}{\wh{\wt{\wt{w_\beta}}}}\,
\frac{B_{-\delta}(\wh{\wt{\xi}})\wh{\wt{w_\beta}}-B_{-\ven}(\wt{\wt{\xi}})\wt{\wt{w_\beta} }}
{(p_{\wt{\xi}}-b_{\wt{\xi}})\wt{\wt{w_\beta}}-(q_{\wt{\xi}}-b_{\wt{\xi}})\wh{\wt{w_\beta}} }-
(p_{\wh{\wt{\wt{\xi}}}}-b_{\wh{\wt{\wt{\xi}}}})\frac{\wh{\wt{\wt{\wt{w_\beta}}}}}{\wh{\wt{\wt{w_\beta}}}} \nn  \\
&~~~ =~(p_\xi-b_\xi) \frac{\wt{w_\beta}}{w_\beta}- (q_\xi-b_\xi) \frac{\wh{w_\beta}}{w_\beta} - (p_{\wh{\wt{\wt{\xi}}}}-b_{\wh{\wt{\wt{\xi}}}}) \frac{\wh{\wt{\wt{\wt{w_\beta}}}}}{{\wh{\wt{\wt{w_\beta}}}}}+ (q_{\wh{\wt{\wt{\xi}}}}-b_{\wh{\wt{\wt{\xi}}}}) \frac{\wh{\wh{\wt{\wt{w_\beta}}}}}{{\wh{\wt{\wt{w_\beta}}}}}   \ .
\end{align}
In section 7 we shall bring this equation in the standard for of lattice modified BSQ and present some explicit 
elliptic type solutions.

\subsection{Trilinear equation of the $\tau$-function for the lattice BSQ system}\label{sec-5-3} 

The $\tau$-function associated with the lattice BSQ system is exactly the same as given for the 
lattice KP system, introduced in subsection 2, namely given by \eqref{eq:tau}, 
but of course taking into account the BSQ reduction discussed in subsection 4.2.  The principal relations 
this $\tau$-function obeys were derived in Appendix B, and are given by te shift relations 
\be\label{eq:tau-shift-th}
\frac{\ut{\tau_\xi }}{\tau_\xi }= v_\dd(\xi) \ ,~~\frac{\wt{\tau_\xi }}{\tau_\xi }= w_{-\dd}(\xi) \  , ~~\frac{{\hypohat 0 \tau_\xi }}{\tau_\xi }= v_\varepsilon(\xi) ,~~\frac{\wh{\tau_\xi }}{\tau_\xi }= w_{-\varepsilon}(\xi) \ ,
\ee
where the overtilde $\wt{\tau_\xi}$ and the undertilde $\ut{\tau_\xi}$ denotes the forward and 
backward shift $T_\delta$ and $T_\delta^{-1}$ respectively, implemented also on the argument $\xi$, and 
similarly for the overhat $\wh{\tau_\xi}$ and underhat ${\hypohat 0 \tau_\xi}$, associated with the forward and 
backward shifts $T_\ven$ and $T_\ven^{-1}$ respectively, acting also on the argument $\xi$. 
As a consequence of \eqref{eq:tau-shift-th}, together with the relations 
\bse
\label{relations_ss}
\begin{align}
 &1-\chi^{(1)}_{\alpha,-\delta}(\xi)  s_{\alpha,\beta}(\xi) - \chi^{(1)}_{\beta,\delta}(\wt \xi) \wt {s_{\alpha,\beta}(\xi)}  = \wt{v_\alpha (\xi) }  w_\beta (\xi)  \ , \\
&  \zeta(\xi+\alpha) -\zeta(\alpha)-\zeta(\dd) +\zeta(\beta)-\zeta(\wt{\xi}+\beta)-\frac{1}{2}\frac{\wp'(\beta)+\wp'(\dd)}{ \eta_{\beta}(\xi)-\eta_{-\dd}(\xi)} s_{\alpha,\beta}
(\xi) - \frac{1}{2}\frac{\wp'(\dd)-\wp'(\alpha)}{ \eta_{\alpha}(\wt \xi)-\eta_{\dd}( \wt \xi)} \wt {s_{\alpha,\beta}(\xi) }
\nn  \\
&~~~ =~ s_\alpha (\xi)  \wt{w_\beta (\xi)  }-  v_\alpha (\xi)  \wt{t_\beta (\xi) } + \eta_{\dd}(\wt  \xi)  {v_\alpha (\xi) } \wt{w_\beta (\xi)}   \ ,
\end{align}
\ese
(which are equivalent to \eqref{relations_s}), setting $\alpha=\delta, \beta=-\ven$, we also have
\be\label{eq:stau}
1-\chi^{(1)}_{\delta,-\ven}(\xi) s_{\delta,-\ven}(\xi)= \frac{\wh{\hypotilde 0 \tau_\xi}}{\tau_\xi}\  . 
\ee 
Furthermore, from \eqref{eq:wwuu-b}, setting $\beta=-\delta$, together with \eqref{eq:tau-shift-th}, 
we get 
\be\label{eq:utau}  p_\xi-q_\xi+\wh{u_{0,0}}-\wt{u_{0,0}}= 
(p_\xi-q_\xi)\frac{\tau_\xi\,\wh{\wt{\tau_\xi}}}{\wh{\tau_\xi}\,\wt{\tau_\xi}}  \ee 
Furthermore, from \eqref{eq:st-a} setting $\alpha=\delta$, which implies $a_{\wt{\xi}}=-p_\xi$, we 
see that 
\[ \frac{\wt{s_\delta}}{\wt{v_\delta}}=-(p_\xi+u_{0,0})\  , \] 
which, inserted into \eqref{eq:vvus-a}, with $\alpha=\delta$, implying $A_\delta(\wt{\xi})=P_\delta(\xi),~A_\ven(\wh{\xi})=P_\ven(\wh{\xi})$, 
gives us:
\[ (p_\xi-q_\xi)\left( p_{\wh{\xi}}+q_\xi+p_{\ut{\xi}} +\ut{u_{0,0}}-\wh{\wt{u_{0,0}}}\right)
\frac{\wh{\wt{\tau_\xi}}}{\wh{\tau_\xi}\,\wt{\tau_\xi}} =  
P_\delta(\xi)\frac{\tau_\xi}{\wt{\tau_\xi}\,\ut{\tau_\xi}}-
P_\ven(\wh{\xi}) \frac{\wh{\hypotilde 0 \tau_\xi}}{\wh{\tau_\xi}\,\ut{\tau_\xi}}\  , \] 
where now the coefficients $A_\delta(\wt{\xi})$ and $A_\ven(\wh{\xi})$ when $\alpha=\delta$ are given by 
\[ P_\delta(\xi):=\left. A_\delta(\wt{\xi})\right|_{\alpha=\delta}= 
\frac{1}{2}\frac{\wp''(\delta)}{\wp(\delta)-\wp(\xi)}\   , \quad  
P_\ven(\wh{\xi}):=\left. A_\ven(\wh{\xi})\right|_{\alpha=\delta}
=\frac{1}{2} \frac{ \wp'(\ven)-\wp'(\delta)}{\eta_\ven(\wh{\xi})-\eta_\delta(\wh{\xi})}  \ .  \]   
Adding to the latter relation the one in which the $\wt{~}$- and $\wh{~}$ shifts and the $\delta$- and 
$\ven$ parameters are interchanged, thereby introducing the coefficients 
\[ Q_\ven(\xi):=\left. A_\ven(\wt{\xi})\right|_{\alpha=\ven}= 
\frac{1}{2}\frac{\wp''(\ven)}{\wp(\ven)-\wp(\xi)}\   , \quad  
Q_\delta(\wt{\xi}):=\left. A_\delta(\wt{\xi})\right|_{\alpha=\ven}
=\frac{1}{2} \frac{ \wp'(\delta)-\wp'(\ven)}{\eta_\delta(\wt{\xi})-\eta_\ven(\wt{\xi})}  \ .  \]   
 and eliminating the $u_{0,0}$ using the relation \eqref{eq:utau}, we 
arrive at a \textit{trilinear} equation for the $\tau$-function, namely  
\be\label{eq:tau-tri}
(p_\xi-q_\xi) (p_{\ut{\xi}}-q_{{\hypohat 0 \xi}}) {\hypohat 0 {\hypotilde 0 {\tau_{\xi}}}} {\tau_{\xi}}
\wh{\wt{\tau_{\xi}}}= P_\delta(\xi) {\hypohat 0{\tau_{\xi}}} {\tau_{\xi}} {\wh{\tau_{\xi}}}  
+ Q_\ven(\xi) \ut{\tau_{\xi}} {\tau_{\xi}} {\wt{\tau_{\xi}}} 
- Q_\delta(\wt{\xi}) \ut{\tau_{\xi}} {\hypohat 0 { {\wt {\tau_\xi}}}} \wh{\tau_\xi} 
-P_\ven(\wh{\xi}) { \hypohat 0{{\tau_{\xi}}}} \ut{\wh{\tau_{\xi}}} \wt{\tau_{\xi}} \, .
\ee
A similar trilinear equation like \eqref{eq:tau-tri} was found in \cite{ZZN} for the extended lattice BSQ 
system in the case of the rational parametrisation of the equation.

\subsection{2-parameter variables and generalised lattice Schwarzian BSQ equation}\label{sec-5-3}

Now we will focus on the relations \eqref{relations_s} for the quantity $s_{\alpha,\beta}(\xi)$ and 
derive a closed-form equation for that quantity.  

First, we want to rewrite the relations in a form that is more suitable for a comparison with the 
standard ABS class of quad-lattice equations, cf. \cite{ABS}. To achieve this we introduce the quantities
\footnote{From \eqref{eq:stau} we see that the quantity $S_{\alpha,\beta}$ could  readily expressed in the 
$\tau$-function if we allow there to lattice shifts associated with the parameters $\alpha$ and $\beta$ in 
lieu of $\delta$ and $\ven$. This would require, by MDC, that we have to extend the plane-wave factors 
$\rho_\kp$ and $\sigma_{\kp'}$ in the DL scheme to contain factors with $\alpha$ and $\beta$ as lattice 
parameters, and hence directions in the lattice with discrete variables associated with these additional 
lattice directions.} 
\begin{equation}\label{eq:S} 
	S_{\alpha,\beta}(\xi) = s_{\alpha,\beta}(\xi) -\left(\chi^{(1)}_{\alpha,\beta}(\xi)\right) ^{-1}~. 
\end{equation}
The corresponding shift relations following from \eqref{relations_s}, or equivalently \eqref{eq:ss}, read 
\bse
\label{relations_St}
\begin{align}
& (p_\xi+a_{\wt{\xi}})S_{\alpha,\beta} - (p_\xi-b_\xi)\wt {S_{\alpha,\beta}}  = \wt{v_\alpha }  w_\beta  \ , 
\label{relations_St-1} \\
&  B_{-\delta}(\xi) S_{\alpha,\beta} - A_\delta( \wt{\xi}) \wt {S_{\alpha,\beta}}
 = p_\xi v_\alpha \wt{w_\beta}+v_\alpha \wt{t_\beta} -s_\alpha \wt{w_\beta}\  , 
 \label{relations_St-2}
\end{align}
\ese
(we omit the arguments $\xi$ in the main functions, while keeping them in the coefficients, where it is 
understood that all the shifts acting on those functions also act on that argument $\xi$ of the functions), 
and similarly the corresponding counterparts in the other lattice direction, namely 
\bse
\label{relations_Sh}
\begin{align}
& (q_\xi+a_{\wh{\xi}})S_{\alpha,\beta} - (q_\xi-b_\xi)\wh {S_{\alpha,\beta}}  = \wh{v_\alpha }  w_\beta  \ , 
\label{relations_Sh-1} \\
&  B_{-\ven}(\xi) S_{\alpha,\beta} - A_\ven( \wh{\xi}) \wh {S_{\alpha,\beta}}
= q_\xi v_\alpha \wh{w_\beta}+v_\alpha \wh{t_\beta} -s_\alpha \wh{w_\beta}\  , 
 \label{relations_Sh-2}
\end{align}
\ese
Next, taking the $\wh{~}$ shift of \eqref{relations_St-2} and using \eqref{eq:vvus-b} and the $\{\wh{~},
\varepsilon\}$ version of \eqref{eq:st-a} to replace $\wh {\wt {t_\beta(\xi)}}$ and ~$\wh{s_\alpha 
(\xi)}$, respectively, and using \eqref{eq:wwuu-b} to replace $p-q+\wh{u_{0,0}}-\wt{u_{0,0}}$, we obtain 
\begin{eqnarray}\label{eq:svw} 
(q_\xi+a_{\wh{\xi}}) v_\alpha\,\wh{\wt{w_\beta}} &=&  A_\delta(\wh{\wt{\xi}}) \wh{\wt{S_{\alpha,\beta}}} 
-B_{-\delta}(\wh{\xi}) \wh{S_{\alpha,\beta}} \nn \\ 
&&+ \wh{v_\alpha}\,w_\beta\,\frac{B_{-\delta}(\wh{\xi}) \wh{w_\beta}-B_{-\ven}(\wt{\xi}) \wt{w_\beta}}
{(p_\xi-b_\xi) \wt{w_\beta} -(q_\xi-b_\xi) \wh{w_\beta} }\  .  
\end{eqnarray} 
Multiplying numerator and denominator in the fraction on the right-hand side by $\wh{\wt{v_\alpha}}$ we can use 
\eqref{relations_Sh-1} to express the entire right-hand side in terms of $S_{\alpha,\beta}$, which leads to 
\begin{equation}\label{eq:quadS} 
 v_\alpha\,\wh{\wt{w_\beta}}= \frac{\mathcal{Q}({S_{\alpha,\beta} },\wt{S_{\alpha,\beta} },\wh{S_{\alpha,\beta} },\wh{ \wt{S_{\alpha,\beta} }}) }
{(p_\xi-b_\xi)(q_{\wt{\xi}}+a_{\wh{\wt{\xi}}})\wt{S_{\alpha,\beta}}-
(q_\xi-b_\xi)(p_{\wh{\xi}}+a_{\wh{\wt{\xi}}})\wh{S_{\alpha,\beta}} }
\end{equation} 
where in the denominator some of the terms have disappeared by virtue of the relation
\[ (p_\xi-b_\xi)(q_{\wt{\xi}}-b_{\wt{\xi}})-(q_\xi-b_\xi)(p_{\wh{\xi}}-b_{\wh{\xi}})=
\chi^{(1)}_{\beta,\delta}(\wt{\xi}) \chi^{(1)}_{\beta,\ven}(\wh{\wt{\xi}})-
\chi^{(1)}_{\beta,\ven}(\wh{\xi}) \chi^{(1)}_{\beta,\delta}(\wh{\wt{\xi}})=0\  ,  \]
and where the quad function in the numerator is given by 
\bse \label{num-den} \begin{align}\label{num-N}
\mathcal{Q}({S_{\alpha,\beta}},\wt{S_{\alpha,\beta}},\wh{S_{\alpha,\beta}},\wh{ \wt{S_{\alpha,\beta}}})= 
& p_\beta ( p_{\wh{\xi}}+a_{\wh{\wt{\xi}}}) (p_{\wh{\xi}}+b_{\wh{\wt{\xi}}})\left({S_{\alpha,\beta}}
\wh{S_{\alpha,\beta}} + \gamma \wt {S_{\alpha,\beta}} \wh{\wt{S_{\alpha,\beta}}}\right)    \nn  \\ 
& -q_\beta ( q_{\wt{\xi}}+a_{\wh{\wt{\xi}}}) (q_{\wt{\xi}}+b_{\wh{\wt{\xi}}})\left({S_{\alpha,\beta}}
\wt{S_{\alpha,\beta}} +\gamma' \wh {S_{\alpha,\beta}} \wh{\wt{S_{\alpha,\beta}}}\right)    \nn  \\ 
& + \frac{1}{2}\left(\wp'(\dd)-\wp'(\varepsilon) \right) \left({S_{\alpha,\beta}} \wh{\wt{S_{\alpha,\beta} }} 
+ \gamma'' \wt{S_{\alpha,\beta} }\wh{S_{\alpha,\beta} }    \right)   \ ,
\end{align}
where we have used \eqref{eq:alt-AB} and where the coefficients $\gamma$, $\gamma'$ and $\gamma''$ are given 
by
\begin{eqnarray}\label{den-D} 
\gamma &=&  \frac{p_{-\alpha}}{p_\beta}\,\frac{(p_\xi-a_\xi)(p_\xi-b_\xi)}
{(p_{\wh{\xi}}+a_{\wh{\wt{\xi}}}) (p_{\wh{\xi}}+b_{\wh{\wt{\xi}}})}\  ,  \\ 
\gamma' &=&  \frac{q_{-\alpha}}{q_\beta}\,\frac{(q_\xi-a_\xi)(q_\xi-b_\xi)}
{(q_{\wt{\xi}}+a_{\wh{\wt{\xi}}}) (q_{\wt{\xi}}+b_{\wh{\wt{\xi}}})}\  , \\ 
\gamma'' &=& \frac{\wp'(\beta)+\wp'(\delta)}{\wp'(\delta)-\wp'(\ven)}\,\frac{(p_\xi-a_\xi)(p_\xi-b_\xi)}
{(p_{\wh{\xi}}-a_{\wh{\xi}})(p_{\wh{\xi}}-b_{\wh{\xi}})}-
\frac{\wp'(\beta)+\wp'(\ven)}{\wp'(\delta)-\wp'(\ven)}\,\frac{(q_\xi-a_\xi)(q_\xi-b_\xi)}
{(q_{\wt{\xi}}-a_{\wt{\xi}})(q_{\wt{\xi}}-b_{\wt{\xi}})}\  , 
\end{eqnarray}
\ese 
where we have used the identities\footnote{These identities are also consequences of the chain of relations 
\[ 
\frac{\chi^{(1)}_{\alpha,-\varepsilon}(\wt \xi) }{\chi^{(1)}_{\alpha,-\varepsilon}( \xi) } =\frac{\chi^{(1)}_{\alpha,\dd}(\wt \xi) }{\chi^{(1)}_{\alpha,\dd}(\wh{\wt \xi) } }  =  \frac{\chi^{(1)}_{\alpha,-\dd}(\wh \xi) }{\chi^{(1)}_{\alpha,-\dd}( \xi) }=\frac{\chi^{(1)}_{\alpha,\ven}(\wh \xi) }{\chi^{(1)}_{\alpha,\ven}(\wh{\wt \xi) } } \ , 
\]
which in turn follow from the expressions for the $\chi{(1)}$ quantities in terms of the quantities $\Phi$ 
as in \eqref{chi-1}, and using the identifications
\[ \chi^{(1)}_{\delta,\alpha}(\wt{\xi})= p_\xi-a_\xi\  , \quad 
\chi^{(1)}_{-\delta,\alpha}(\xi)= -(p_\xi+a_{\wt{\xi}})\  ,  \quad 
\chi^{(1)}_{\delta,\beta}(\wt{\xi})= p_\xi-b_\xi\  , \quad 
\chi^{(1)}_{-\delta,\beta}(\xi)= -(p_\xi+b_{\wt{\xi}})\  , 
 \] 
and similarly in the other direction 
 \[ \chi^{(1)}_{\ven,\alpha}(\wt{\xi})= q_\xi-a_\xi\  , \quad 
\chi^{(1)}_{-\ven,\alpha}(\xi)= -(q_\xi+a_{\wh{\xi}})\  ,  \quad 
\chi^{(1)}_{\ven,\beta}(\wt{\xi})= q_\xi-b_\xi\  , \quad 
\chi^{(1)}_{-\ven,\beta}(\xi)= -(q_\xi+b_{\wh{\xi}})\  .  
 \] 
}  
\bse\label{eq:pqids}\be\label{eq:pqids-a}
 \frac{p_{\wh{\xi}}-a_{\wh{\xi}}}{p_\xi-a_\xi}=\frac{q_{\wt{\xi}}-a_{\wt{\xi}}}{q_\xi-a_\xi}=
\frac{p_\xi+a_{\wt{\xi}}}{p_{\wh{\xi}}+a_{\wh{\wt{\xi}}}}
=\frac{q_\xi+a_{\wh{\xi}}}{q_{\wt{\xi}}+a_{\wh{\wt{\xi}}}}\ ,   
\ee 
(and similarly with $a$'s replaced by $b$'s).  
The latter are a consequence of the relations 
\be\label{eq:pqids-b}  
(p_\xi-a_\xi)(p_\xi+a_{\wt{\xi}})=\wp(\delta)-\wp(\alpha)\  ,  \quad 
(q_\xi-b_\xi)(q_\xi+b_{\wh{\xi}})=\wp(\ven)-\wp(\beta)\  , 
\ee\ese  
and similar relations with $a$ and $b$ and $\delta$ and $\ven$ interchanged. 

We can bring the quadrilateral $\mathcal{Q}$ into a more standard form by performing a further 
(gauge) transformation, namely  
\bse\label{eq:gauge} 
 \be
 S_{\alpha,\beta}(\xi) =  \phi_{\alpha,\beta}(\xi) \frak{u}_{\alpha,\beta}(\xi)    \ , 
 \ee
 where the function $\phi_{\alpha,\beta}(\xi)$ solves the compatible system of ordinary first order 
 difference equations 
\begin{equation}\label{eq:gauge-2} 
\wt{\phi_{\alpha,\beta}(\xi)} = \left( \frac{p_\beta}{p_{-\alpha}}
\frac{(p_\xi+a_{\wt{\xi}})(p_\xi+b_{\wt{\xi}})}{(p_\xi-a_\xi)(p_\xi-b_\xi)}\right)^{1/2}
\phi_{\alpha,\beta}(\xi)\  , \quad 
\wh{\phi_{\alpha,\beta}(\xi)} = \left( \frac{q_\beta}{q_{-\alpha}}
\frac{(q_\xi+a_{\wh{\xi}})(q_\xi+b_{\wh{\xi}})}{(q_\xi-a_\xi)(q_\xi-b_\xi)}\right)^{1/2}
\phi_{\alpha,\beta}(\xi)\  . 
\end{equation} 
 \ese
Implementing the change of variables \eqref{eq:gauge} the relation \eqref{eq:quadS} changes into 
\bse \label{eq:Quad}\begin{equation}\label{eq:quadu} 
 2 \frac{(p_\xi+a_{\wt{\xi}})(q_{\wt{\xi}}+a_{\wh{\wt{\xi}}})}{P_\alpha^- Q_\alpha^- \wh{\wt{\phi_{\alpha,\beta}(\xi)}}}\,
 v_\alpha\,\wh{\wt{w_\beta}}= \frac{\bar{\mathcal{Q}}({\frak{u}_{\alpha,\beta} },\wt{\frak{u}_{\alpha,\beta} },\wh{\frak{u}_{\alpha,\beta} },\wh{ \wt{\frak{u}_{\alpha,\beta} }}) }
{Q_\alpha^- P_\beta^+\wt{\frak{u}_{\alpha,\beta}}-
P_\alpha^- Q_\beta^+ \wh{\frak{u}_{\alpha,\beta}} }
\end{equation}
where now
 \begin{align}\label{num-N}
\bar{\mathcal{Q}}({\frak{u}_{\alpha,\beta}},\wt{\frak{u}_{\alpha,\beta}},\wh{\frak{u}_{\alpha,\beta}},\wh{ \wt{\frak{u}_{\alpha,\beta}}})= 
& P_\alpha^- P_\beta^+\left({\frak{u}_{\alpha,\beta}}
\wh{\frak{u}_{\alpha,\beta}} + \wt {\frak{u}_{\alpha,\beta}} \wh{\wt{\frak{u}_{\alpha,\beta}}}\right)    \nn  \\ 
& -Q_\alpha^- Q_\beta^+ \left({\frak{u}_{\alpha,\beta}}
\wt{\frak{u}_{\alpha,\beta}} + \wh {\frak{u}_{\alpha,\beta}} \wh{\wt{\frak{u}_{\alpha,\beta}}}\right)    \nn  \\ 
& +\left(\wp'(\dd)-\wp'(\varepsilon) \right) \left({\frak{u}_{\alpha,\beta}} \wh{\wt{\frak{u}_{\alpha,\beta} }} 
+ \wt{\frak{u}_{\alpha,\beta} }\wh{\frak{u}_{\alpha,\beta} }    \right)   \ ,
\end{align}
\ese 
where 
\be \label{nota:PQ} 
 (P_\alpha^\pm)^2= \wp'(\delta)\pm\wp'(\alpha)\ , \quad (P_\beta^\pm)^2= \wp'(\delta)\pm\wp'(\beta)\ , \quad 
(Q_\alpha^\pm)^2= \wp'(\ven)\pm\wp'(\alpha)\ , \quad (Q_\beta^\pm)^2= \wp'(\ven)\pm\wp'(\beta)\ . 
\ee
Furthermore, rewriting \eqref{relations_St-1} in terms of $\frak{u}_{\alpha,\beta}$ we have 
\begin{equation}\label{eq:vw-urel} 
P_\alpha^- \frak{u}_{\alpha,\beta}-P_\beta^+\wt{\frak{u}_{\alpha,\beta}}=
\frac{ P_\alpha^- }{(p_\xi+a_{\wt{\xi}}) \phi_{\alpha,\beta}(\xi)} \wt{v_\alpha}\,w_\beta\  . 
\end{equation}
Eq. \eqref{eq:Quad}, together with \eqref{eq:vw-urel} and its counterpart in the other direction (with $\delta$ 
replaced by  $\ven$ and $\wt{~}$ replaced by $\wh{~}$) forms a closed quadrilateral system for the 
the three variables $\frak{u}_{\alpha,\beta}$, $v_\alpha$ and $w_\beta$, which is the most general form of 
the lattice BSQ system. However, by using the identity
\[ \frac{ \wh{(v_\alpha\wh{\wt{w_\beta}})}}{\wt{(v_\alpha\wh{\wt{w_\beta}})}}= 
\frac{ \wh{\wh{\wt{(\wt{v_\alpha}w_\beta)}}} }{ \wh{\wt{\wt{(\wh{v_\alpha}w_\beta)}}}  }\,\frac{\wh{v_\alpha} w_\beta}{\wt{v_\alpha} w_\beta} \  ,   \] 
and inserting the expressions for the quantities in brackets in terms of $\frak{u}_{\alpha,\beta}$ one 
can write down a closed form 9-point equation in terms of  $\frak{u}_{\alpha,\beta}$ alone, see \cite{ZZN}. 
In the elliptic case that equation takes exactly the same form, except for the parametrisation of the lattice  
parameters involved, namely 
\begin{eqnarray}\label{eq:NQC-BSQ} 
&& \frac{\bar{\mathcal{Q}}(\wh{{\frak{u}_{\alpha,\beta}}},\wh{\wt{\frak{u}_{\alpha,\beta}}},
\wh{\wh{\frak{u}_{\alpha,\beta}}},\wh{\wh{ \wt{\frak{u}_{\alpha,\beta}}}})}
{\bar{\mathcal{Q}}({\wt{\frak{u}_{\alpha,\beta}}},\wt{\wt{\frak{u}_{\alpha,\beta}}},\wh{\wt{\frak{u}_{\alpha,\beta}}},\wh{ \wt{\wt{\frak{u}_{\alpha,\beta}}}})}= \nn \\ 
&& = \frac{\left( Q_\alpha^- \frak{u}_{\alpha,\beta}-Q_\beta^+\wh{\frak{u}_{\alpha,\beta}}  \right) 
\left( Q_\alpha^- P_\beta^+ \wh{\wt{\frak{u}_{\alpha,\beta}}}-
P_\alpha^- Q_\beta^+ \wh{\wh{\frak{u}_{\alpha,\beta}}} \right) \left( P_\alpha^- \wh{\wh{\wt{\frak{u}_{\alpha,\beta}}}}-P_\beta^+\wh{\wh{\wt{\wt{\frak{u}_{\alpha,\beta}}}}}  \right) }{
\left( P_\alpha^- \frak{u}_{\alpha,\beta}-P_\beta^+\wt{\frak{u}_{\alpha,\beta}}  \right) 
\left( Q_\alpha^- P_\beta^+ \wt{\wt{\frak{u}_{\alpha,\beta}}}-
P_\alpha^- Q_\beta^+ \wt{\wh{\frak{u}_{\alpha,\beta}}} \right) \left(Q_\alpha^- \wh{\wt{\wt{\frak{u}_{\alpha,\beta}}}}-Q_\beta^+\wh{\wh{\wt{\wt{\frak{u}_{\alpha,\beta}}}}}  \right)} \ .
\end{eqnarray} 
The latter equation, which can be thought of as a generalisation of the lattice Schwarzian Boussinesq equation, \cite{N1}, 
contains both the 'regular' lattice BSQ equation,  \eqref{eq:BSQ} as well as the 
lattice modified BSQ equation as special coalescence limits on the parameters $\alpha$ and $\beta$. Thus, the 9-point is the 
most general lattice BSQ system known so far.

\section{Explicit examples of elliptic lattice BSQ solutions}\label{sec-7}
\setcounter{equation}{0}

Here we present the explicit seed solutions for the lattice BSQ systems in standard form, 
and show how DL approach reduces to the Cauchy-matrix scheme for elliptic $N$-soliton 
solutions. 

\subsection{Seed and elliptic one-soliton solution of the lattice BSQ system} \label{sec-7-1}

Consider the following transformation
\bse
\label{trans:lBSQ}
\begin{align}
 u_{0,0} &~=~ x_0-u~, \label{trans:lBSQ-u}\\
u_{1,0} &~=~y_{0}  - v - x_0 u_{0,0}~, \label{trans:lBSQ-v}\\
u_{0,1}& ~=~ z_{0} - w - x_0 u_{0,0}~, \label{trans:lBSQ-w}
\end{align}
\ese
with
\bse
\begin{align}
x_0 &= \zeta(\xi)+n\zeta(\delta)+m\zeta(\varepsilon)-\zeta(\xi_0)~,\label{u0}\\
y_0 &= \frac{1}{2}x^2_0-\frac{1}{2}\wp(\xi)+\frac{1}{2}(n\wp(\delta)+m\wp(\varepsilon)+\wp(\xi_0))~,\label{v0}\\
z_0
&=\frac{1}{2}x^2_0-\frac{1}{2}\wp(\xi)-\frac{1}{2}(n\wp(\delta)+m\wp(\varepsilon)+\wp(\xi_0))~,\label{w0}
\end{align}
\label{0SS}
\ese
and $u=u_{n,m},~v=v_{n,m},~w=w_{n,m}.$ Imposing this transformation on \eqref{eq:u-t}, \eqref{eq:u-h} and \eqref{BSQ_uvw}, rewritten as
\bse
\label{def:uvw}
\begin{align}
& p_\xi \wt {u_{0,0}}+\wt {u_{0,1} }  = p_\xi u_{0,0} - u_{1,0} - \wt {u_{0,0}} u_{0,0} \ ,\\
& q_\xi \wh {u_{0,0}}+\wh {u_{0,1} }  = q_\xi u_{0,0} - u_{1,0} - \wh {u_{0,0}} u_{0,0}\ , \\
& \frac{1}{2}\frac{\wp'(\dd)-\wp'(\ven)}{ p_\xi-q_\xi+\wh{u_{0,0}}-\wt{u_{0,0}}}  =\frac{1}{2}\frac{\wp'(\dd)-\wp'(\ven)}{p_\xi-q_\xi} +\wh{ \wt{u_{1,0}}} +u_{0,1}+ u_{0,0}\wh{ \wt{u_{0,0}}}
+(p_{\wh{\xi}}+q_\xi) (\wh{ \wt{u_{0,0}}}- u_{0,0} )  \ ,
\end{align}
\ese
 we obtain the following system which is called (B-2) system in Ref. \cite{H}
\bse \label{DB2}
\begin{align}
&\wt w - u\wt u+v=0~,\label{DB-a}\\
& \wh w - u\wh{u}+v=0~,\label{DB-b}\\
& \frac{1}{2}\frac{\wp'(\dd)-\wp'(\ven)}{\wh{u}-\wt{u}} = w - u \wh{ \wt{u}}+\wh {\wt{v}}~. \label{DB-c}
\end{align}
\ese 
 By eliminating $v$ and $w$ in the above system, we also get the lattice BSQ equation  \eqref{eq:BSQ} with $P-Q=  \frac{1}{2}(\wp'(\dd)-\wp'(\ven))$. It is noted that \eqref{eq:BSQ}  can be transformed into  \eqref{eq:lBSQ-u}  by the transformation \eqref{trans:lBSQ-u} and \eqref{u0}.  
Moreover, from the seed solution of the lattice system \eqref{DB2}, we can construct  the elliptic one-soliton solution through the B\"acklund transformations, written as (see \cite{S-2017})  
\begin{subequations}\label{sol-DBu}
\begin{align}
 & u_{n,m}^{1SS} 
=x_0+\frac{\eta_\xi(\kappa)\Phi_\xi(\kappa)+\eta_{\xi}(\omega_1(\kappa))\Phi_{\xi}(\omega_1(\kappa))\rho_1
+\eta_{\xi}(\omega_2(\kappa)) \Phi_{\xi}(\omega_2(\kappa))\rho_2}{\Phi_\xi(\kappa)+\Phi_{\xi}(\omega_1(\kappa))\rho_1+\Phi_{\xi}(\omega_2(\kappa))\rho_2}~ , \label{1ss-u}\\
 & v_{n,m}^{1SS} 
= y_0+  x_0\frac{\eta_\xi(\kappa)\Phi_\xi(\kappa)+\eta_{\xi}(\omega_1(\kappa))\Phi_{\xi}(\omega_1(\kappa))\rho_1+
                 \eta_{\xi}(\omega_2(\kappa)) \Phi_{\xi}(\omega_2(\kappa))\rho_2}{\Phi_\xi(\kappa)
                 +\Phi_{\xi}(\omega_1(\kappa))\rho_1+\Phi_{\xi}(\omega_2(\kappa))\rho_2}  \nonumber\\
&~~~~~~~+\frac{ (\wp(\kappa)+\wp(\xi))\Phi_\xi(\kappa)+(\wp(\omega_1(\kappa))+\wp(\xi))\Phi_{\xi}(\omega_1(\kappa))\rho_1
               +(\wp(\omega_2(\kappa))+\wp(\xi))\Phi_{\xi}(\omega_2(\kappa))\rho_2}
               {\Phi_\xi(\kappa)+\Phi_{\xi}(\omega_1(\kappa))\rho_1+\Phi_{\xi}(\omega_2(\kappa))\rho_2}~,
\label{1ss-v} \\
& w_{n,m}^{1SS}=z_0+ x_0\frac{\eta_\xi(\kappa)\Phi_\xi(\kappa)+\eta_{\xi}(\omega_1(\kappa))\Phi_{\xi}(\omega_1(\kappa))\rho_1
+\eta_{\xi}(\omega_2(\kappa)) \Phi_{\xi}(\omega_2(\kappa))\rho_2}
{\Phi_\xi(\kappa)+\Phi_{\xi}(\omega_1(\kappa))\rho_1+\Phi_{\xi}(\omega_2(\kappa))\rho_2}~.
\label{1ss-w1}
\end{align}
\end{subequations}
Here
\begin{equation}
\kappa+\omega_1(\kappa)+\omega_2(\kappa)\equiv 0(\textrm{mod\
\ period\ \ lattice})~,
\label{root}
\end{equation}
and $\rho_i~(i=1,2)$
are the plain wave factors
\begin{align} \label{rho_nm}
  \rho_i=\rho_i(n,m;\kappa)&=\biggl(\frac{\Phi_{\delta}(-\omega_i(\kappa))}
 {\Phi_{\delta}(-\kappa)}\biggr)^n\biggl(\frac{\Phi_{\varepsilon}(-\omega_i(\kappa))}{\Phi_{\varepsilon}(-\kappa)}\biggr)^m \cdot \frac{\rho_i^0}{\rho_0^0}~,
~~(i=1,2)~,
\end{align}
where $\kappa$ is the soliton parameter.

\subsection{Elliptic seed solution for the lattice modified BSQ equation} \label{sec-7-2}

To derive the standard form of the lattice modified BSQ equation, we consider the following transformation
\bse
\label{trans:lmodified BSQ}
\begin{align}
 u_{0,0} &~=-U+x_0~, \label{trans:lBSQ-U}\\
v_\alpha (\xi) &~=\biggl(\Phi_{\alpha}(-\delta)\biggr)^n\biggl(\Phi_{\alpha}(-\varepsilon)\biggr)^m \frac{1}{\Phi_{\alpha}(\xi)}  V ~, \label{trans:lBSQ-V}\\
s_\alpha (\xi)& ~=\biggl(\Phi_{\alpha}(-\delta)\biggr)^n\biggl(\Phi_{\alpha}(-\varepsilon)\biggr)^m \frac{1}{\Phi_{\alpha}(\xi)}\biggl( Y-(x_0+2C)V \biggr) ~, \label{trans:lBSQ-Y}
\end{align}
\ese
where $U=U_{n,m},~V=V_{n,m},~Y=Y_{n,m},$ and
\begin{align} 
x_0 = \zeta(\xi)+n\zeta(\delta)+m\zeta(\varepsilon)-\zeta(\xi_0)+nC+mC~,\label{x0-m}
\end{align}
and $C$ is an arbitrary complex constant. 
By using this transformation, \eqref {eq:st-a} and its
$\{\wh{~},\varepsilon\}$ counterparts and \eqref{eq:vvus-a}  are turned to the following system corresponding to the (A-2) system in Ref. \cite{H}
\bse \label{mDB}
\begin{align}
&\wt Y = U\wt V-V~,~~~\wh Y = U\wh V-V~,\label{mDB-ab} \\
&Y=V \wh {\wt{U}} +\frac{ \frac{1}{2}(P_\alpha^- )^2 \wt V -\frac{1}{2}(Q_\alpha^- )^2\wh{V} }{\wh{U}- \wt{U}}~, \label{mDB-c}
\end{align}
\ese
in which we have used the notation in \eqref{nota:PQ}. Eliminating $U$ and $Y$ in the above system, we obtain the lattice 
modified BSQ equation, written as 
\begin{eqnarray} \label{eq:modified BSQ-v}
\bigg(\frac{\frac{1}{2}(P_\alpha^- )^2 \wh{\wt V}-\frac{1}{2}(Q_\alpha^- )^2 \wh{\wh V}}{\wh{\wh V}-\wh{\wt V}} \bigg) 
\frac{\wh{\wh{\wt{V}}}}{\wh{V}} - 
\bigg(\frac{\frac{1}{2}(P_\alpha^- )^2\wt{\wt V}-\frac{1}{2}(Q_\alpha^- )^2 \wh{\wt V}}{\wh{\wt V}-\wt{\wt V}} \bigg) 
\frac{\wh{\wt{\wt{V}}}}{\wt{V}} =  \frac{V}{\wt V}- \frac{V}{\wh V}- \frac{\wh{\wh{\wt{V}}}}{\wh{\wh{\wt{\wt V}}}}+
\frac{\wh{\wt{\wt{V}}}}{\wh{\wh{\wt{\wt V}}}}~.
\end{eqnarray} 

It is noted that by setting $(\bU_\xi)_{0,0}=0$, $V= \biggl(\Phi_{\alpha}(-\delta)\biggr)^{-n}\biggl(\Phi_{\alpha}(-
\varepsilon)\biggr)^{-m} \Phi_{\alpha}(\xi) $ from \eqref{trans:lBSQ-V} and \eqref{BSQ_u-v} provides a seed solution for the 
lattice BSQ equation. 

Alternatively, for the system \eqref{eq:vvus-b} and its
$\{\wh{~},\varepsilon\}$ counterparts and \eqref{eq:st-b}, 
we employ the following transformation
\bse
\label{trans:lmodified BSQ}
\begin{align}
 u_{0,0} &~=-U+x_0~, \label{trans:lBSQ-U2}\\
w_\beta (\xi) &~=\biggl(\frac{1}{\Phi_{\beta}(\delta)}\biggr)^n\biggl(\frac{1}{\Phi_{\beta}(\varepsilon)}\biggr)^m \frac{1}
{\Phi_{\beta}(\xi)}  W ~, \label{trans:lBSQ-W}\\
t_\beta (\xi)& ~=\biggl(\frac{1}{\Phi_{\beta}(\delta)}\biggr)^n\biggl(\frac{1}{\Phi_{\beta}(\varepsilon)}\biggr)^m \frac{1}
{\Phi_{\beta}(\xi)} \biggl( Z-(x_0-2C)V \biggr) ~, \label{trans:lBSQ-Z}
\end{align}
\ese
and obtain
\bse \label{mDB-2}
\begin{align}
&Z = \wt U W- \wt W~,~~~Z = \wh U W- \wh W~,,\label{mDB2-ab} \\
&\wh {\wt{Z}}=U \wh {\wt{W}} +\frac{ \frac{1}{2}(P_\beta^+)^2 \wh W -\frac{1}{2}(Q_\beta^+)^2\wt{W} }{\wh{U}- \wt{U}}~, \label{mDB2-c}
\end{align}
\ese
with $U=U_{n,m},~W=W_{n,m},~Z=Z_{n,m},$ and constant $C$, $x_0 $ in \eqref{trans:lmodified BSQ} defined by \eqref{x0-m}.
Eliminating $U$ and $Z$ in \eqref{mDB-2} or noticing the reversal symmetry (see \cite{H}) of the system \eqref{mDB} we have an alternate form of the lattice modified BSQ equation, written as 
\begin{eqnarray} \label{eq:modified BSQ-W}
\bigg(\frac{\frac{1}{2}(P_\beta^+)^2 \wh{\wh W} -\frac{1}{2}(Q_\beta^+)^2 \wh{\wt W}}{\wh{\wh W}-\wh{\wt W}} \bigg) \frac{\wh{W}}{\wh{\wh{\wt{W}}}} - 
\bigg(\frac{\frac{1}{2}(Q_\beta^+)^2 \wt{\wt W}-\frac{1}{2}(P_\beta^+)^2  \wh{\wt W} }{\wt{\wt W}-\wh{\wt W}} \bigg) \frac{\wt{W}}{\wh{\wt{\wt{W}}}} =  \frac{\wt W}{ W}- \frac{\wh W}{ W}- \frac{\wh{\wh{\wt{\wt W}}} }{\wh{\wh{\wt{W}}}}+\frac{\wh{\wh{\wt{\wt W}}}}{\wh{\wt{\wt{W}}}}~,
\end{eqnarray} 
which is related to the equation \eqref{eq:modified BSQ-w} through the transformation \eqref{trans:lBSQ-W}.

\subsection{Elliptic seed solution for the lattice Schwarzian BSQ equation} \label{sec-7-3}

Now we want to present the lattice Schwarzian BSQ equation from the three-component lattice system, i.e, \eqref{relations_St-1}, \eqref{relations_Sh-1} and \eqref{eq:svw}. Employing the following transformation to this system
\bse
\label{trans:lsBSQ}
\begin{align}
v_\alpha (\xi) &~=\biggl(\Phi_{-\alpha}(\delta)\biggr)^n\biggl(\Phi_{-\alpha}(\varepsilon)\biggr)^m \frac{1}{\Phi_{\alpha}(\xi)}  V ~, \label{trans:lsBSQ-V}\\
w_\alpha (\xi) &~=\biggl(\Phi_{\beta}(\delta)\biggr)^{-n}\biggl(\Phi_{\beta}(\varepsilon)\biggr)^{-m} \frac{1}{\Phi_{\beta}(\xi)}  W ~, \label{trans:lsBSQ-W}\\
S_{\alpha,\beta} (\xi)& ~=\biggl(\frac{\Phi_{-\alpha}(\delta)}{\Phi_{\beta}(\delta)}\biggr)^n \biggl(\frac{\Phi_{-\alpha}(\ven)}{\Phi_{\beta}(\ven)}\biggr)^m \frac{1}{\Phi_{\alpha}(\xi) \Phi_{\beta}(\xi)} H ~, \label{trans:lsBSQ-H}
\end{align}
\ese
where $V=V_{n,m},~W=W_{n,m},~H=H_{n,m}$, we get the following system corresponding to the (C-3) system in Ref. \cite{H}
\bse \label{DB}
\begin{align}
&H-\wt H = \wt V W~,~~~H-\wh H = \wh V W~,\label{sDB-ab} \\
&V \wh {\wt{W}} = \frac{\frac{1}{2}(Q_\beta^+)^2 \wh V \wt W -\frac{1}{2}(P_\beta^+)^2 \wt V \wh{W} } {\wt{W}- \wh{W}}  W + \frac{1}{2}(\wp'(\alpha)+\wp'(\beta))\wh{ \wt H}~, \label{sDB-c}
\end{align}
\ese
which leads to the lattice Schwarzian  BSQ equation, written as 
\begin{eqnarray} \label{eq:lsBSQ}
&&
\frac{(\wh H - \wh {\wt H}) (\wh{\wh H}-\wh{\wh {\wt{H}}})(P_\beta^+)^2- (\wh H - \wh {\wh H}) (\wh{\wt H}-\wh{\wh{ \wt{H}}})(Q_\beta^+)^2 -(\wp'(\alpha)+\wp'(\beta) )\wh{\wh {\wt{H}}}(\wh{\wt H}-\wh{\wh {H}}) }{(\wt H - \wt {\wt H}) (\wh{\wt H}-\wh{\wt {\wt{H}}})(P_\beta^+)^2- (\wt H - \wh {\wt H}) (\wt{\wt H}-\wh{\wt {\wt{H}}})(Q_\beta^+)^2 -(\wp'(\alpha)+\wp'(\beta) )\wh{\wt {\wt{H}}}(\wt{\wt H}-\wh{\wt {H}}) } ~ \nonumber \\
&& \quad\quad\quad\quad =~ \frac{(H-\wh H)(\wh {\wt H}-\wh{\wh H})(\wh{\wh {\wt H}}-\wh{\wh{\wt{\wt H}}})}{(H-\wt H)(\wt {\wt H}-\wh{\wt H})(\wh{\wt {\wt H}}-\wh{\wh {\wt{\wt H}}})}~.
\end{eqnarray} 
From the transformations \eqref{trans:lsBSQ-H} and \eqref{eq:S} and the definition of $s_{\alpha,\beta}(\xi)$ given by \eqref{BSQ_u-s}, we know that $H= -\biggl(\frac{\Phi_{\beta}(\delta)}{\Phi_{-\alpha}(\delta)}\biggr)^n \biggl(\frac{\Phi_{\beta}(\ven)}{\Phi_{-\alpha}(\ven)}\biggr)^m \Phi_{\alpha+\beta}(\xi) $ is  an elliptic seed solution of the lattice Schwarzian  BSQ equation \eqref{eq:lsBSQ}.

\subsection{Cauchy matrix scheme for elliptic $N$-soliton solutions} \label{sec-7-2} 

We now consider the elliptic $N$-soliton solutions, starting from
choosing a particular measure in \eqref{eq:inteqa} written as 
\be \label{measure-BSQ}
d\mu(\ell ,\ell ^\prime)=\frac{1}{2\pi  \mathrm{i}}\frac{1}{2\pi \mathrm{i}} \sum_{j=0}^2 \sum_{j'=1}^{N_j} \Lambda_{j,j'}
 \left(\frac{d\ell~ d\ell'}{(\ell- \kappa_{j,j'})(\ell'+\omega_j(\kappa_{j,j'}))}   \right),
\ee
where $\Lambda_{j,j'}$ are residues
of the measures $d\mu_j(\ell)$  in \eqref{eq:inteqBSQ}, and $\mathrm{i}$ is the imaginary unit. 
In this case the linear integral equation \eqref{eq:inteqa} is reduced to 
\begin{equation}\label{eq:uksoliton-1}
\bu_\kappa +\sum_{j=0}^2\sum_{j'=1}^{N_j} \Lambda_{j,j'}\,\rho_\kappa\, \Phi_{\xi}(\kappa-\omega_j(\kappa_{j,j'}) )\, \sg_{-\oa_j(\kappa_{j,j'})} \bu_{\kappa_{j,j'}}=\rho_\kappa{\Phi_\xi(\Ld)}  \bc_\kappa \ .
\end{equation}

Setting $\kappa=\kappa_{i,i'}$, where $i=0,1,2$, $i'=1,\dots, N_{i}$ in eq. \eqref{eq:uksoliton-1}, we then have a linear
system for the quantities $u_{\kappa_{i,i'}}$ from the equation \eqref{eq:uksoliton-1}, i.e.,
\begin{align}
 & \left(\bu_{\kappa_{0,1}},\dots,\bu_{\kappa_{0,N_0}},\bu_{\kappa_{1,1}},\dots,\bu_{\kappa_{1,N_1}},\bu_{\kappa_{2,1}},\dots,\bu_{\kappa_{2,N_2}}\right) ({\boldsymbol 1}+\bM)  \nn \\
& ~~~~~~= \brr^T   {\rm diag}  \left(\bc_{\kappa_{0,1}},\dots,\bc_{\kappa_{0,N_0}},\bc_{\kappa_{1,1}},\dots,\bc_{\kappa_{1,N_1}},\bc_{\kappa_{2,1}},\dots,\bc_{\kappa_{2,N_2}}\right) \ , \label{eq:uksoliton-sq}
\end{align}
where 
\begin{align}
 & \brr= \Big(\rho_{\kappa_{0,1}} \Phi_\xi(\kappa_{0,1}),\dots,\rho_{\kappa_{0,N_0}}\Phi_\xi(\kappa_{0,N_0});\rho_{\kappa_{1,1}}\Phi_\xi(\kappa_{1,1}),\dots,\rho_{\kappa_{1,N_1}}\Phi_\xi(\kappa_{1,N_1}); \nn \\
&~~~~~~~~~~~~~ \rho_{\kappa_{2,1}}\Phi_\xi(\kappa_{2,1}),\dots,\rho_{\kappa_{2,N_2}}\Phi_\xi(\kappa_{2,N_2})\Big)^T \ ,
\end{align}
and $\bM$ is a 3$\times$3 block-Cauchy matrix with rectangular blocks of size $N_i\times N_j~(i,j=0,1,2)$  with elements
\begin{equation}\label{eq:MCauchy}
\bM_{I,J}=\left( M_{(i,i'),(j,j')}\right)_{i,j=0,1,2;i'=1,\dots,N_i,j'=1,\dots,N_j}:=
\rho_{\kappa_{i,i'}}\Lambda_{j,j'}\sg_{-\oa_j(\kappa_{j,j'})} \Phi_{\xi}({\kappa_{i,i'}-\oa_j(\kappa_{j,j'})})\ .
\end{equation}
We have used the capital compound indices
$I=(i,i')$ and $J=(j,j')$ to simplify the notation. 

Now we can make the soliton solutions now explicit by assuming that the coefficients $\Ld_{j,j'}$ are chosen such that the
matrix ${\boldsymbol 1}+\bM$ is invertible, in which case solutions to \eqref{eq:uksoliton-1} or \eqref{eq:uksoliton-sq} could be written out explicitly. Making use of \eqref{eq:U-in} and \eqref{measure-BSQ} we can also obtain the explicit expression for $\bU_\xi $ and make the following identifications for the quantities introduced in
 \eqref{BSQ_u}:
\bse
\label{def:BSQ_u}
\begin{align}
&s_{\alpha} (\xi)= \eta_{\alpha}(\xi) - \brr ^T \left(\chi^{(1)}_{\alpha,\xi}(\bK)\right) ^{-1} ({\boldsymbol 1}+\bM)^{-1}  \eta_\xi(\bL)\, \bs \ ,  \\
&t_{\beta} (\xi)= \eta_{\beta}(\xi) -\brr ^T  \, \eta_\xi(\bK) \,({\boldsymbol 1}+\bM)^{-1}   \left(\chi^{(1)}_{\beta,\xi}(\bL)\right) ^{-1} \, \bs \ , \\
&r_{\alpha} (\xi)= \wp(\alpha) - \brr ^T \left(\chi^{(1)}_{\alpha,\xi}(\bK)\right) ^{-1} ({\boldsymbol 1}+\bM)^{-1}  \wp(\bL)\, \bs \ , \\ 
&z_{\beta} (\xi)= \wp(\beta) -\brr ^T \wp(\bK)  \, ({\boldsymbol 1}+\bM)^{-1}   \left(\chi^{(1)}_{\beta,\xi}(\bL)\right) ^{-1} \, \bs  \ ,\\
&v_{\alpha} (\xi)= 1 - \brr ^T \left(\chi^{(1)}_{\alpha,\xi}(\bK)\right) ^{-1} ({\boldsymbol 1}+\bM)^{-1}  \, \bs \ ,  \\
&w_{\beta} (\xi)=1 -\brr ^T ({\boldsymbol 1}+\bM)^{-1}   \left(\chi^{(1)}_{\beta,\xi}(\bL)\right) ^{-1} \, \bs  \ , \\
&s_{\alpha,\beta}(\xi)=  \brr ^T  \left(\chi^{(1)}_{\alpha,\xi}(\bK)\right) ^{-1}  ({\boldsymbol 1}+\bM)^{-1}  \left(\chi^{(1)}_{\beta,\xi}(\bL)\right) ^{-1} \, \bs \ ,
\end{align}
\ese
in which $\bK$ and $\bL$ are the block diagonal matrices:
\bse\label{eq:KK'}\begin{eqnarray}
\bK&=&{\rm diag}\left(\kappa_{1,1},\dots, \kappa_{1,N_1};\kappa_{2,1},\dots,\kappa_{2,N_2};\kappa_{3,1},\dots,\kappa_{3,N_3}\right) \  , \label{eq:KK'a} \\
\bL&=&{\rm diag}\left(-\oa_1(\kappa_{1,1}),\dots,-\oa_1(\kappa_{1,N_1});-\oa_2(\kappa_{2,1}),\dots,
-\oa_2(\kappa_{2,N_2});-\oa_3(\kappa_{3,1}),\dots,-\oa_3(\kappa_{3,N_3})\right)\  , \nn \\
&& \label{eq:KK'b}
\end{eqnarray}\ese
and $\bs$ is a colummn vector
\be\label{eq:sc}
\bs = \left(\sg_{-\oa_1(\kappa_{1,1})},\dots,\sg_{-\oa_1(\kappa_{1,N_1})};\sg_{-\oa_2(\kappa_{2,1})},\dots,
\sg_{-\oa_2(\kappa_{2,N_2})};\sg_{-\oa_3(\kappa_{3,1})},\dots,\sg_{-\oa_3(\kappa_{3,N_3})}\right) ^T \  . 
\ee
Note that in eqs. \eqref{def:BSQ_u} the matrices $\wp(\bK)$, $\eta_\xi(\bL)$, $\chi^{(1)}_{\alpha,\xi}(\bK)$ and 
$\chi^{(1)}_{\beta,\xi}(\bL)$ denote the diagononal marices with as entries the corresponding elliptic functions 
evaluated at the entries of the corresponding matrices $\bK$ or $\bL$, e.g. 
\[ \wp(\bK)= {\rm diag}\left( \wp(\kappa_{1,1}),\dots, \wp(\kappa_{1,N_1});\wp(\kappa_{2,1}),\dots,\wp(\kappa_{2,N_2});
\wp(\kappa_{3,1}),\dots,\wp(\kappa_{3,N_3}) \right)\ ,  \] 
etc. 
Note that all these expressions \eqref{def:BSQ_u} can be written out explicitly after choosing the various
parameters of the solution. In this sense, $\tau$-function can also be expressed by 
\be\label{eq:tau-BSQ-CM}
\tau_\xi = \det\,_{N \times N}\left(
{\boldsymbol 1}+\bM \right)\ , \quad N=N_1+N_2+N_3\ ,  
\ee
where the explicit expression can be computed by means of the usual expansion methods for determinants.

\section{Conclusions} 
\setcounter{equation}{0} 

In this paper we constructed a rich class of elliptic solutions of a family of Boussinesq type partial difference equations, in particular the 
lattice BSQ equation \eqref{eq:BSQ}, the lattice modified BSQ equations \eqref{eq:modified BSQ-v} and \eqref{eq:modified BSQ-w}, and the lattice Schwarzian BSQ equation 
\eqref{eq:lsBSQ}, while the most general equation is the lattice BSQ form \eqref{eq:NQC-BSQ}. The various equations, which in their scalar form 
are difference equations on a 9-point stencil, can also be written in their multi-component forms, see subsection 7.1. From the construction of the 
solutions, using the Direct linearisation (DL) approach with an elliptic Cauchy kernel, are in their most general form arising from the integral equations 
\eqref{eq:inteqBSQ}, where the main quantities are constructed from the entries of the infinite matrix \eqref{eq:UBSQ}. In the simplest case, when the 
measures and integration contours lead to  contributions from bound states, we recover elliptic seed and $N$-soliton solutions as described in subsection 7.2, 
but for more general measures and contours the DL formulae also comprise in principle also the formulae for the inverse scattering type solutions 
which asymptotically tend to elliptic seed solutions. The structures were developed from dimensional reductions of the corresponding KP class of elliptic 
solutions (cf. \cite{Y-KN} for the  corresponding elliptic soliton solutions), but whereas the reduction from the KP class to the KdV class is relatively straightforward, 
the one to the BSQ class (and higher Gel'fand-Dikii systems) requires in the elliptic case the novel notion of the elliptic cube (viz. $N^{\rm th}$) root of unity, 
which was introduced and treated in section 3. The latter notion may well have significance beyond the confines of the present application to solutions of 
integrable lattice equations. 

Whereas we restricted ourselves in this paper to solutions of the BSQ class of lattice systems, the main underlying structural relations  were given for 
the entire lattice Gel'fand-Dikii hierarchy, and thus the results obtained may be readily generalised to the higher 
order systems in this hierarchy, although the development of the explicit formulae and corresponding computations may fast become quite overwhelming. 
What the present paper has made clear is that the KdV case ($N=2$) of the GD hierarchy is not really representative of what goes on in the general case, while the 
BSQ class ($N=3$) provides a good insight into the real challenges that are encountered in the present kind of constructions. Thus, in spite of the 
importance of the famous ABS classification results of \cite{ABS}, the present work demonstrates that there is 'integrable life beyond ABS' and that in many respects the results on 
the latter systems do not really provide the necessary insights for multicomponent and higher-order integrable lattice systems. 

A particular generalisation that we have not considered (yet) in the present paper, is that of the so-called \textit{extended BSQ systems}. These involve parameter-generalisations 
of the lattice BSQ systems of \cite{GD} found by Hietarinta in \cite{H}. It was shown in \cite{ZZN}, in the rational case, how these parameter generalisations 
fit into the DL framework, thereby providing a rich class of solutions of those extended BSQ systems. In fact, those extensions arise from an unfolding of 
the dispersion curve from $G(k,\oa)=\oa^3-k^3=0$ to $G(k,\oa)=k^3-\oa^3+\alpha_2(\oa^2-k^2)+\alpha_1(\oa-k)=0$ where the parameters $\alpha_1$ and $\alpha_2$ 
are the extension parameters, the inclusion of which effectively embeds the ($N=2$) KdV system into the ($N=3$) BSQ system. 
In the elliptic case, this extension would require an unfolding of the elliptic cube root of unity condition, to a more general condition on the roots 
$\oa_i(\delta)$, ($i=0,1,\cdots,N-1$), 
namely 
\[ \prod_{j=0}^{N-1}\Phi_\kp(\oa_j(\delta))=\sum_{j=0}^{N-2} \alpha_j\left(\wp^{(j)}(-\kp)-\wp^{(j)}(\delta)\right)\  ,  \] 
(possibly with $\alpha_{N-2}=1$) but we leave the development of the relevant explicit formulae involving the extension parameters 
$\alpha_j$ corresponding to a future publication.

\section*{Acknowledgements} 
This project is supported by the NSF of China (Nos.11875040 and 11631007). FWN acknowledges the hospitality of the 
Department of Mathematics of Shanghai University for various visits during which this paper was initiated and finalised.


\begin{appendix}

\subsection*{Appendix A:  Formulae for elliptic functions}
\def\theequation{A.\arabic{equation}}
\setcounter{equation}{0}

Here, we collect some useful formulae for elliptic functions, see also
the standard textbooks e.g. \cite{WW}.
The Weierstrass sigma-function is defined by
\be
\sigma(x) = x \prod_{(k,\ell) \ne (0,0)} (1-\frac{x}{\omega_{k\ell}})
 \exp\left[ \frac{x}{\omega_{k\ell}} + \frac{1}{2}
( \frac{x}{\omega_{k\ell}})^2\right]\ ,
\ee
with $\oa_{kl}=2k\oa_1 + 2\ell \oa_2$ and
$2\omega_{1,2}$  being a fixed pair of the primitive periods.
The relations between the Weierstrass elliptic functions are given by
\be
\zeta(x) =  \frac{\sigma^\prime(x)}{\sigma(x)}\   \  , \   \
\wp(x) = - \zeta^\prime(x)\   ,
 \ee
where $\sg(x)$ and $\zeta(x)$ are odd functions and $\wp(x)$ is an
even function of its argument.
We recall also that the $\sg(x)$ is an entire function, and
$\zeta(x)$ is a meromorphic function having simple poles at
$\omega_{kl}$, both being quasi-periodic, obeying
\[
\zeta(x+2\omega_{1,2}) = \zeta(x) + 2\xi_{1,2}\    \ ,\    \
\sigma(x+2\omega_{1,2}) = -\sigma(x)
e^{2\xi_{1,2}(x+\omega_{1,2})}\  , 
\]
in which $\xi_{1,2}$ satisfy ~$\xi_1\omega_2 - \xi_2\omega_1
= \frac{\pi i}{2}$~, whereas $\wp(x)$ is doubly periodic.
{}From an algebraic point of view, the most important property of
these elliptic functions is the existence of a number of functional
relations, the most fundamental being
\be
\label{eq:zs}
\zeta(\ar) + \zeta(\bb) + \zeta(\gm) - \zeta(\ar +\bb + \gm)
  = \frac{  \sigma(\ar + \bb )  \sigma(\bb + \gm )
\sigma( \gm + \ar) }{ \sigma(\ar) \sigma(\bb) \sigma(\gm)
\sigma(\ar + \bb + \gm )}~ ,
\ee
which can also be cast into the following form
\be\label{eq:12}
\Phi_\kp(x)\Phi_\kp(y) =
\Phi_{\kappa}(x+y)\left[ \zeta(\kp) +\zeta(x) +\zeta(y) -\zeta(\kp +x+y)
\right] \   .
\ee
The well-known three-term relation for $\sigma(x)$ is a consequence of
(\ref{eq:zs})
\bea
\sigma(x+y) \sigma(x-y) \sigma(a+b) \sigma(a-b) &=& \sigma(x+a) \sigma(x-a)
\sigma(y+b) \sigma(y-b)  \nn  \\
&& ~ - \sigma(x+b) \sigma(x-b) \sigma(y+a) \sigma(y-a)\   ,  \label{eq:8}
\eea
and this equation can be cast into the following convenient form
\be\label{eq:14}
\Phi_\kappa(x)\Phi_\ld(y) =
\Phi_{\kappa}(x-y)\Phi_{\kappa+\ld}(y) + \Phi_{\kappa +\ld}(x)
\Phi_{\ld}(y-x)\   ,
\ee
which is obtained from the elliptic analogue of the partial
fraction expansion, i.e. eq. (\ref{eq:12}). Another important
relation is given by
\bea
&&\left[ \zeta(\ar)+\zeta(\bb)+\zeta(\gm)-\zeta(\ar+\bb+\gm)\right]\,
\left[ \zeta(\ar+\bb)-\zeta(\bb)-\zeta(\ar+\gm)
+\zeta(\gm)\right]  \nn \\
&&~~~~ = -\Phi_\gm(\bb)\Phi_{-\gm}(\bb)= \wp(\gm)-\wp(\bb)\   ,
\label{eq:21}
\eea
as well as
\be\label{eq:22}
\eta_\alpha(\beta):=\zeta(\ar+\bb) -\zeta(\ar)-\zeta(\bb)=\frac{1}{2}
\frac{\wp^\prime(\ar)-\wp^\prime(\bb)}{\wp(\ar)-\wp(\bb)}\   .
\ee
The well-known addition formula for the Weierstrass elliptic function
can be written in the form
\be\label{eq:addform}
\eta_\alpha(\beta)^2=\Big(\zeta(\ar+\bb) -\zeta(\ar)-\zeta(\bb)\Big)^2= \wp(\ar)+\wp(\bb)+\wp(\ar+\bb)\  , 
\ee
which together with the relation 
\be\label{eq:FSrel} 
\left| \begin{array}{ccc} 
1 & \wp(\ar) & \wp'(\ar) \\ 1 & \wp(\beta) & \wp'(\beta) \\ 1 & \wp(\gm) & \wp'(\gm) \end{array}
\right| =0 \ , \quad {\rm provided}\quad \ar+\beta+\gm\equiv 0({\rm mod\ \ root\ \ lattice )}\ , 
\ee 
characterises the addition on the elliptic curve. 
An important special relation used in the derivation is:
\bea\label{eq:special}
&&\frac{\zeta(\xi)+\zeta(\ar)+\zeta(\ld)-\zeta(\xi+\ar+\ld)}{
\zeta(\xi)+\zeta(\bb)+\zeta(\ld)-\zeta(\xi+\bb+\ld)} \nn \\
&& ~~ +
\frac{\zeta(\xi+\ar)-\zeta(\ar)+\zeta(\bb)-\zeta(\xi+\bb)}{
\zeta(\xi+\ar)+\zeta(\ld)+\zeta(\bb)-\zeta(\xi+\ar+\bb+\ld)}
= 1\  , \nn \\
\eea
which is used in the derivation of the various versions of
the discrete KP equation for elliptic solutions.
Furthermore, the following identity used several times in the BSQ reduction case: 
\be\label{eq:specialBSQ} 
\wp(\alpha)+\wp(\beta)+\wp(\lambda+\alpha)-\eta_{-\alpha}(\lambda+\alpha)\eta_\beta(\lambda+\alpha)
=\frac{1}{2} \frac{\wp'(\alpha)-\wp'(\beta)}{\eta_\alpha(\lambda)-\eta_\beta(\lambda)}\ , 
\ee 
which can be deduced by using \eqref{eq:21}-\eqref{eq:addform}.

\subsection*{Appendix B: Derivation of the fundamental KP relations}
\def\theequation{B.\arabic{equation}}
\setcounter{equation}{0}

Here we derive the fundamental relations of subsection 2.1, eqs. \eqref{eq:tau}-\eqref{eq:ukrels}. Let 
$\bC$ be an infinite matrix as in \eqref{eq:C}, obeying the linear relations \eqref{eq:CC}. Let us 
introduce the infinite matrix $\bU^0_\xi$ defined by the relation 
\be\label{eq:U0} 
\bU^0_\xi=\left( \boldsymbol{1}-\bU^0_\xi\, \bOm_\xi\right)\bC\  , \ee 
where the infinite matrix $\bOm_\xi$ is given by 
\be\label{eq:bOm} 
\bOm_\xi= {\Large :}\,\Phi_\xi(\Ld+\tLd)\,\bO\,{\Large :}\  , 
\ee 
where $\Phi_\xi(x)$ is the function \eqref{eq:Phi}. Eq. \eqref{eq:bOm} should be viewed as 
a formal series in both the operators $\Ld$ and $\tLd$ where the normal ordering symbol ~${\Large :}
~{\Large :}$~ means that all operators $\tLd$ are ordered on the left of the projection matrix 
$\bO$ and all operators $\Ld$ on the right of $\bO$. Then, as a consequence of the addition formula 
\eqref{eq:14}, we have the operator identity \eqref{eq:Omega}. 

Since in \eqref{eq:U0} the dependence of $\bU^0_\xi$ only stems from the matrix $\bOm_\xi$ and the 
dependence on the discrete variables $n,m,l$ via the matrix $\bC$, we can derive the elementary 
shift relation of $\bU^0_\xi$ as follows: 
\begin{eqnarray*} 
\wt{\bU}^0_\xi\,\Phi_\delta(-\tLd) &=& \left( \boldsymbol{1}-\wt{\bU}^0_\xi\, \bOm_\xi\right) 
\Phi_\delta(\Ld)\bC  \\ 
&=& \Phi_\delta(\Ld)\,\bC-\wt{\bU}^0_\xi\,\left[ \Phi_\delta(-\tLd)\,\bOm_{\xi+\dd}+ 
\Phi_\xi(\tLd)\bO\Phi_{\xi+\dd}(\Ld)\right]\bC \\ 
\Rightarrow &&  \wt{\bU}^0_\xi\,\Phi_\delta(-\tLd)\left(\boldsymbol{1}+\bOm_{\xi+\delta}\bC\right) 
=\Phi_\delta(\Ld)\,\bC-\wt{\bU}^0_{\xi}\Phi_\xi(\tLd)\bO\Phi_{\xi+\delta}(\Ld)\bC\  , \\ 
\Rightarrow &&  \wt{\bU}^0_\xi\,\Phi_\delta(-\tLd) 
=\left[\Phi_\delta(\Ld)-\wt{\bU}^0_{\xi}\Phi_\xi(\tLd)\bO\Phi_{\xi+\delta}(\Ld)\right]\bC
\left(\boldsymbol{1}+\bOm_{\xi+\delta}\bC\right)^{-1}\  , \\
\end{eqnarray*} 
and since from \eqref{eq:U0} we have 
\[ \bC\left(\boldsymbol{1}+\bOm_{\xi+\delta}\bC\right)^{-1}=\bU^0_{\xi+\delta}\ , \]
we arrive at: 
\be\label{eq:U0dyn} 
\wt{\bU}^0_\xi\,\Phi_\delta(-\tLd)=\left[\Phi_\delta(\Ld)-\wt{\bU}^0_{\xi}\Phi_\xi(\tLd)\bO\Phi_{\xi+\delta}(\Ld)\right]\bU^0_{\xi+\delta}\  . 
\ee  
Multiplying \eqref{eq:U0dyn} from the left by $\Phi_\xi(\Ld)$ and from the right by 
$\Phi_{\xi+\delta}(\tLd)$ and setting 
\[ \Phi_\xi(\Ld)\bU^0_\xi\Phi_\xi(\tLd)=:\bU_\xi  \, \] 
in accordance with \eqref{eq:U-al}, we get 
\be\label{eq:PhiU0} \wt{\bU}_\xi \frac{\Phi_\delta(-\tLd)\Phi_{\xi+\delta}(\tLd)}{\Phi_\xi(\tLd)} = 
\frac{\Phi_\xi(\Ld)\Phi_\delta(\Ld)}{\Phi_{\xi+\delta}(\Ld)}\bU_{\xi+\delta}-\wt{\bU}_\xi\bO 
\bU_{\xi+\delta}\  . \ee
The coefficients in the first and second term of \eqref{eq:PhiU0} can be cast in the form 
of the functions $-\chi^{(1)}_{-\delta,\xi+\delta}(\tLd)$ and $\chi^{(1)}_{\delta,\xi}(\Ld)$ respectively. 

Next, we derive the relations for infinite vectors 
\be\label{eq:uk0}
\bu_\kp^0(\xi):= \left( \boldsymbol{1}-\bU^0_\xi\, \bOm_\xi\right)\bc_\kp\rho_\kp\ , \quad 
\tuk^0:=\sg_{\kp'}\tck\left( \boldsymbol{1}-\bOm_\xi\bU^0_\xi\right)\ , 
\ee 
with $\rho_\kp$ and $\sg_{\kp'}$ obeying \eqref{eq:rho}. Deriving the shift relation for 
$\bu^0_\kp(\xi)$ proceeds as follows 
\begin{eqnarray*} 
\wt{\bu}^0_\kp(\xi) &=& \left( \boldsymbol{1}-\wt{\bU}^0_\xi\, \bOm_\xi\right) 
\Phi_\delta(\Ld)\bc_\kp\rho_\kp  \\ 
&=& \Phi_\delta(\Ld)\,\bc_\kp\rho_\kp-\wt{\bU}^0_\xi\,\left[ \Phi_\delta(-\tLd)\,\bOm_{\xi+\dd}+ 
\Phi_\xi(\tLd)\bO\Phi_{\xi+\dd}(\Ld)\right]\bc_\kp\rho_\kp \\ 
&=& \Phi_\delta(\Ld)\,\bc_\kp\rho_\kp-\wt{\bU}^0_\xi\,\Phi_\xi(\tLd)\bO\Phi_{\xi+\dd}(\Ld)\bc_\kp\rho_\kp \\ 
&& \qquad -\left[\Phi_\delta(\Ld)\bU^0_{\xi+\delta}-\wt{\bU}^0_\xi\Phi_\xi(\tLd)\bO
\Phi_{\xi+\delta}(\Ld)\bU^0_{\xi+\delta} \right] \bOm_{\xi+\delta}\bc_\kp\rho_\kp\   \\ 
&=& \Phi_\delta(\Ld)\,\bu^0_\kp(\xi+\delta)-\wt{\bU}^0_\xi\Phi_\xi(\tLd)\bO
\Phi_{\xi+\delta}(\Ld)\bu^0_\kp(\xi+\delta)\  , 
\end{eqnarray*} 
where in the third step we have made use of \eqref{eq:U0dyn}. Setting 
\[ \Phi_\xi(\Ld)\bu^0_\kp(\xi)=: \bu_\kp(\xi)\ , \quad {\rm and \ \ similarly}\quad 
\tuk^0(\xi)\,\Phi_\xi(\tLd)=:\tuk(\xi)\  , \] 
we obtain from the latter result 
\be\label{eq:uk0dyn} 
\wt{\bu}_\kp(\xi)=\frac{\Phi_\xi(\Ld)\Phi_\delta(\Ld)}{\Phi_{\xi+\delta}(\Ld)}\bu_\kp(\xi+\delta) 
-\wt{\bU}_\xi\bO\,\bu_\kp(\xi+\delta)\  , 
\ee 
where the coefficient in the first term on the r.h.s. equals $\chi_{\delta,\xi}^{(1)}(\Ld)$. 
In a similar way the shift relation for the adjoint vector $\tuk^0(\xi)$ can be derived, namely 
\begin{eqnarray*} 
\tuk^0(\xi+\delta) &=& \wt{\sg}_\kp \tck\Phi_\delta(-\tLd)
\left( \boldsymbol{1}-\bOm_{\xi+\delta}\bU^0_{\xi+\delta}\right)  \\  
&=& \wt{\sg}_{\kp'} \tck \Phi_\delta(-\tLd) - \wt{\sg}_{\kp'} \tck\left[ \bOm_\xi\Phi_\delta(\Ld) 
- \Phi_\xi(\tLd)\bO\Phi_{\xi+\dd}(\Ld)\right]\bU^0_{\xi+\delta} \\ 
&=& \wt{\sg}_{\kp'} \tck \Phi_\delta(-\tLd)+ \wt{\sg}_{\kp'} \tck\Phi_\xi(\tLd)\bO\Phi_{\xi+\dd}(\Ld)
\bU^0_{\xi+\delta}  \\ 
&& \qquad - \wt{\sg}_{\kp'} \tck \bOm_\xi\left[ \wt{\bU}^0_\xi\Phi_\delta(-\tLd)+
\wt{\bU}^0_{\xi}\Phi_\xi(\tLd)\bO\Phi_{\xi+\delta}(\Ld)\bU^0_{\xi+\delta} \right]   \\ 
&=& \wt{\tuk^0}(\xi)\Phi_\delta(-\tLd)+\wt{\tuk^0}(\xi)\Phi_\xi(\tLd)\bO
\Phi_{\xi+\delta}(\Ld)\bU^0_{\xi+\delta}\  .  
\end{eqnarray*} 
Multiplying the latter result from the right by $\Phi_{\xi+\delta}(\tLd)$ we get 
\be\label{eq:tuk0dyn} 
\tuk(\xi+\delta)=\wt{\tuk}(\xi)\frac{\Phi_\delta(-\tLd)\Phi_{\xi+\delta}(\tLd)}{\Phi_\xi(\tLd)}+ 
\wt{\tuk}(\xi)\bO\bU_{\xi+\delta}\  ,
\ee 
where the coefficient in the first terms on the r.h.s. is identified with $-\chi^{(1)}_{-\delta,\xi}(\tLd)$. 

Finally, we derive the relations from the $\tau$-function defined by \eqref{eq:tau}. Applying the 
$\wt{\phantom{a}}$-shift we have
\begin{eqnarray*} 
\wt{\tau_\xi} &=& \det\left( \bun+\bOm_{\wt{\xi}}\wt{\bC}\right) 
= \det\left( \bun +\bOm_{\wt{\xi}}\Phi_\delta(\Ld)\bC(\Phi_\delta(-\tLd))^{-1} \right) \\
&=& \det\left( \bun + \left[ \bOm_{\wt{\xi}}\Phi_\delta(\Ld)-\Phi_\delta(-\tLd)\bOm_\xi\right]
\bC(\Phi_\delta(-\tLd))^{-1} + \Phi_\delta(-\tLd)\bOm_\xi\bC(\Phi_\delta(-\tLd))^{-1}  \right) \\ 
&=& \det\left( \bun + \Phi_{\wt{\xi}}(\tLd)\bO\Phi_\xi(\Ld)\bC(\Phi_\delta(-\tLd))^{-1} 
+  \Phi_\delta(-\tLd)\bOm_\xi\bC(\Phi_\delta(-\tLd))^{-1} \right) \\ 
&=& \det\left(\bun+\bOm_\xi\bC+(\Phi_\delta(-\tLd))^{-1} \Phi_{\wt{\xi}}(\tLd)\bO\Phi_\xi(\Ld)\bC  \right) \\ 
&=& \det\left( \bun+\bOm_\xi\bC\right)\,\det\left(\bun+(\Phi_\delta(-\tLd))^{-1} \Phi_{\wt{\xi}}(\tLd)\bO\Phi_\xi(\Ld)\bC\left(\bun+\bOm_\xi\bC \right)^{-1}  \right) 
\end{eqnarray*} 
where we have subsequently used the shift relation \eqref{eq:CCa}, the relation \eqref{eq:Omega} for the Cauchy 
kernel and the invariance of the determinant under similarity transformations, as well as the 
definition \eqref{eq:U0}. Thus, we get 
\begin{eqnarray*}
\wt{\tau_\xi}/\tau_\xi &=& \det\left(\bun+(\Phi_\delta(-\tLd))^{-1} \Phi_{\wt{\xi}}(\tLd)\bO\Phi_\xi(\Ld)\bU^0_\xi \right) \\ 
&=& 1+ \left(\Phi_\xi(\Ld)\bU^0_\xi(\Phi_\delta(-\tLd))^{-1} \Phi_{\wt{\xi}}(\tLd) \right)_{0,0} \  , 
\end{eqnarray*} 
where the latter step uses the fact that $\bO$ is a rank 1 projector matrix and we apply the famous 
Weinstein-Aroszajn formula:
\[ \det\left(\bun+\boldsymbol{a}\,\boldsymbol{b}^t\right) = 1+\boldsymbol{b}^t\cdot\boldsymbol{a}\ , \] 
for determinants involving a rank 1 perturbation from the unit matrix.

\subsection*{Appendix C: Some higher-order elliptic identities}
\def\theequation{C.\arabic{equation}}
\setcounter{equation}{0}

The following higher addition rules were established in Appendix C of \cite{DN}, which for the sake of 
self-containedness we reiterate here. They may be used, in future, for deriving the 
higher reductions from the lattice KP system to the the lattice GD hierarchy, and they play a role in establishing the 
defining relations for the higher elliptic roots of unity.   
First, we have the following generalisation of \eqref{eq:zs} 
\begin{eqnarray*}
&& \quad\sigma(\kappa+x)\,\sigma(\lambda+x)\,\sigma(\mu+x)
\sigma(\kappa+\lambda+\mu+y)\,\sigma^2(y) \\
&& \quad\quad -\sigma(\kappa+y)\,\sigma(\lambda+y)\,\sigma(\mu+y)\sigma(\kappa+\lambda+\mu+x)\,\sigma^2(x) \\
&& = \sigma(\kappa)\,\sigma(\lambda)\,\sigma(\mu)\,\sigma(x)\,\sigma(y)\,\sigma(\kappa+\lambda+\mu+x+y)\,\sigma(y-x) \\
&& \quad\quad \times \left[\zeta(\kappa)+\zeta(\lambda)+\zeta(\mu)+\zeta(x)+\zeta(y)-\zeta(\kappa+\lambda+\mu+x+y) \right]
\end{eqnarray*}
which derives from:
\begin{eqnarray*}
&& \zeta(\kp)+\zeta(\ld)+\zeta(\mu)+\zeta(x)+\zeta(y)-\zeta(\kp+\ld+\mu+x+y)=  \\
&& = \frac{\Phi_\kp(x)\Phi_\ld(x)\Phi_\mu(x)\Phi_{\kp+\ld+\mu}(y) - \Phi_\kp(y)
\Phi_\ld(y)\Phi_\mu(y) \Phi_{\kp+\ld+\mu}(x)}
{\Phi_{\kp+\ld+\mu}(x+y)\,(\wp(x)-\wp(y))} \  .
\end{eqnarray*}

Furthermore, we have \cite{DN}
\begin{eqnarray}\label{eq:tripleprod}
\Phi_\kp(x)\Phi_\kp(y)\Phi_\kp(z)&=& \tfrac{1}{2}\Phi_\kp(x+y+z)\,
\left[ \left(\zeta(\kp)+\zeta(x)+\zeta(y)+\zeta(z)-\zeta(\kp+x+y+z)\right)^2 \right. \nn \\
                     && \left. \quad +\wp(\kp)-\left(\wp(x)+\wp(y)+\wp(z)+\wp(\kp+x+y+z) \right)\right] \  .
\end{eqnarray}

At the next level we get the following identity:
\begin{eqnarray}\label{eq:4ordrel}
\Phi_\kp(x)\Phi_\ld(y)\Phi_\mu(z)\Phi_\nu(w)&=&
\Phi_{\kp+\ld+\mu+\nu}(x)\Phi_\ld(y-x)\Phi_\mu(z-x)\Phi_\nu(w-x) \nn \\
&& + \Phi_\kp(x-y)\Phi_{\kp+\ld+\mu+\nu}(y)\Phi_\mu(z-y)\Phi_\nu(w-y)  \nn \\
&& + \Phi_\kp(x-z)\Phi_\ld(y-z)\Phi_{\kp+\ld+\mu+\nu}(z)\Phi_\nu(w-z)  \nn \\
&& + \Phi_\kp(x-w)\Phi_\ld(y-w)\Phi_\mu(z-w)\Phi_{\kp+\ld+\mu+\nu}(w)\ .    \nn
\end{eqnarray}
Setting $w=z+\delta$ and taking the limit $\delta\to 0$ in \eqref{eq:4ordrel} we obtain:
\begin{eqnarray}\label{eq:4ordprerel}
&& \Phi_\kp(x)\Phi_\ld(y)\Phi_\mu(z)\Phi_\nu(z)
-\Phi_{\kp+\ld+\mu+\nu}(x)\Phi_\ld(y-x)\Phi_\mu(z-x)\Phi_\nu(z-x) \nn \\
&& -\Phi_\kp(x-y)\Phi_{\kp+\ld+\mu+\nu}(y)\Phi_\mu(z-y)\Phi_\nu(z-y)
= \Phi_\kp(x-z)\Phi_\ld(y-z)\Phi_{\kp+\ld+\mu+\nu}(z) \nn \\
&& \times [\zeta(\nu)+\zeta(\mu)+\zeta(z)+\zeta(\kp+x-z)+\zeta(\ld+y-z)-\zeta(x-z)
-\zeta(y-z)  \nn \\
&&  -\zeta(\kp+\ld+\mu+\nu+z) ]\  , 
\end{eqnarray}
which, after a renaming of the arguments, can be cast into the following form:
\begin{eqnarray}\label{eq:4ordderrel}
&& \Phi_{\kp+\ld+\mu+\nu}(x+y+z)\,
\frac{\sg(x+y+z)\,\sg(x-y)\,\sg(x-z)\,\sg(y-z)}{\sg^3(x)\,\sg^3(y)\,\sg^3(z)} \nn \\
&& \times \left[\zeta(\kp)+\zeta(\ld)+\zeta(\mu)+\zeta(\nu)+\zeta(x)+\zeta(y)
+\zeta(z)-\zeta(\kp+\ld+\mu+\nu+x+y+z)\right]=~~~~~~~~  \nn \\
&=&  \Phi_\kp(x)\Phi_\ld(x)\Phi_\mu(x)\Phi_\nu(x)\left(\wp(z)-\wp(y)\right)\Phi_{\kp+\ld+\mu+\nu}(y+z) \nn \\
&&  + \Phi_\kp(y)\Phi_\ld(y)\Phi_\mu(y)\Phi_\nu(y)\left(\wp(x)-\wp(z)\right)\Phi_{\kp+\ld+\mu+\nu}(x+z) \nn \\
&&  + \Phi_\kp(z)\Phi_\ld(z)\Phi_\mu(z)\Phi_\nu(z)\left(\wp(y)-\wp(x)\right)\Phi_{\kp+\ld+\mu+\nu}(x+y)  \  .
\end{eqnarray}
In the next stage we set $z=y+\ven$ in \eqref{eq:4ordprerel}, and take the limit $\ven\to 0$, which yields the relation
\begin{eqnarray}\label{eq:4ord2derrel}
&& ~~~~\Phi_\kp(x)\Phi_\ld(y)\Phi_\mu(y)\Phi_\nu(y)-\Phi_{\kp+\ld+\mu+\nu}(x)
\Phi_\ld(y-x)\Phi_\mu(y-x)\Phi_\nu(y-x)= \nn \\
&&= \tfrac{1}{2}\Phi_{\kp+\ld+\mu+\nu}(y)\Phi_\kp(x-y)
\Big[ \wp(y)+\wp(x-y)-\wp(\ld)-\wp(\mu)-\wp(\nu) \nn \\
&&~~~ + \Big(\zeta(\ld)+\zeta(\mu)+\zeta(\nu)+\zeta(\kp+x-y)-\zeta(x-y)+\zeta(y)
-\zeta(\kp+\ld+\mu+\nu+y)\Big)^2 \nn \\
&&~~~~ -\wp(\kp+x-y)
-\wp(\kp+\ld+\mu+\nu+y)  \, \Big]  \  ,
\end{eqnarray}
which generalises \eqref{eq:tripleprod}. In the final stage, however, setting $y=x+\gm$ and letting $\gm\to 0$ (which requires expansions up to third order
in $\gm$), we get the identity
\begin{eqnarray}\label{eq:4ord3derrel}
&& \Phi_\kp(x)\Phi_\ld(x)\Phi_\mu(x)\Phi_\nu(x)= \nn \\
&=&  \tfrac{1}{6}\Phi_{\kp+\ld+\mu+\nu}(x)\Big\{ \Big(\zeta(\kp)
+\zeta(\ld)+\zeta(\mu)+\zeta(\nu)+\zeta(x)-\zeta(\kp+\ld+\mu+\nu+x)\Big)^3 \nn \\
&& -3\Big(\zeta(\kp)+\zeta(\ld)+\zeta(\mu)+\zeta(\nu)+\zeta(x)-\zeta(\kp+\ld+\mu+\nu+x)\Big) \nn \\
&& \qquad \times  \Big(\wp(\kp)+\wp(\ld)+\wp(\mu)+\wp(\nu)+\wp(\kp+\ld+\mu+\nu+x)-\wp(x)\Big)  \nn \\
&& -\Big(\wp'(\kp)+\wp'(\ld)+\wp'(\mu)+\wp'(\nu)+\wp'(x)-\wp'(\kp+\ld+\mu+\nu+x)\Big)  \Big\}\  .
\end{eqnarray}

The previous relations can be obviously generalised to arbitrary order. Thus, the general form of the basic identity \eqref{eq:14}
(3-term relation for the $\sg$-function, or the elliptic partial fraction expansion formula) is:
\begin{equation}
\prod_{i=1}^n\,\Phi_{\kp_i}(x_i) =
\sum_{i=1}^n\,\Phi_{\kp_1+\cdots+\kp_n}(x_i)\,\prod_{j=1\atop j\neq i}^n\,\Phi_{\kp_j}(x_j-x_i)\  .
\end{equation}
Extending this identity to $n+1$ variables, including a $\kp_0$ and $x_0$, and subsequently taking the
limit $x_0=x_1+\varepsilon$, with $\varepsilon\to 0$, we obtain the following identity (after some obvious
relabelling of parameters and changes of variables):
\begin{eqnarray*}
&& (-1)^{n-1}\Phi_{\kp_0+\kp_1+\cdots+\kp_n}(x_1+\cdots+x_n)\,
\frac{\sg(x_1+\cdots+x_n)}{\prod_{j=1}^n\,\sg(x_j)} \\
&& \times\left[ \zeta(\kp_0)+\sum_{j=1}^n\left(\zeta(\kp_j)+\zeta(x_j)\right)-
\zeta(\kp_0+\kp_1+\cdots+\kp_n+x_1+\cdots+x_n)\right] \\
&=& \sum_{i=1}^n\,\Phi_{\kp_0+\kp_1+\cdots+\kp_n}(x_1+\cdots+\slashed{x_i}+\cdots+x_n)\,
\frac{\sg(x_1+\cdots+\slashed{x_i}+\cdots+x_n)\,\sg^{n-1}(x_i)}{\prod_{j=1\atop j\neq i}^n\,\sg(x_i-x_j)}\,
\prod_{j=0}^n\,\Phi_{\kp_j}(x_i)\  .
\end{eqnarray*}
These identities are associated with the famous Frobenius-Stickelberger determinantal formula.


\subsection*{Appendix D:  Lax pairs}
\def\theequation{D.\arabic{equation}}
\setcounter{equation}{0}

The Lax pairs for the BSQ type lattice equations are derived from the fundamental system of relations 
\eqref{eq:fundBSQsyst-b} and \eqref{eq:fundBSQsyst-d}, or equivalently \eqref{2ndfundBSQsyst-b} and 
\eqref{2ndfundBSQsyst-d}, for the vectors $\buk(\xi)$ together with \eqref{eq:BSQalgconstrsb}, 
and their counterparts in the other lattice direction. By choosing specific components of these infinite 
vectors we construct 3-component vectors which will yield the eigenvectors for the corresponding 3$\times$3 
matrix Lax pairs. There are two different choices of entries that we will consider, leading to the 
Lax pairs for the lattice BSQ system and lattice modified BSQ system respectively. 

\paragraph{a) Lax pair for the lattice BSQ system} $\phantom{a}$ \\ 
\vspace{.2cm} 

\noindent 
In this case we specify the entries: 
\[
 (u_\kp(\xi))_0:= \left(\buk(\xi) \right) _0\  , \quad (u_\kp(\xi))_1:=
 \left(\Ld_\xi\buk(\xi)\right)_0\  ,
\quad  (u_\kp(\xi))_2:=\left(\wp(\Ld)\buk(\xi)\right)_0\  ,
 \]
and consider the relations \eqref{eq:fundBSQsyst}, together with
\eqref{eq:BSQalgconstrs}, to derive a Lax pair for the 3-component vector
\be\label{eq:Laxphi} 
\bphi(\xi):= \Big((u_\kp(\xi))_0\,,\,(u_\kp(\xi))_1\,,\,(u_\kp(\xi))_2\Big)\   .
\ee 
In fact, the dynamical relations \eqref{eq:fundBSQsyst}, using the above notation, can be rewritten as:
\bse
\bea
&& \wt{\buk(\xi)} = \Big( p_\xi-\Ld_\xi-\wt{\bU_\xi}\bO\Big)\,\buk(\xi) \  , \label{rel:Lax-pairBSQ} \\
&& \Ld_{\wt{\xi}}\wt{\buk(\xi)} = \Big( p_\xi\Ld_\xi-\wp(\wt{\xi})-\wp(\xi)
-\wp(\Ld)\Big)\buk(\xi)-\Ld_{\wt{\xi}}\wt{\bU_\xi}\,\bO\,\buk(\xi)\  , \\
&& -\tfrac{1}{2}\Big(\wp'(\dd)+\wp'(\kp)\Big)\buk(\xi)=\Big(\wp(\wt{\xi})
+\wp(\dd)+\wp(\Ld)+p_\xi\Ld_{\wt{\xi}}\Big)\wt{\buk(\xi)} \nn \\ 
&& \qquad\qquad \qquad \qquad\qquad\qquad + \bU_\xi\big(p_\xi\bO
+\bO\Ld_{\wt{\xi}}-\tLd_\xi\bO\Big)\wt{\buk(\xi)} \  ,
\eea
\ese
and similar relations for the other lattice direction. Taking the 0-component of these relations and 
their counterparts in the other direction we constitute the Lax pair as the shift
relations for the vector $\bphi(\xi)$, as given by 
\bse
\be
\label{eq:Lax-pairBSQ}
\wt{\bphi(\xi)}=\bL_\kp(\xi)\bphi(\xi)\  , \qquad
\wh{\bphi(\xi)}=\bM_\kp(\xi)\bphi(\xi)\  ,
\ee
with Lax matrices
$\bL_\kp(\xi)$ and $\bM_\kp(\xi)$ given by
\be
\label{eq:Lax-matricesBSQ}
 \bL_\kappa(\xi):= \left(
\begin{array}{ccc}
p_\xi-\wt{u_{0,0}} & -1 & 0 \\
-\wp(\wt{\xi})-\wp(\xi)-\wt{u_{1,0}} & p_\xi  & -1 \\
\clubsuit  & -\wp(\xi)-u_{0,1} & p_\xi+u_{0,0}
\end{array}\right)\  ,
\ee
\ese
in which
\begin{eqnarray*}
 \clubsuit &:=& -\tfrac{1}{2}\Big(\wp'(\dd)+\wp'(\kp)\Big)+(p_\xi-\wt{u_{0,0}})\left(u_{0,1}-\wp(\wt{\xi})-\wp(\dd)-p_\xi u_{0,0}\right)\\
&& +(p_\xi+u_{0,0})\left(\wp(\wt{\xi})+\wp(\xi)+\wt{u_{1,0}}\right) \  ,
\end{eqnarray*}
and $\bM_\kp(\xi)$ similarly  replacing $\wt{\phantom{a}}$ by $\wh{\phantom{a}}$ and $\delta$ by $\ven$ everywhere.
The entry $\clubsuit$ is such that $\det(\bL_\kp(\xi))=-\tfrac{1}{2}(\wp'(\dd)+\wp'(\kp))$~.
The compatibility of the Lax pair, ~$ \wt{\bL_\kp(\xi)}\bM_\kp(\xi)= \wh{\bM_\kp(\xi)} \bL(\kp(\xi)$~, 
yields the following coupled set of relations
\bse
\bea
\wt{u_{1,0}}-\wh{u_{1,0}} &=& \left(p_\xi-q_\xi+\wh{u_{0,0}}-\wt{u_{0,0}}\right)\wh{\wt{u_{0,0}}}
-p_{\wh{\xi}}\wh{u_{0,0}}+q_{\wt{\xi}}\wt{u_{0,0}}\  , \\
\wt{u_{0,1}}-\wh{u_{0,1}} &=& \left(p_\xi-q_\xi+\wh{u_{0,0}}-\wt{u_{0,0}}\right)u_{0,0}
-p_{\xi}\wt{u_{0,0}}+q_{\xi}\wh{u_{0,0}}\  , \\
\tfrac{1}{2}\frac{\wp'(\dd)-\wp'(\ven)}{ p_\xi-q_\xi+\wh{u_{0,0}}-\wt{u_{0,0}}} &=&
\wh{\wt{u_{1,0}}}+u_{0,1}+ (p_{\wh{\xi}}+q_\xi+u_{0,0})(\wh{\wt{u_{0,0}}}-p_\xi-q_{\wt{\xi}}) \nn  \\
&&  +
\tfrac{1}{2}\left(\wp(\xi)+\wp(\wh{\wt{\xi}})-\wp(\dd)-\wp(\ven)\right)
+\tfrac{1}{2}\left(p_\xi+q_{\wt{\xi}}\right)^2\ ,
\eea
\ese
where a number of identities are used, such as \eqref{eq:pq1} and \eqref{eq:pq2}, as well as 
\[
(p_{\wh{\xi}}+q_\xi)\left(\wp(\wh{\xi})-\wp(\wt{\xi}) \right)
=(p_\xi-q_\xi)\left(\wp(\wh{\wt{\xi}})-\wp(\xi) \right)
=(p_{\wh{\xi}}-p_\xi)\left(\wp(\dd)-\wp(\ven)\right)\  ,
\]
This coupled set of equations is similar to the one appearing in
\cite{DIGP,TN2} and by elimination of the quantities $u_{1,0}$ and
$u_{0,1}$ yields the lattice BSQ system \eqref{eq:lBSQ-u}. 
In principle, the Lax pair can be used to investigate initial-boundary value problems for 
the lattice BSQ equation in a similar was as the inverse scattering scheme for the 
BSQ equation. That approach to solutions is beyond the scope of the present paper.

\paragraph{b) Lax pair for the lattice modified BSQ system} $\phantom{a}$ \\ 
\vspace{.2cm} 

\noindent 
In the case of the lattice modified BSQ equation we can chose an eigenvector in the form  
\be\label{eq:Laxpsi} 
  \psi(\xi)=\left(\left(\,\left(\chi^{(1)}_{\alpha,\Ld}( \xi)\right)^{-1}\buk (\xi) \right)_0 \ , \left(\buk(\xi) \right) _0\  ,\left(\Ld_\xi\buk(\xi)\right)_0 \right )^T\ ,  
  \ee 
depending on the additional parameter $\alpha$. For this choice of eigenvector we can make use of the 
identities \eqref{eq:chichi} and \eqref{eq:wpchi-a} to derive the corresponding relations for the entries of 
$\psi_\kp(\xi)$ from the fundamental relations \eqref{eq:fundBSQsyst-b} and \eqref{eq:fundBSQsyst-d}. 
Thus, we can derive the Lax pairs 
\bse \label{eq:modified BSQLax} 
\be \label{eq:modified BSQLax-a} 
\wt{\psi(\xi)}=\mathcal{L}_\kp(\xi)\psi(\xi),~~\wh{\psi(\xi)}=\mathcal{M}_\kp(\xi) \psi(\xi) 
\ee 
where the modified BSQ Lax matrices are given by 
 \be\label{eq:modified BSQLax-b} 
\mathcal{L}_\kp(\xi)=
\left( \begin{array}{ccccc}
p_\xi + a_{\wt{\xi}} && \wt{v_\alpha (\xi)} && 0 \\
0 && p_\xi-\wt{u_{0,0}} && -1 \\
\tfrac{1}{2}\frac{\wp'(\alpha)+\wp'(\kappa)}{{v}_\alpha (\xi)} && \spadesuit && p_\xi-\frac{s_\alpha(\xi)}{v_\alpha(\xi)}
\end{array}\right)\ ,
\ee in which
\be \spadesuit =-(p_\xi-\wt{u_{0,0}})\left(p_\xi-\frac{s_\alpha(\xi)}{v_\alpha(\xi)}\right)-\tfrac{1}{2} \frac{\wp'(\dd)-\wp'(\alpha)}{p_\xi+a_{\wt{\xi}} }\frac{\wt{v_\alpha(\xi)}}{{v_\alpha(\xi)}}\ ,
\ee
\ese 
and similarly $\mathcal{M}_\kp$ obtained from \eqref{eq:modified BSQLax-b} by replacing $\delta$ by $\ven$ (and hence 
$p_\xi$ by $q_\xi$) and $\wt{~}$ by $\wh{~}$. The quantity indicated by $\spadesuit$ is such that 
~$\det(\mathcal{L}_\kp)=-\tfrac{1}{2}(\wp'(\delta)+\wp'(\kp))\wt{v_\alpha}/v_\alpha$~ and 
~$\det(\mathcal{M}_\kp)=-\tfrac{1}{2}(\wp'(\ven)+\wp'(\kp))\wh{v_\alpha}/v_\alpha$~. 

The compatibility relation ~$\wh{\mathcal{L}_\kp}\,\mathcal{M}_\kp=\wt{\mathcal{M}_\kp}\,\mathcal{L}_\kp$~ leads to 
the relations \eqref{eq:wwuu-a}, \eqref{eq:vvus-a} and \eqref{eq:wwss-a}, as well as to the relation 
\be\label{eq:compat2} 
A_\delta(\wh{\wt{\xi}})\frac{\wh{\wt{v_\alpha}}}{\wh{v_\alpha}}-
A_\ven(\wh{\wt{\xi}})\frac{\wh{\wt{v_\alpha}}}{\wt{v_\alpha}}= 
\left(p_{\wh{\xi}}+q_\xi-\wh{\wt{u_{0,0}}}-\frac{s_\alpha}{v_\alpha}\right) 
\left(p_{\wh{\xi}}-q_{\wt{\xi}}+\frac{\wt{s_\alpha}}{\wt{v_\alpha}}-
\frac{\wt{s_\alpha}}{\wt{v_\alpha}}\right)\  , 
\ee 
which can be obtained from combining \eqref{relations_strz-r} and \eqref{eq:rvs}. The relations thus obtained 
from the Lax pair allow us to reconstruct the lattice modified BSQ equation \eqref{eq:lmodified BSQ}. Similarly, a Lax pair can be set up 
for the modified BSQ equation in the form \eqref{eq:modified BSQ-w}, starting from the relations in section 4  for the adjoint vector $\tuk(\xi)$, and involving the 
quantities $w_\beta$. We will omit the details here.

\end{appendix}

\end{document}